\documentclass[preprint2]{emulateapj}

\usepackage{longtable}
\usepackage{cancel}	
\usepackage{color}
\usepackage{graphicx, subfigure}

\usepackage{placeins}
\usepackage{amsmath}
\usepackage{tabulary}
\usepackage{natbib}

\def\lesssim{\mathrel{\hbox{\rlap{\hbox{\lower4pt\hbox{$\sim$}}}\hbox{$<$}}}}
\def\gtrsim{\mathrel{\hbox{\rlap{\hbox{\lower4pt\hbox{$\sim$}}}\hbox{$>$}}}}

\newcommand{\grizy}{\ensuremath{grizy_{\rm P1}}}

\newcommand{\PS}{\protect \hbox {Pan-STARRS1}}

\shorttitle{The Pan-STARRS1 Database and Data Products}
\shortauthors{H. A. Flewelling}

\begin{document}
\title{The Pan-STARRS1 Database and Data Products}
\author{
H.~A.~Flewelling\altaffilmark{1}, 
E.~A.~Magnier\altaffilmark{1}, 
K.~C.~Chambers\altaffilmark{1},
J.~N.~Heasley\altaffilmark{8},
C.~Holmberg\altaffilmark{1},
M.~E.~Huber\altaffilmark{1},
W.~Sweeney\altaffilmark{1}, 
C.~Z.~Waters\altaffilmark{1},
A.~Calamida\altaffilmark{4},
S.~Casertano\altaffilmark{4},
X.~Chen\altaffilmark{10},
D.~Farrow\altaffilmark{5}
G.~Hasinger\altaffilmark{1},
R.~Henderson\altaffilmark{11},
K.~S.~Long\altaffilmark{4},
N.~Metcalfe\altaffilmark{2},
G.~Narayan\altaffilmark{4},
M.~A.~Nieto-Santisteban\altaffilmark{4},
P.~Norberg\altaffilmark{6,7},
A.~Rest\altaffilmark{4},
R.~P.~Saglia\altaffilmark{5},
A.~Szalay\altaffilmark{3},
A.~R.~Thakar\altaffilmark{3},
J.~L.~Tonry\altaffilmark{1}, 
J.~Valenti\altaffilmark{4},
S.~Werner\altaffilmark{3},
R.~White\altaffilmark{4},
L.~Denneau\altaffilmark{1},
P.~W.~Draper\altaffilmark{2},
K.~W.~Hodapp\altaffilmark{1},
R.~Jedicke\altaffilmark{1},
N.~Kaiser\altaffilmark{1},
R.~P.~Kudritzki\altaffilmark{1},
P.~A.~Price\altaffilmark{9},
R.~J.~Wainscoat\altaffilmark{1},
P.~S.~Builders\altaffilmark{PS1},
S.~Chastel\altaffilmark{1},
B.~McLean\altaffilmark{4},
M.~Postman\altaffilmark{4},
B.~Shiao\altaffilmark{4}.
}

\altaffiltext{1}{Institute for Astronomy, University of Hawaii, 2680 Woodlawn Drive, Honolulu, Hawaii 96822, USA}
\altaffiltext{2}{Department of Physics, Durham University, South Road, Durham DH1 3LE, UK}
\altaffiltext{6}{Institute for Computational Cosmology, Department of Physics, Durham University, South Road, Durham DH1 3LE, UK}
\altaffiltext{7}{Centre for Extragalactic Astronomy,  Department of Physics, Durham University, South Road, Durham DH1 3LE, UK}
\altaffiltext{3}{Department of Physics and Astronomy, The Johns Hopkins University, 3400 North Charles Street, Baltimore, MD 21218, USA}
\altaffiltext{4}{Space Telescope Science Institute, 3700 San Martin Drive, Baltimore, MD 21218, USA}
\altaffiltext{5}{ Max-Planck Institut f\"ur extraterrestrische Physik, Giessenbachstra\ss e 1, D-85748 Garching, Germany}
\altaffiltext{8}{Back Yard Observatory, P.O. BOX 68856, Tucson, AZ 85737}
\altaffiltext{9}{Department of Astrophysical Sciences, Princeton University, Princeton, NJ 08544, USA}
\altaffiltext{10}{Google Inc., 1600 Amphitheatre Pkwy, Mountain View, CA 94043}
\altaffiltext{11}{Spire Global, Sky Park 5,45 Finnieston Street, Glasgow, G3 8JU, UK }
\altaffiltext{PS1}{Pan-STARRS1 Builders}






\begin{abstract}
\noindent This paper describes the organization of the database and the catalog data products from the \PS\ $3\pi$ Steradian Survey. The catalog data products are available in the form of an SQL-based relational database from MAST, the Mikulski Archive for Space Telescopes at STScI. The database is described in detail, including the construction of the database, the provenance of the data, the schema, and how the database tables are related. Examples of queries for a range of science goals are included. The catalog data products are available in the form of an SQL-based relational database from MAST, the Mikulski Archive for Space Telescopes at STScI.
\end{abstract}


\section{Introduction}\label{sec:introduction}

\noindent This is the sixth in a series of seven papers that describe the Pan-STARRS1 Surveys, the data processing algorithms, calibration,
and the resulting data products. \citet[][Paper I]{Chambers2017} describe the \PS\ Surveys, an overview of the \PS\ System, the resulting image and catalog data products, a discussion of the overall data quality and basic characteristics, and a summary of important results. \citet[][Paper II]{Magnier2017a} describe how the various data processing stages of the Pan-STARRS Image Processing Pipeline (IPP) are organised and implemented, \citet[][Paper III]{Waters2017} describe the details of the pixel processing algorithms, including detrending, warping, and adding (to create stacked images) and subtracting (to create difference images) and resulting image products and their properties. \citet[][Paper IV]{Magnier2017b} describe the details of the source detection and photometry, including point-spread-function and extended source fitting models, and the techniques for forced photometry measurements. \citet[][Paper V]{Magnier2017c} describe the final calibration process, and the resulting photometric and astrometric quality. This paper (Paper VI) describes the \PS\ database, the data products, and details of their organization in the \PS\ database. 
\citet[][Paper VII]{Huber2017} will describe the Medium Deep Survey in detail, including the unique issues and data products specific to that survey.

\section{Background}
\label{sec:background}
The \PS\ Project teamed with Alex Szalay's database development group at The Johns Hopkins University to undertake the task of providing a publicly accessible hierarchical database for \PS\ \citep{Heasley2008}. The JHU team was the major developer of the SDSS database \citep{Thakar2003}, and it is useful to reuse as much of the software developed for the SDSS as possible. However, due to the Pan-STARRS's data having a larger intrinsic size and more complicated dataset, which covers a larger area of sky than SDSS and which includes measurements on the stacks, single exposures, and mean properties of each, major changes were required. The system developed is called the {\em Published Science Products Subsystem}, or PSPS \citep{Heasley2006}.

The key to moving from the SDSS database to a system capable of dealing with the large volume of \PS\ data is the design of the Data Storage layers. It was immediately clear that a single monolithic database design (like SDSS) would not work for the challenges posed by the PS1 data. Our approach has been to use several features available within the Microsoft SQL Server product line to implement a system that would meet our requirements. While SQL Server does not have (at present) a cluster implementation, a bespoke version can be crafted using a combination of distributed partition views and slices~\citep{Heasley2008}. Partitioning data into smaller databases spread over multiple server machines allows us to still treat the information as a unified table (from the users' perspective). Further, by staying with SQL Server we retain a wealth of software tools developed for SDSS, including the use of Hierarchical Triangular Mesh \citep{Szalay2007}.

\section{Data Releases}

\noindent This paper covers multiple data releases.  The first several \PS\ data releases will be dedicated to the 3$\pi$ survey, covering the sky North of $ -30\deg $ \citep[See ][for a description of the various surveys carried out with the PS1 telescope]{Chambers2017}, with plans to release the Medium Deep Survey (MD) data thereafter.  The first \PS\ data release (DR1, Dec 2016) covers the 3$\pi$ {\em stack} images and the static sky catalog.  The available pixel data products for DR1 include the PS1 {\em stack} image products from the 3$\pi$ survey.  The DR1 image products are deep {\em stacked} images along with ancillary data including signal, masks, variance, and number maps.  The available catalog data products, called Published Science Products Subsystem (PSPS), include the PS1 static sky 3$\pi$ catalog. Source properties are organized into several tables, as described in Table~\ref{table:pspstables}; only tables referring to the static results, without time domain information, are included in DR1. 

Data Release 2 (DR2), scheduled for January 2019, will add more of the PS1 image products from the 3$\pi$ survey, including the single epoch {\em warp} images and their ancillary data, such as signal, masks, and variance maps. DR2 will add the {\em Detection} tables and {\em Forced*} tables, containing the single epoch source detections and the forced photometry.  DR2 also contains numerous improvements to the data products in DR1, which it supersedes. Future data releases will provide the 3$\pi$ {\em diff} image products and catalogs, and analogous data products for the MD surveys.

\section{Overview of the Data Products}\label{sec:overview}

\noindent Public access to the Pan-STARRS data is located at \texttt{http://panstarrs.stsci.edu} and is provided for by the {\em Barbara A. Mikulski Archive for Space Telescopes} (MAST) at STScI. MAST provides the access point for downloading different pixel data products and their associated metadata and source catalogs. This includes FITS images, FITS and JPEG image cutouts, scriptable image access, color JPEG images, and an interactive image browser with catalog overlays through the MAST portal and the MAST PS1 image cutout server. In addition, MAST provides a simple web-based interface to access the Pan-STARRS catalog database. Full database access to the Pan-STARRS tables is available through the Catalog Archive Server Jobs System (CasJobs) interface (see description at \texttt{http://mastweb.stsci.edu/mcasjobs}. CasJobs emulates local free-form SQL access in a web environment, and provides both synchronous and asynchronous query execution. The interface can execute complex, large queries of the PS1 (DR1/DR2) catalogs, with results saved to a private space allocated to each registered user. The Pan-STARRS catalog database accessible through CasJobs contains calibrated catalogs of photometric and astrometric parameters for single epoch exposures, stacks, difference images, and forced photometry. The database schema for the Pan-STARRS catalog database is briefly described in Section~\ref{sec:schemaintro}, and is fully expanded in Appendix~\ref{sec:schema}. Examples of queries are described in Appendix~\ref{sec:query}.

The Pan-STARRS1 catalog database schema is organized into four sections:
\begin{enumerate}
\item Fundamental Data Products. These are attributes that are calculated from either detrended but untransformed pixels or warped pixels. It should be noted that, once in the database, the instrumental fluxes and magnitudes have been subject to re-calibration, as have the sky coordinates. Because of these re-calibrations on the catalogs, the catalog values are to be preferred to making a new measurement directly from the available released pixel data, and care should be taken when using the recalibrated astrometry with the original images (see Table \ref{table:fundamentalipp}).

\item Derived Data Products. These are higher order science products that have been calculated from the Fundamental data products, such as proper motions and photometric redshifts. These data products are not yet available and will come in later data releases. 

\item Observational Metadata. These are metadata that provide detailed information about the individual exposures (e.g. information like exposure time, filter used, etc.) or about which exposures went into an image combination (stacks and diffs), 
as well as information such as detection efficiencies (see Table \ref{table:observationalmetadata}).

\item System Metadata. These tables have fixed information about the system and the database itself, primarily descriptions of various flags and their bits, but also for other metadata such as filter information (see Table \ref{table:systemmetadata}).
\end{enumerate}
Various database ``Views'' are also constructed as an aid to the user for standard types of queries. Views act like tables, and primarily consist of joins of different commonly used tables, in order to simplify queries. Views are also used to join slices of tables (sliced by area of sky) into a full sky view. For example, ``Detection'' is a view of 32 Detection tables, but the individual tables are hidden from the user. For more information on views, including the currently defines ones, see Table \ref{table:views}.

This paper covers the data products and schema for the 3$\pi$ data releases, though most details also apply to the Medium Deep fields. Additional documentation is available together with the data products through MAST.

\section{Flow of Data From Pipeline to the Pan-STARRS Catalog Database}

\label{sec:creation}
\begin{figure*}
\centerline{\includegraphics[width=0.8\textwidth,angle=0]{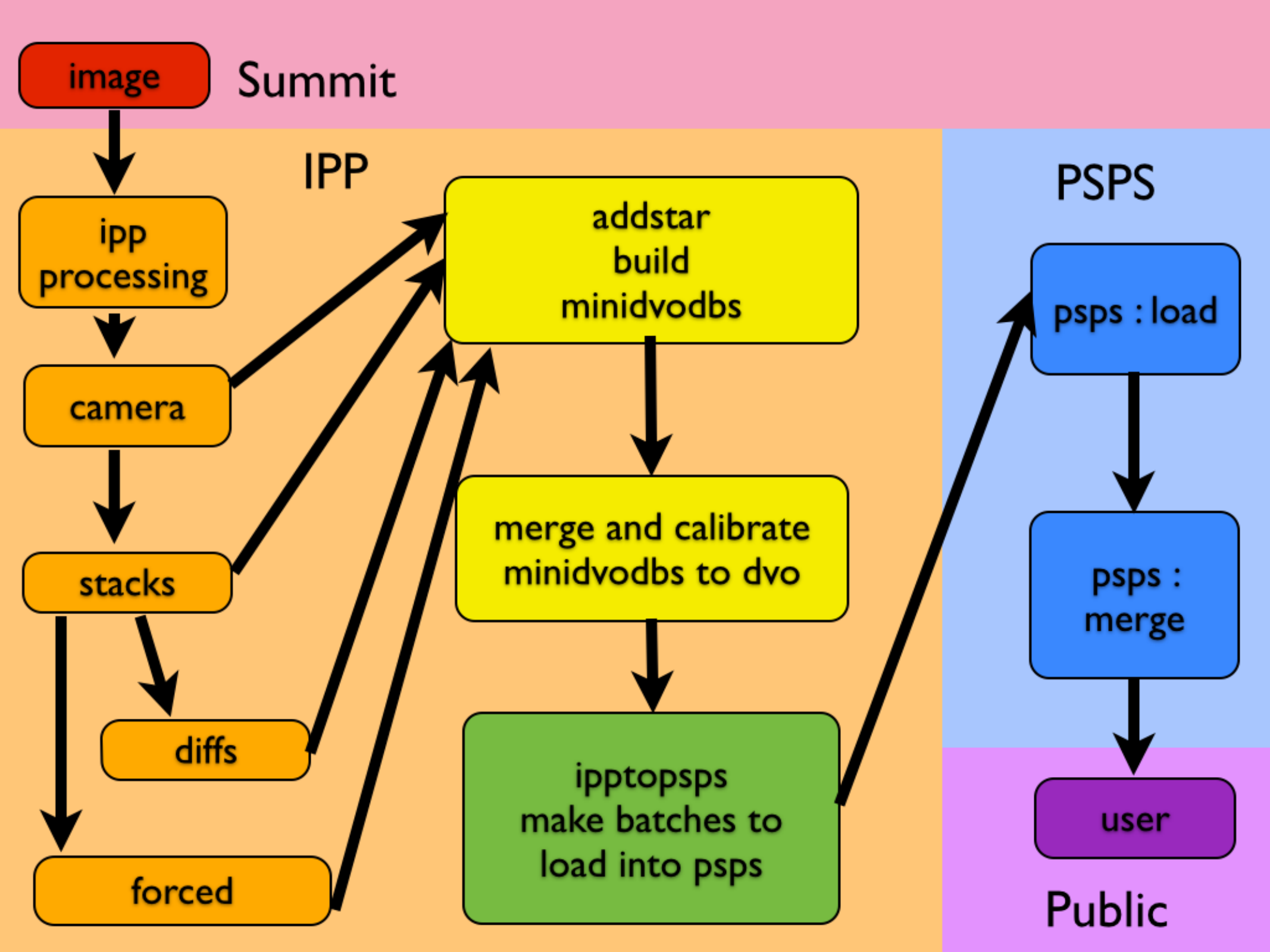}}
\caption{An overview of the steps necessary to create publicly accessible \PS\ data.  The first step is to take exposures from the summit, process them via the image processing pipeline (IPP), ingest the data into the PSPS, and then provide public access to the user.  The IPP has many steps of processing, not all are shown here. The {\em camera}, {\em stacks}, {\em difference images} and {\em forced photometry} stages produce binary catalog FITS files which are the foundation of building the DVO database, which is then calibrated.  The final step of IPP processing is to use {\em IppToPsps} to generate small batches of data in the appropriate database schema to be ingested into PSPS. This paper primarily focuses on the PSPS and the database schema. The other steps are explained in enough detail to describe known and potential sources of inconsistencies within the database.}
\label{fig:revisedipptopsps}
\end{figure*}

\noindent Here we present a condensed version of the flow of data starting with raw image processing with the IPP, and ending with the steps used to generate the catalog database. This flow provides readers with an overview of the various terms used to describe different data products. The full description of these steps are in \citet[][Paper II]{Magnier2017a}. During the various processing steps some data are removed or flagged out, requiring updates to the list of available data. We describe these procedures here to provide additional information about the process that will help the user determine the best data products to use. A flowchart of the whole process can be seen in Figure~\ref{fig:revisedipptopsps}.  

First, exposures are taken at the summit, then they undergo IPP processing which produces catalogs files from the {\em camera} stage, {\em stack} stage, {\em diff} stage and {\em forced} stage. These catalog files (in the form of binary FITS files called {\em cmf/smf} files) are then ingested into a ``Desktop Virtual Observatory'' (DVO, described briefly in section~\ref{sec:DVOdatabase} and more extensively in \citet[][Paper V]{Magnier2017c}) database.  The DVO database is then calibrated \citet[][Paper V]{Magnier2017c}.  Next, {\em IppToPsps} creates ``batches'' or small chunks of sky with the appropriate database schema, from the DVO and {\em smf/cmf} files to be loaded into the Pan-STARRS catalog database. The Pan-STARRS catalog database is made available to the user.

\subsection{Processing Stages}
\label{sec:processingstages}

\noindent The first stage of data products involves processing raw images into detrended (images with flats/biases/darks and other corrections applied) and warped images, described in \citet[][Paper II]{Waters2017}, followed by the generation of {\em stacks}, {\em diff images} and catalogs. Finally, forced photometry on the {\em warps} is performed using positions of sources detected in the stacks. This processing is done by the {\em Image Processing Pipeline} (IPP). This first iteration of processing is called nightly science processing.  It was designed to be fast to allow the \PS\ {\em Moving Object Processing System} \citep[MOPS]{Denneau2013} and the PS1 science consortium access to the data a few hours after observation in order to detect moving objects, supernovae, and other time-sensitive transients.  The 3$\pi$ data spans 4 years of observations (2010-2014, with some observations in 2009 and 2015), a time during which the IPP was actively being developed and improved. To make a consistent set of data, both for the internal catalog database as well as for the public release, all of the raw images were reprocessed from scratch all using a specific revision number of the IPP code. The data released to the public in DR1 and DR2 are internally called Processing Version 3 (PV3).  

We provide a short description of the different stages of this pipeline, separated into sections based on final available image data products and products that are used to create the large relational database below. 

\subsubsection{Download and registry of images}
\label{sec:ippmetadata}

Before processing of images can take place, they must first be transferred to the IPP cluster from the summit, and registered in a metadata database to ensure that all of the $>$ 1.3 million images taken by Pan-STARRS have been properly handled.  This is represented as the ``image" $\rightarrow$ ``ipp processing" step in Figure~\ref{fig:revisedipptopsps}. The IPP uses an internal database to track all parts of the image processing.  This database keeps basic metrics relevant to each stage, including details on what type of image was taken,  when an image was processed, how long processing took, flags and other metrics on the quality, etc.  While this database is internal to Pan-STARRS, it is referenced frequently in this paper as ``GPC1.''

\subsubsection{Chip and Camera Stages}
\label{sec:chipandcamera}

\noindent The first step, called ``chip'' stage, takes the raw images, generally 60 FITS files, one FITS file per OTA, and detrends them, one chip per computing job. Dark, flat, bias, background and other corrections, as described in the detrend paper \citep{Waters2017}, are applied to each chip image, followed by finding the sources and doing photometry on them using {\em psphot} \citep{Magnier2017b}. The next step, called ``camera'' stage, is to mosaic the {\em chip} processing together, to do basic astrometry on the image, and to generate a binary FITS table, called an {\em smf} file, which holds the catalog information for the image. The end product of {\em camera} produces the catalog {\em smf} files which are later ingested into a relational database (DVO) for internal use. These 2 stages are represented as the ``ipp processing" $\rightarrow$ ``camera" steps in Figure~\ref{fig:revisedipptopsps}. Camera stage products are available to the user as the 'Detection' tables, starting with DR2.

\subsubsection{Warp Stage}
\label{sec:fakeandwarp}

\noindent The next steps is the {\em warp} stage, represented as ``camera" $\rightarrow$ ``stacks" in Figure~\ref{fig:revisedipptopsps}. The {\em warp} stage geometrically transforms and bins the {\em chip} outputs to be on a tangential RA/Dec plane, with 0.25\arcsec\ pixels, with the images chopped into $6242 \times 6254$ pixel ``skycells.''  The skycells themselves are tessellated across the sky, and thus any image can be split and projected onto a common layout on the sky. For 3$\pi$, the skycell tessellation is \texttt{Rings.V3}, based on a Budavari-Magnier Rings tessellation \citep{Magnier2017a}. This tessellation subdivides the sky into projections cell rings, each projection cell is $4.0 \times 4.0$ degrees, subdivided into $10 \times 10$ skycells, each with 60\arcsec\ of overlap on a side. All image data products beyond {\em warp} ({\em stacks}/{\em forced warps}/{\em diffs}/ etc.) are laid out in skycells as well.  

The warp image products are available to users via MAST for DR2.

\subsubsection{Stack, StaticSky, Skycal Stages}
\label{sec:stackstages}

\noindent There are 3 stack-related stages: stack, staticsky, and skycal. The stack stage generates the images, and staticsky and skycal generate the catalogs files. All of the stack related stages are represented as ``stacks" in Figure~\ref{fig:revisedipptopsps}.

\noindent {\em Stacks} are generated from adding together {\em warp} exposures with the same skycell id and filter; the process described in more detail in \citet{Waters2017}.  There are several {\em stack} types, which are listed in the {\em StackType} table in the PSPS database.  For the 3$\pi$ database, the {\em stack} type is set to \texttt{DEEP\_STACK}, i.e., all available and good quality {\em warps} for a given skycell and filter within 3$\pi$ are used to generate the best and deepest {\em stack} possible. The 3$\pi$ deep {\em stacks} consist of one deep {\em stack} per skycell per filter. This step produces unconvolved {\em stacks}, {\em stacks} with input {\em warps} convolved to 6 pixels, and {\em stacks} with input {\em warps} convolved to 8 pixels. Masks, weights, variance, and number maps are also generated for each deep stack.   

Once each {\em stack} image is created, photometry and astrometry are determined in the {\em staticsky} and {\em skycal} stages. For the {\em staticsky} stage, photometric analysis is run on all 5 filters at once, on the unconvolved stacks. Catalog files, one per filter, of matched sources within a 5 pixel radius are generated. The {\em skycal} stage calibrates the {\em staticsky} catalogs relative to the reference catalog. The {\em skycal} catalog files are later ingested into the DVO database and then into the PSPS database. Due to the overlap between skycells, sources that land in the overlaps can be reported 2, 3, or 4 times in the DVO and PSPS database. Use the \texttt{BestDetection} or \texttt{PrimaryDetection} flag to select one of these.

Stack data products are available, starting with DR1. Stack image products are available to users via MAST, and Stack* tables are available in the PSPS database.

\subsubsection{Diff Stage}
\label{sec:diffstage}

\noindent In general, for the IPP, there are several different types of {\em diffs}. These are \texttt{WARP\_WARP}, \texttt{WARP\_STACK}, \texttt{STACK\_WARP} and \texttt{STACK\_STACK} {\em diffs}.  For 3$\pi$, difference images are generated between {\em warps} and deep {\em stacks}, i.e., \texttt{WARP\_STACK} diffs.  For each exposure within the 3$\pi$ dataset, a corresponding diff is generated, subtracting the appropriate  3$\pi$ deep {\em stack}s from each good quality {\em warp}s for that exposure (same filter, and skycell\_ids). %
The results from this stage of processing include diff catalog files, which will be available in DR3. The diff images can be reconstructed from available data products hosted at STScI, but at this time we anticipate they will not be stored there due to space constraints.

\subsubsection{Forced Photometry Stage}
\label{sec:forcedphotom}

\noindent Most of the phootmetry done in earlier stages, with the exception of some stack photometry, is unforced. That is, significant sources are identified on the images, and then photometry is performed.  Forced photometry is done with prior knowledge of source positions measured from the stacks.  There are 2 steps of forced photometry for the IPP: One performs forced photometry on the individual {\em warp} exposures, and the other calculates forced galaxy model fits on the stacks. 

The forced warp photometry stage takes the positions of sources located in the deep {\em stacks} and then measures the PSF model flux at those positions on the individual warps.  The catalogs generated by this process are first ingested into the DVO database, then translated using {\em IppToPsps} into the {\em ForcedWarp*} tables for the PSPS, and are available starting with DR2. 

For extended sources, galaxy models are fitted on the {\em stack} images. These models are then used as a seed to determine galaxy models for each warp image. The position, aspect ratio, and (where appropriate) Sersic radius are kept fixed to the values determined for the stack image, while the major and minor axis values are fitted for each warp image, along with the model normalization. All the models include PSF convolution.  

Catalog files are produced for this stage of processing, and these are later ingested into the PSPS database, as the {\em ForcedGalaxy*} tables. 




\subsection{DVO Database Steps}
\label{sec:DVOdatabase}

\subsubsection{Building and calibrating a DVO Database}
\label{sec:buildingdvo}


The Desktop Virtual Observatory (DVO) \citet{Magnier2017c}, \citet{Magnier2004} is a database that takes a subset of photometric, astrometric, flags, metadata, etc. from the catalog FITS files {\em smf/cmf} from various IPP stages and puts them into a relational database built from FITS files. Within this database, we determine the connection between detections and objects, and we perform the necessary astrometric and photometric calibrations. The smf/cmf files are the core binary fits catalog files generated by IPP processing, and include measurements of all the sources for an exposure or skycell (or other unit) for a given stage, detector efficiency, and for the smf files (for the {\em camera} stage, chip level astrometric transformation information.  

Catalog files are ingested into the DVO database via the \texttt{addstar} program (part of the IPP code). The output from several stages of IPP processing is ingested into the DVO database.  These are the catalogs produced by the  {\em skycal}, {\em camera}, forced photometry on warps, and forced galaxy model fit stages.  Difference image catalogs are ingested into the diff DVO database. Measurements from 2MASS, WISE, and Gaia are merged into the DVO database, and there are flags within the external Pan-STARRS database (and within dvo) that note the presence of these surveys. Finally, the DVO database (and the external Pan-STARRS database) is tied to the Gaia calibration \citet{Magnier2017c}. 

Once the DVO database is built, it is then calibrated. Steps include relative photometry and astrometry.  Average properties are also calculated. Full details are in \citet{Magnier2017c}. 

Several categories of DVO files are used by {\em IppToPsps} to populate the PSPS database.  
Here we give a short summary of the subset of DVO files that are most relevant for {\em IppToPsps} (see \citet{Magnier2017c} for more details).

\begin{description}

\item[.cpt] Object information - each .cpt table has one entry for each object in that region of the sky. It summarizes the average properties of that object as long as those properties can be derived independently of the filter used. Information such as mean RA and Dec are listed in these files.  

\item[.cpm] Measurements - each .cpm table contains all of the measurement information for each object in the .cpt file. Contains measurement information for detections from the stack/skycal cmf, camera smfs, and forced warp smfs.

\item[.cps] Mean properties - the .cps table has filter-dependent average property information for each object listed in the .cpt file.  Information such as mean magnitudes are located in these files.

\item[.cpx] Lensing measurements - the .cpx files contains lensing parameters measured from all the forced warp cmfs. 

\item[.cpy] Lensing Objects - the .cpy table has one entry per filter for each object in that region of the sky, same object ids as for objects in the .cpt file. It summarizes the average properties of the lensing measurements.  

\item[.cpq] Forced Galaxy - the .cpq table has one entry per filter for each object in that region of the sky, same object ids as for objects in the .cpt files. It summarizes the extended source galaxy shape measurements.

\end{description}










\subsection{IppToPsps Steps} 
\label{sec:ipptopsps}

\noindent The DVO database contains a subset of information, primarily average properties of sources, and precisely calibrated astrometric and photometric information.  It is necessary, when building the PSPS database, to combine the DVO database as well as extra columns from the individual {\em smf} and {\em cmf} files.  The {\em IppToPsps} stage does this for the mean properties (split into mean properties for single exposures, calibrations against the Gaia catalogue \citep{Gaia2016}, difference exposures, and {\em forcedwarp} exposures), subdivided by DVO FITS files with RA/Dec boundaries, and again for the individual stages, which represent either single exposures or skycells from other stages ({\em camera}, {\em stacks}, {\em forced warps}, {\em difference images}). This stage therefore combines information from the various IPP data products that are stored in DVO, and propagates this information into a more homogeneous form in the PSPS database.

\begin{figure*}
\centerline{\includegraphics[width=0.8\textwidth,angle=0]{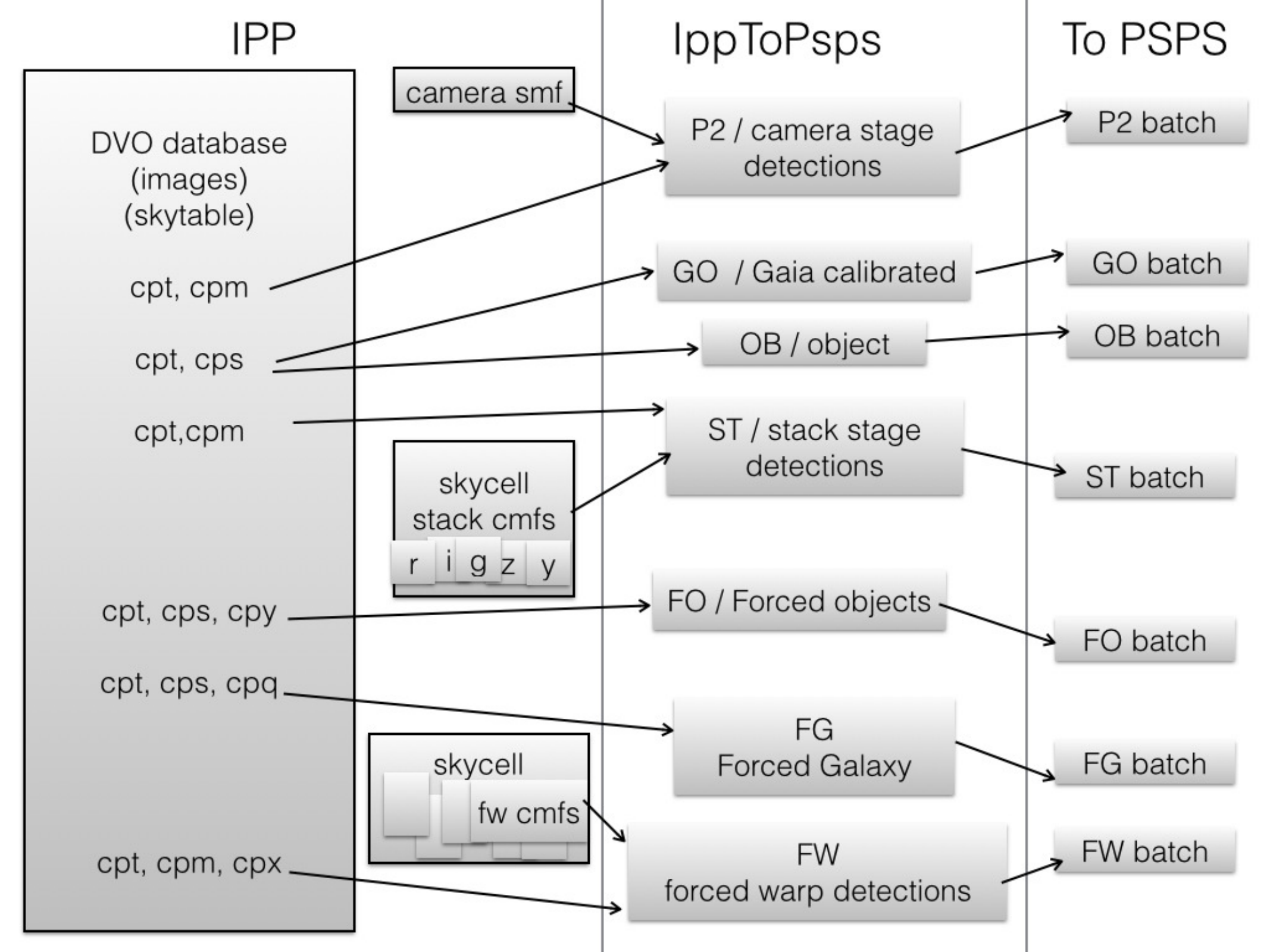}}
\caption{This figure shows a flowchart of how data flows from the IPP side into batches for PSPS, using {\em IppToPsps}. On the IPP side, the DVO database shows cpt/cpm/cps/cpx/cpy/cpq files, organized and grouped by which {\em IppToPsps} batch type uses them. The IPP side also has the smf/cmf files from the camera stage, forced warp stage, and stack (skycal) stages, these smf/cmf files are also needed for {\em IppToPsps}.  {\em IppToPsps} has several different batch types, extracting data from different sources, and generating batches for ingest into PSPS. Batches related to diffs are not shown here, it is a similar process (cpt,cpm) files from the diff DVO and cmf files from the diff skycells go through {\em IppToPsps} to create DF batches (analagous to P2 or ST but using diff cmfs).  DO batches are created using cpt,cps files from the diff DVO (similar to how OB or GO batches are created).}
\label{fig:ipptopsps}
\end{figure*}

The {\em IppToPsps} translates the DVO and various catalog files associated with processing ({\em camera}/{\em stacks}/{\em forced warp}/{\em diff}) into smaller batches which are loaded onto data stores to be ingested by PSPS.  This is the stage that transforms the data into the same schema used by PSPS, and which users interact with. The {\em IppToPsps} is written in python/jython, STILTS \citep{Taylor2006}, and uses MySQL to track the processing and for temporary scratch databases.  This process also queries the IPP processing database, retrieves files from the IPP cluster, and reads data from the DVO database.

There are multiple types of batches that are generated by {\em IppToPsps}.  These include: {\em Init} (IN), {\em Object} (OB), {\em stack} (ST), {\em Detection} (P2), {\em Forced Mean Objects} (FO), {\em Forced warp} (FW), {\em Forced Galaxy} (FG), {\em Gaia Objects} (GO), {\em diff Detection Objects} (DO), and {\em diff Detections} (DF).  For DR1, IN, OB, ST, FO and GO batches are processed through {\em IppToPsps} and ingested into PSPS.  DR2 has P2, FW, FG batches. An overview of the different batch types and associated DVO files and smf/cmf files is shown in the flowchart in Figure~\ref{fig:ipptopsps}.

The init batch (IN) is the first batch created by {\em IppToPsps}, and it is first to be ingested into PSPS.  This includes the system metadata tables described in Section~\ref{sec:schemameta}, with flag bits listed in Appendix~\ref{sec:FlagTables}.  

The object batches (OB) create the {\em ObjectThin} and {\em MeanObject} tables, described in more detail in Section~\ref{sec:schemaobject}.  Each batch represents individual DVO files which are subdivided into small rectangular patches of sky. Columns are filled from the DVO database (cpt and cps files).

The detection batches (P2), create the {\em Detection} tables, described in more detail in Section~\ref{sec:schemap2}.  Each batch corresponds to a single exposure from \PS.  Columns within are filled from the DVO database (cpt and cpm files) as well as the {\em camera} stage catalog file (smf).  

The {\em stack} batches (ST) create the {\em stack} tables, described in more detail in Section~\ref{sec:schemast}.  Each batch corresponds to a skycell from the skycal stage.  Columns are filled from the DVO database (cpt and cpm files ) as well as from the corresponding skycal catalog files (cmf) for all 5 filters (or what is available).

The {\em forced mean objects} (FO) create the {\em ForcedMeanObject} and {\em ForcedMeanLensing} tables, described in more detail in Section~\ref{sec:schemaobject}.  Forced warp processing is ingested into a DVO, forced objects are determined and their mean properties are calculated and contained withing the DVO. Each batch contains data from individual DVO files (cpt, cps, cpy).  

The {\em forced galaxy} batches (FG) create the {\em ForcedGalaxyShape} table, described in more detail in Section~\ref{sec:schemaobject}. Forced galaxy model fits calculated from the stacks are ingested into the DVO, the forced galaxy objects are determined within the DVO. Each batch contains data from individual DVO files (cpt, cps, cpq).

The {\em forced warp} batches (FW) create the {\em ForcedWarp*} tables, described in more detail in Section~\ref{sec:schemafw}.  Each batch corresponds to a different skycell, and contains all the {\em forced warp} catalogs for that skycell.  Each batch contains data from individual DVO files (cpm, cpt, cpx) as well as from the corresponding {\em forced warp} catalog files ({\em cmf}). 

The {\em diff} object batches (DO) create the {\em DiffDetObject} table, described in more detail in Section~\ref{sec:schemaobject}.  Diff detection catalog images are ingested into a DVO, {\em diff} objects are defined and mean properties are calculated for the {\em diff} objects.  Columns are filled from the DVO {\em diff} database (cpt and cps files). 
 
The {\em diff detection} batches (DF) create the {\em DiffDetection} table, described in more detail in Section~\ref{sec:schemadiff}.  Each batch corresponds to a difference image catalog file created in the {\em diff} stage, and will contain all the skycells for a given exposure.  Columns are filled from the DVO database (cpt and cpm files), and from the corresponding {\em diff} catalog file ({\em cmf}). 

Gaia DR1 \citep{Gaia2016} was released after all of the object tables were ingested into the PSPS database.  We used the Gaia data to recalibrate the DVO object positions, which improved the astrometry significantly. Rather than dump the database and start over (with corrected RA and Dec positions), we added a new type of batch, (GO), that contains minimal metadata information, the \texttt{ObjID} for the objects, the {\em objectInfoFlags}, and the corrected RA and Dec as well as the errors.  It is based on exactly the same DVO files as OB batches, has updated RA and Dec calibrated to Gaia, and ignores the rest of the DVO columns.

Within {\em IppToPsps} it is possible to verify that the expected batches were generated, and to re-queue and regenerate batches that failed.  Batches fail for a variety of reasons, but none of the failures are terminal.  Batches can fail if any of the associated {\em mysql} databases time out or are unavailable, if there are disk I/O glitches or other disk problems.  The DVO database sets the expected number of batches to generate, and failures are investigated and retried until they are resolved. Within {\em IppToPsps} it is also possible to poll the PSPS to verify if batches have been ingested.  



\section{PSPS}
\label{sec:overviewpsps}
\label{sec:PSPS}

\noindent We present an overview of the Published Science Products Subsystem (PSPS), both to help users understand better how the PSPS database is constructed in order to optimize the queries, and to describe the currently known issues within the PSPS database.

\begin{figure*}
\centerline{\includegraphics[width=0.8\textwidth,angle=0]{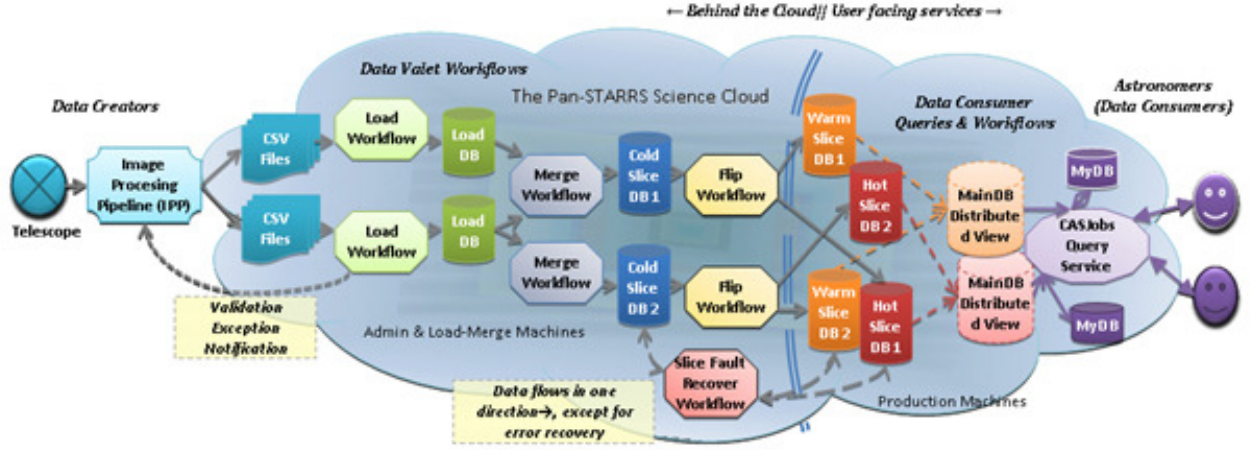}}
\caption{This figure shows a flowchart of how data flows from the IPP (via IppToPsps) into the load merge machines, which is then copied to the slice machines to allow for users to query the data (via a modified CasJobs)}
\label{fig:odm_data_flow}
\end{figure*}

\begin{figure*}
\centerline{\includegraphics[width=0.8\textwidth,angle=0]{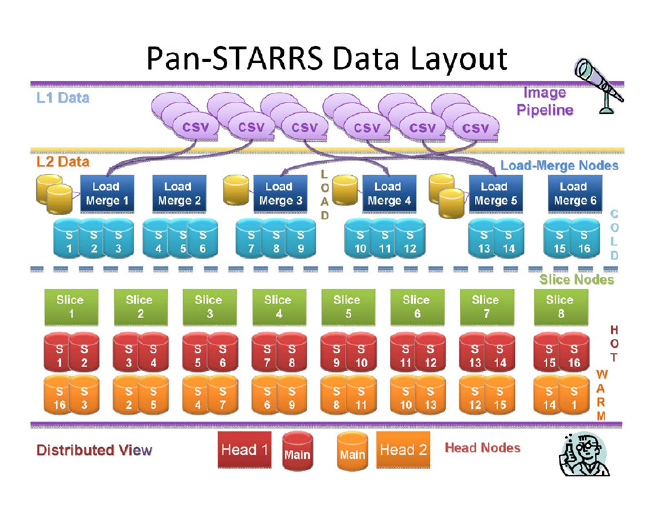}}
\caption{This shows how the data (L1 data/csv files/Image Pipeline) is loaded into L2 data (the load merge machines - responsible for loading the data and merging it into the 'cold' part of the database.  In this figure there are 8 slice machines which hold hot and warm copies of the database.  At the bottom is the head nodes and the main database.  The hot database serves the fast response queue and the warm database serves the slow queue.  The fast queue is specifically for queries that take less than one minute to complete.  The cold database is never accessible by users. }
\label{fig:odm_data_layout}
\end{figure*}

\subsection{Introduction} The PSPS consists of several parts: the data transformation layer (DXLayer), the Object Database Manager (ODM), the Workflow Manager Database (WMD), and the data retrieval layer (DRL).  The user accesses the data through the DRL, using either scripts, the STScI CasJobs interface, or if the user is a Pan-STARRS1 Consortium member, the Published Science Interface (PSI). The DXLayer polls the {\em IppToPsps} datastores for new batches and prepares them for loading.  The ODM is the software that all loading nodes run to load, merge, copy and publish the PSPS databases.  The WMD is the database containing all the logs about the PSPS databases.  The DRL is the intermediate layer between the client and the PSPS database.  The PSI is the web based interface for PS1 consortium members, for interacting with the DRL. Each of these components is described in more detail below, and a diagram of the process is shown in Fig~\ref{fig:odm_data_flow}

\subsection{Partitioning the PSPS} The PSPS uses Distributed Partitioned Views, a mechanism that allows tables to be partitioned to reside on different files on different linked servers. The tables are partitioned into slices, with cuts based on declination, and with each slice containing a similar amount of data. Partition slices are customized for each database ($3\pi$ vs MD);  
Fig~\ref{fig:odm_data_layout} shows how the data is partitioned across a subset of the machines. 

\subsection{The DXLayer} 
\begin{figure*}
\centerline{\includegraphics[width=0.5\textwidth,angle=0]{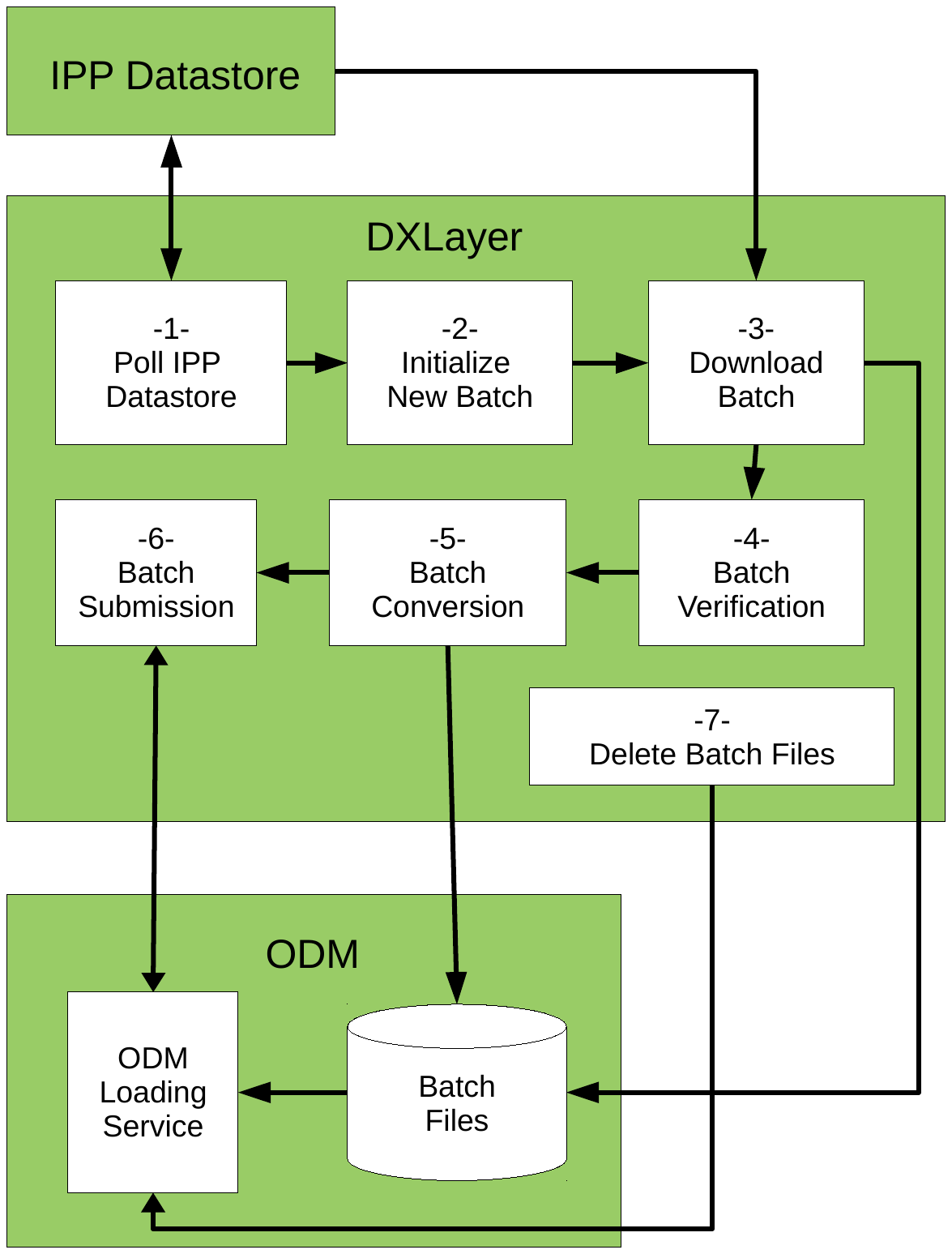}}
\caption{A flowchart of the DXLayer process, showing how batches are loaded into the DXLayer, verified, and submitted to the ODM. The shaded rectangles refer to different systems, and the white boxes and white cylinder refer to difference steps for the systems.}
\label{fig:dxlayerprocess}
\end{figure*}

\begin{figure*}[htpb]
\vskip -8em
\begin{center}
    \includegraphics[width=0.75\textwidth,angle=0]{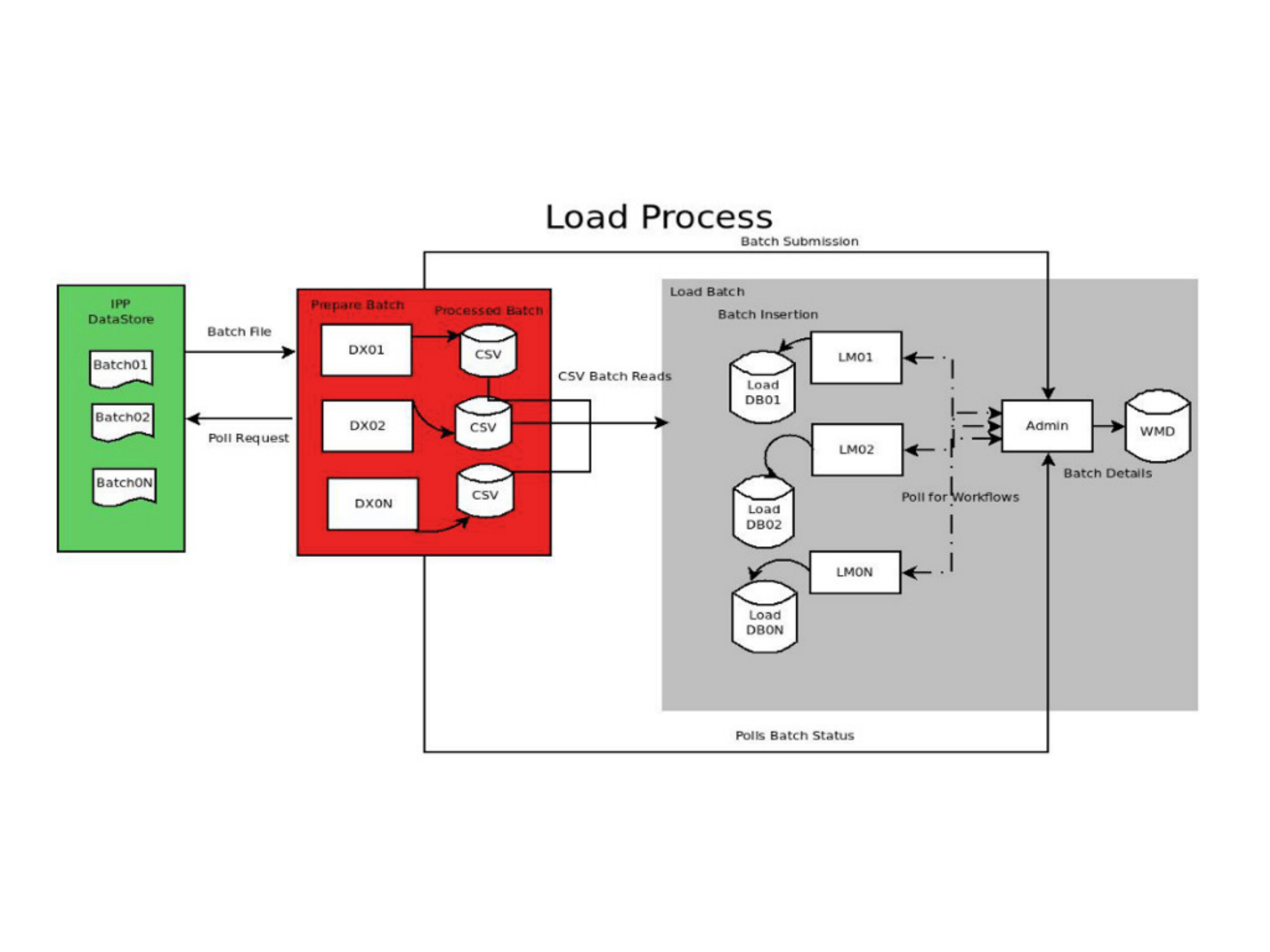}
\vskip -6em
\caption{\noindent A flowchart of loading data from {\em IppToPsps}/IPP as batches into the DXLayer and ODM.}
\label{fig:psps_loadprocess}
\end{center}
\end{figure*}

The DXLayer is the first stage in the PSPS. This is the stage that polls for new batches to load and preps them for the next step (ODM). Fig~\ref{fig:dxlayerprocess} shows the flowchart of the DXLayer process, and Fig~\ref{fig:psps_loadprocess} shows a more detailed flowchart of how batches are loaded and verified within the DXLayer and ODM. 
PSPS loads batches created by the {\em IppToPsps}.  A batch corresponds to either products from different processing stages ({\em camera}, {\em stack/skycal}, {\em diff}, {\em forced warp}) or from different DVO files (arranged by area of sky). Batches contain a manifest file that describes the batch information such as type of batch, min/max \texttt{ObjID}, MD5 checksum, and the tables to load. Batches data is stored in FITS files, which are transformed into csv files in the DXLayer. The batch area cannot exceed two PSPS slices, else it will not load correctly.  The PSPS slices are chosen so that this does not happen.

\subsection{The ODM} The nodes within the ODM have naming conventions for their roles: load/merge (lm), slice (s), head (h) and admin (a).  All ODM processes are named workflows (load, merge, copy, flip).  All logs, processes, and requests are inserted into an administration database called the Workflow Manager Database (WMD).  Databases are named by roles: Load, Cold, Warm, Hot.  These databases are MS-SQL Server and are broken into four volumes with 96 file partitions each. All of the metadata tables and {\em ObjectThin}, {\em GaiaFrameCoordinate}, {\em StackObjectThin} are merged into a head database to provide faster queries.


\subsection{The DRL} 
\begin{figure*}
\centerline{\includegraphics[width=0.7\textwidth,angle=0]{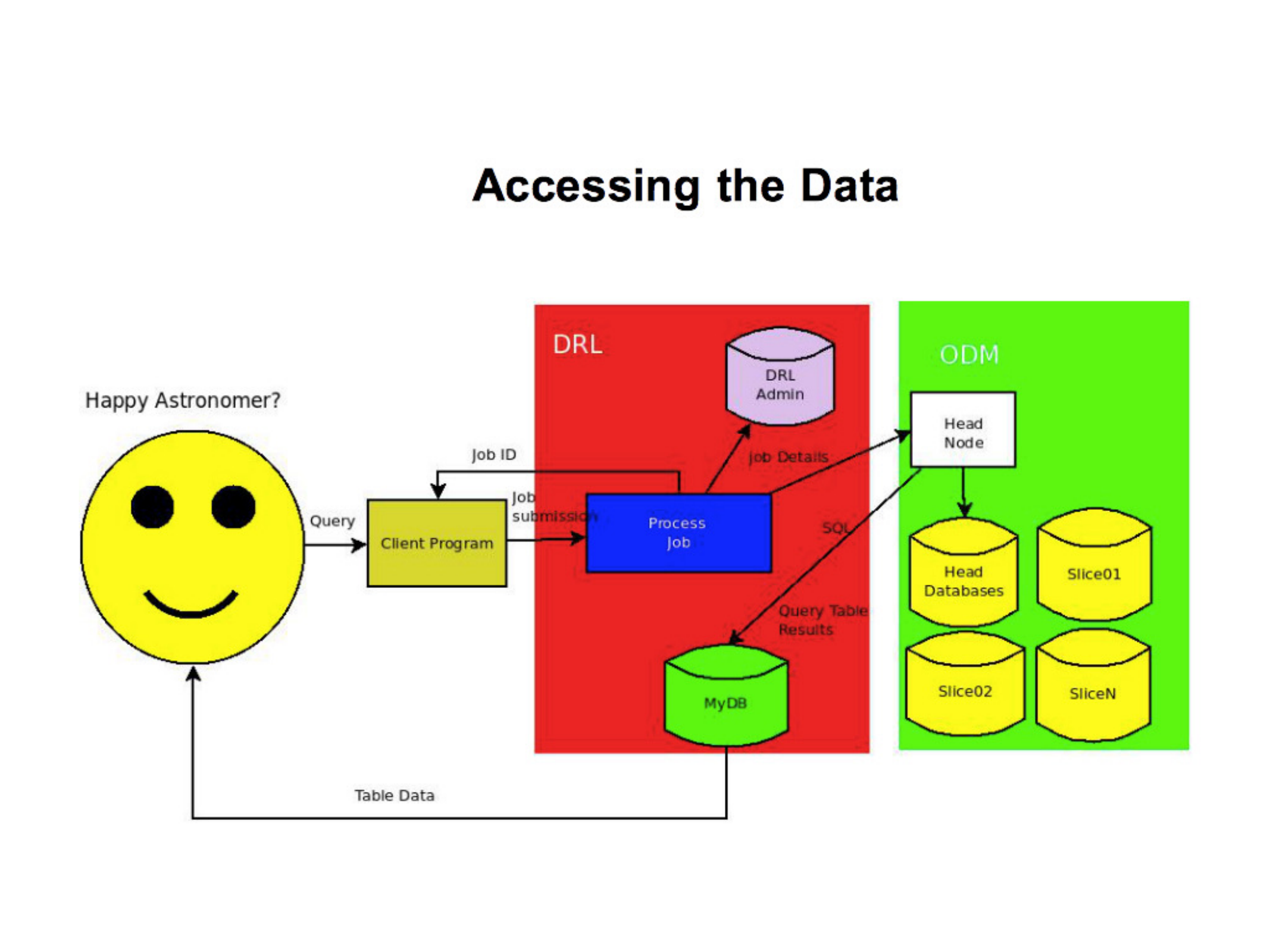}}
\vskip -1.0cm
\caption{\noindent A flowchart explaining the DRL, the layer that provides the user access to the database.}
\label{fig:psps_drl}
\end{figure*}

The DRL is the layer between the user and the PSPS database.  The DRL is responsible for management of queries that the user submits via the DRL API, is based from CasJobs, and has many similar features. It primarily keeps track of all user queries and provides progress of those queries in a secure way. It also kills queries that use too many resources or take too long. The DRL API is accessed via Simple Access Object Protocol (SOAP), allowing for multiple ways for the user to access the database.  Consortium members used the PSI (a web user interface), but it is also possible for the users to query the database via SOAP calls from command line scripts.  A flowchart of the DRL can be seen in Figure~\ref{fig:psps_drl}.

\subsection{PSI} The PSI is the web user interface, but only for consortium members. As it was part of the design of PSPS it is included here for reference. This interface provides many useful features including a query request page, information on query progress, MyDB management tools, graphing tools, access to the pixel data products, and interactive help.  The query request page allows for the user to easily submit queries to a variety of databases (3pi/MD/MyDB), to upload query files or to check the syntax, to name MyDB results tables and to select the queue to submit to. The MyDB management tools allow the user to easily select which MyDB tables to purge as well as well as methods  to extract to csv, FITS or xml files to download.  Some of the interactive features include an interactive schema browser, a query builder to easily create a query with multiple joins and conditions, and a flag generator to create bitmasks for the different types of flag tables. The plotting tool allowed the user to quickly plot data from the PSPS databases or their own MyDB within the browser.  

\begin{figure*}[htpb]
\centerline{\includegraphics[width=0.8\textwidth,angle=0]{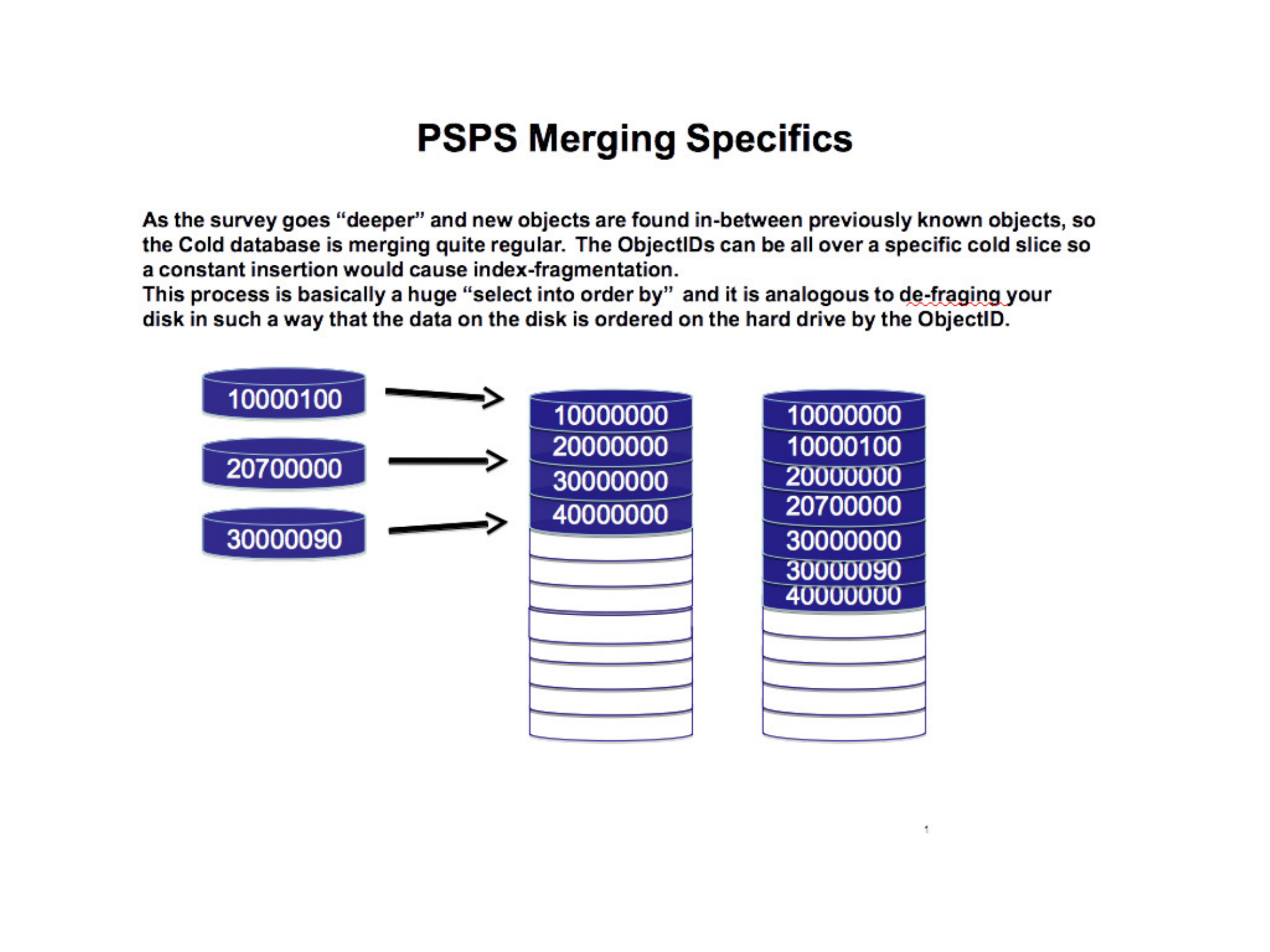}}
\vskip -1.0cm
\caption{\noindent A graphical representation of how the database is re-indexed as more data is ingested. }
\label{fig:psps_merging}
\end{figure*}

\begin{figure*}
\centerline{\includegraphics[width=0.8\textwidth,angle=0]{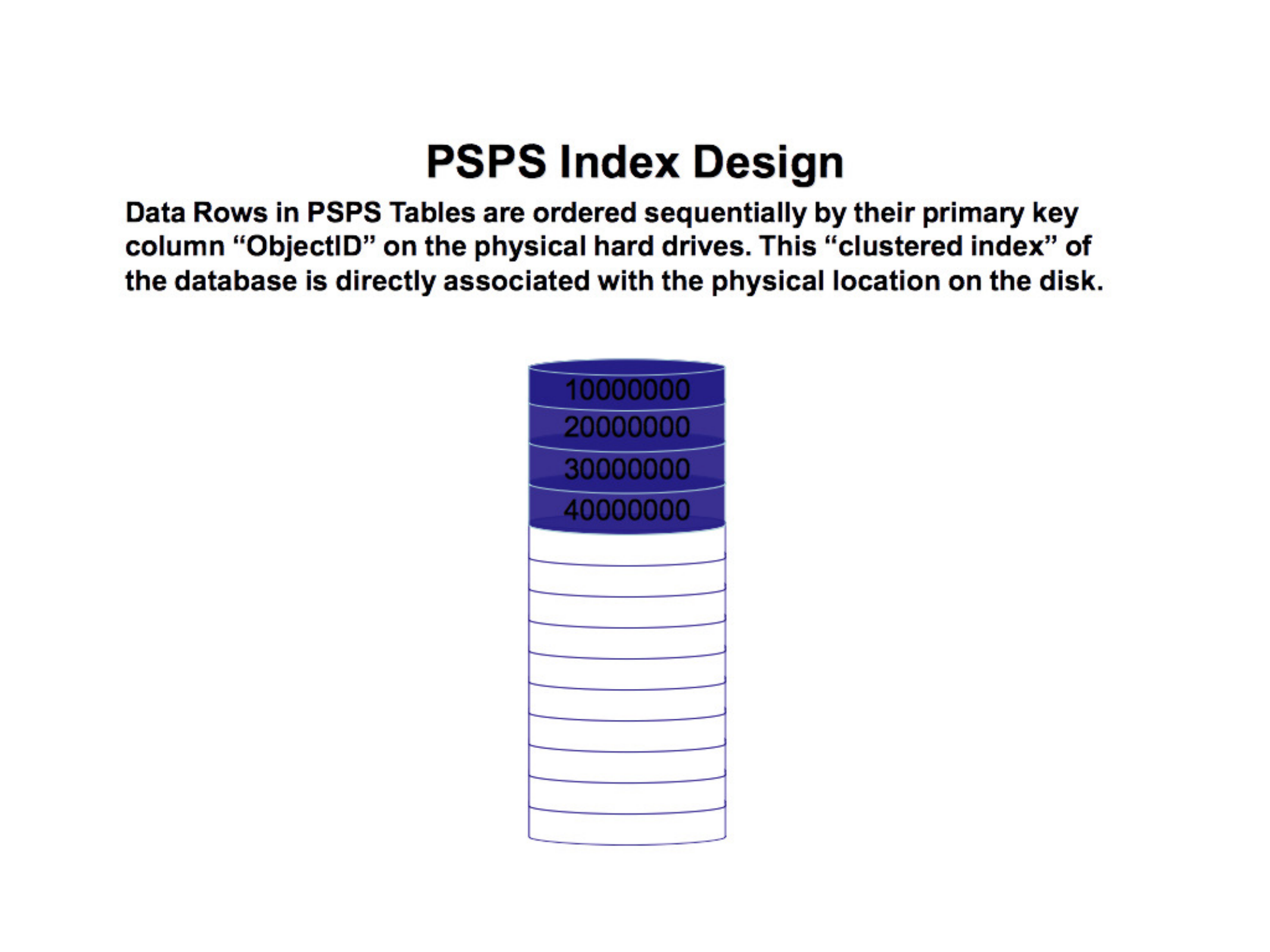}}
\vskip -1.0cm
\caption{\noindent PSPS uses \texttt{ObjID} as the index, with specific \texttt{ObjID} ranges associated with physical locations such as different hard disks}
\label{fig:psps_index}
\end{figure*}

\begin{figure*}
\centerline{\includegraphics[width=0.9\textwidth,angle=0]{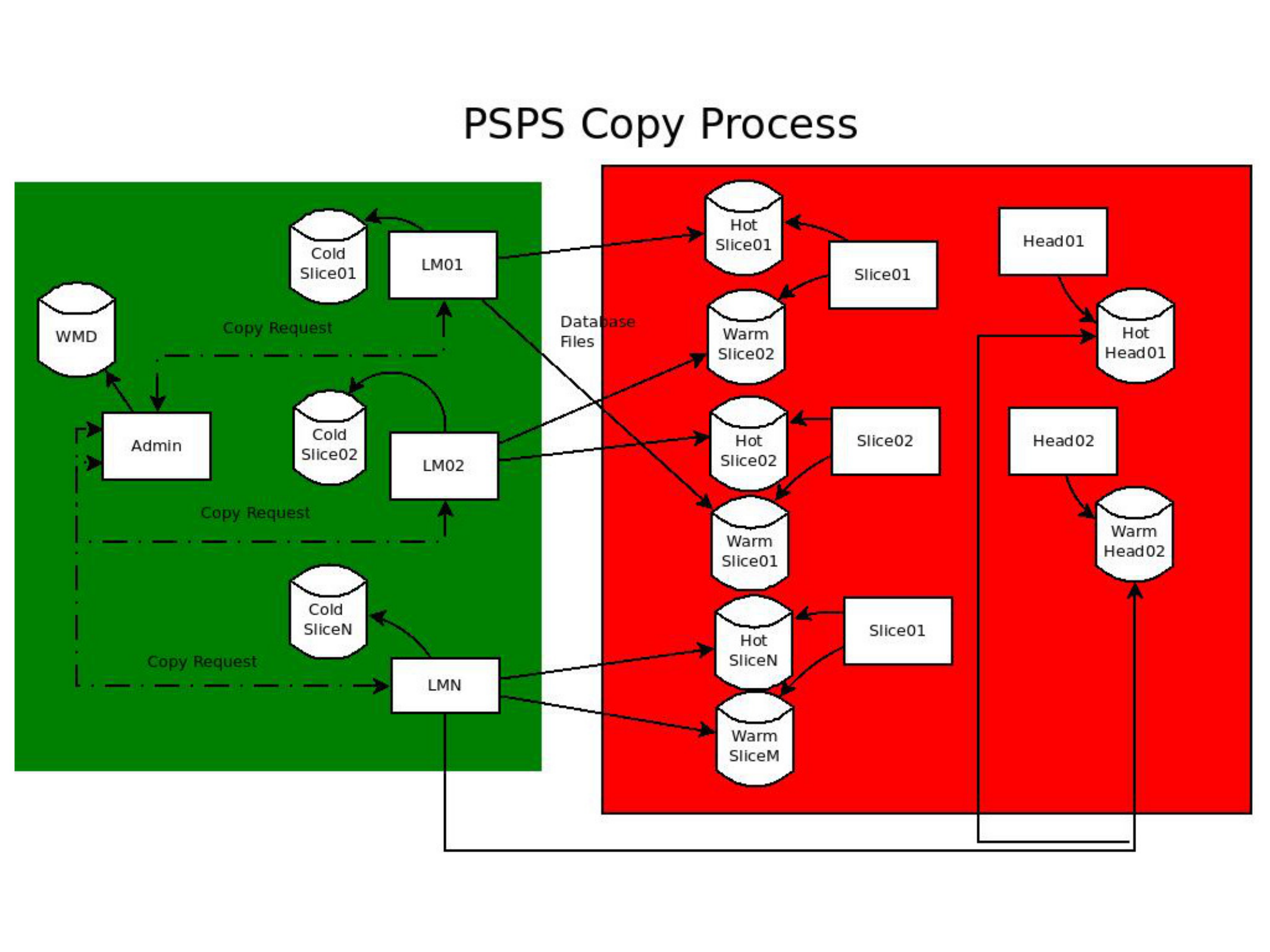}}
\vskip -0.5cm
\caption{\noindent A flowchart showing the steps necessary to do the copy workflow.}
\label{fig:psps_copy}
\end{figure*}

\clearpage

\begin{figure*}
\centerline{\includegraphics[width=0.7\textwidth,angle=0]{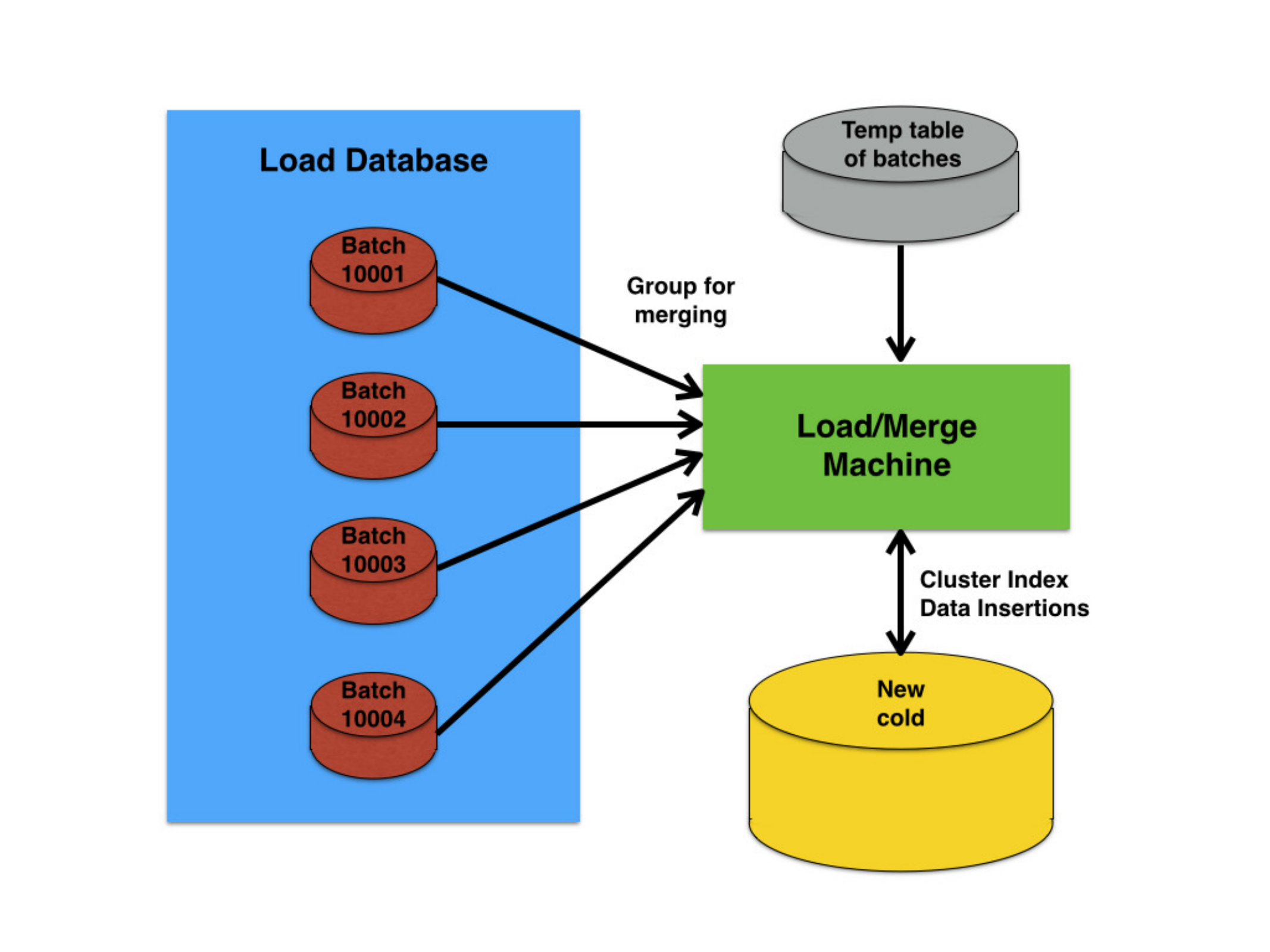}}
\caption{\noindent Flowchart showing how data from IPP/{\em IppToPsps} batches is loaded into the system and then merged into a new cold database.}
\label{fig:psps_mergecriteria}
\end{figure*}



\section{Overview of the PSPS Database Schema}
\label{sec:schemaintro}

\noindent There are over 50 different tables that make up the PSPS schema. Here we give a brief overview of the tables and indexes, to help aid the user in selecting the most appropriate table for queries. The core concept is that the database has a unique \texttt{ObjID} for each object detected within \PS\ data. An object is defined to be a source which has measurable flux at a given R.A. and Dec. In general, multiple detections of an object will all share the same \texttt{ObjID}, as well as multiple detections within 1\arcsec\ of that object (which might not be associated with that object, for example, blended sources). A detailed description of the source deblending algorithm and its properties is beyond the scope of this work. This \texttt{ObjID} is the core index used to join the object and detection tables.  For example, {\em ObjectThin} has the astrometric information for the objects; one would join against the {\em Detection} table, using \texttt{ObjID}, in order to get the individual photometric attributes for all the detections of that object within the single exposures (at a given RA and Dec).  
\subsection{The Main Categories of Tables}
\label{sec:schemacategories}

\noindent There are 4 main types of tables within the PSPS database: Fundamental Data Product tables, Observational Metadata tables, Derived Data Product tables, and System Metadata tables (see also Section~\ref{sec:overview}).
Fundamental Data Product tables can be categorized into either Object Tables (summarized in Section~\ref{sec:schemaobject}) or Detection Tables (summarized in Section~\ref{sec:schemadetections}). 
 Object tables contain basic information on each source in the sky, including the mean photometric and astrometric information.  Detection tables contain photometric and astrometric information on the individual exposures. Detection tables contain detections from individual exposures ({\em Detection}), stacked images ({\em Stack*} tables), Forced {\em warp} images ({\em forcedWarp*} tables) that contain forced photometry of individual exposures using positional knowledge of sources detected on {\em stacked images}, {\em Difference Images} ({\em stack} - individual exposures). 
 Observational Metadata contains information about the different data products and how they are mapped to the individual exposures. Observational Metadata tables are summarized in Sections~\ref{sec:schemadetections}. 
 Derived Data Products will come later and are not described in this paper. System Metadata tables are summarized in Section~\ref{sec:schemameta}; they contain hardwired information about the PSPS, including tables about flags, filters, tessellations and other useful fixed metadata.    
 
This section will start by briefly describing the System Metadata tables (Section~\ref{sec:schemameta}), followed by Object tables (Section~\ref{sec:schemaobject}). The Detection tables (Section~\ref{sec:schemadetections}) are organized by stage of IPP processing and contain a mix of Fundamental Data Product tables and Observational Metadata tables as this seemed the more natural way to organize this part of the schema.


\begin{figure*}
\centerline{\includegraphics[width=0.8\textwidth,angle=0]{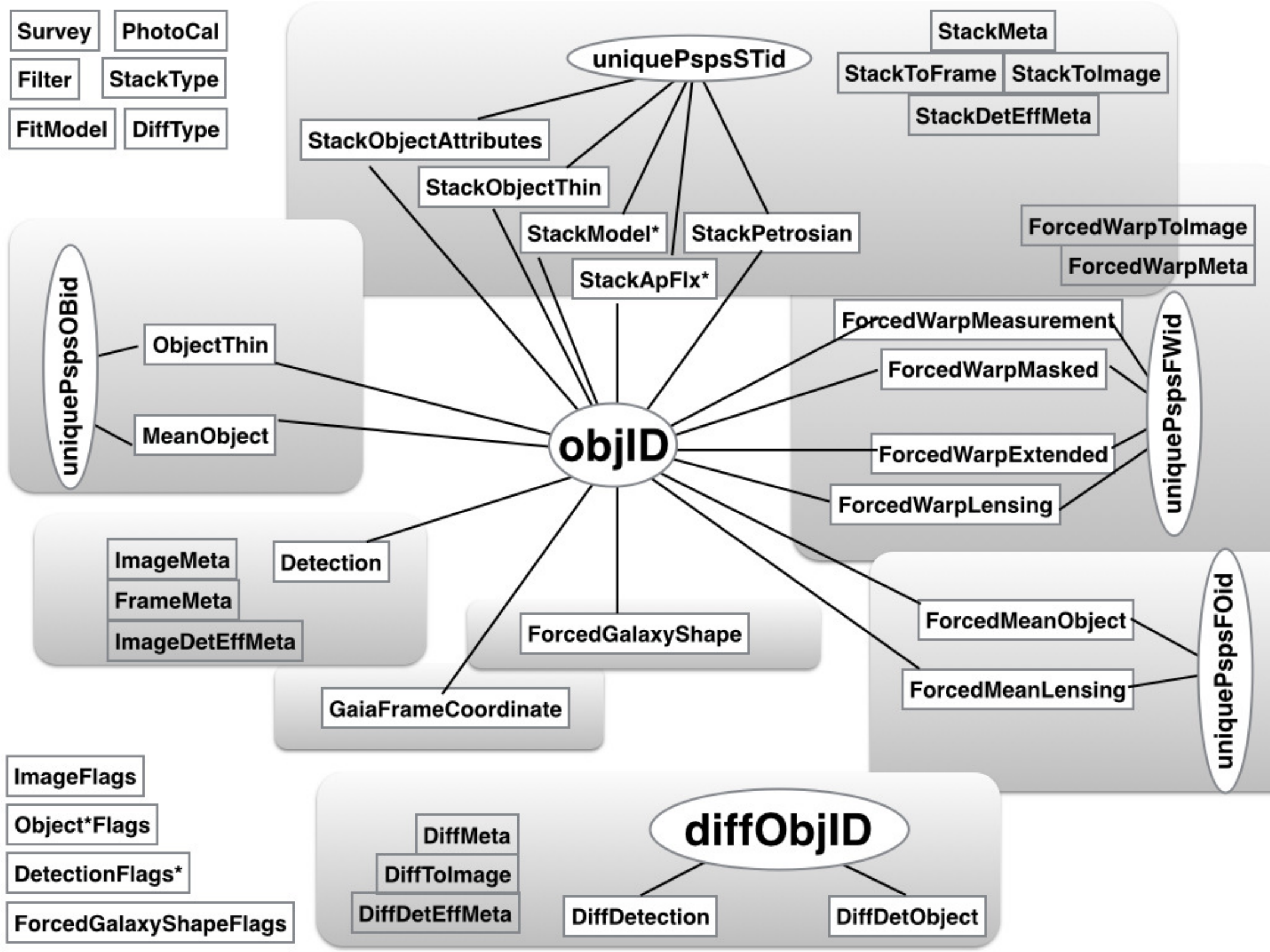}}
\caption{\noindent A summary of how the different tables are related.  The rectangular boxes with words inside represent the different table names.  The ovals with words inside represent the column names to use to join the tables.  Black lines connect table names to columns (i.e., {\em Detection} has a line to \texttt{ObjID} which has a line to {\em StackObjectThin}) - this shows that {\em Detection} can be joined to {\em StackObjectThin} using \texttt{ObjID}).  The grey rounded boxes represent different stages of data processing, which corresponds to different stages of loading into the database.  Tables within the grey boxes are related; connections to the (grey) metadata tables are shown in Figure~\ref{fig:objidmap}.  The tables that are not in grey rounded boxes represent system metadata, metadata that describes the \PS\ system as well as flag information.}
\label{fig:objidmap}
\end{figure*}

\begin{figure*}
\centerline{\includegraphics[width=0.8\textwidth,angle=0]{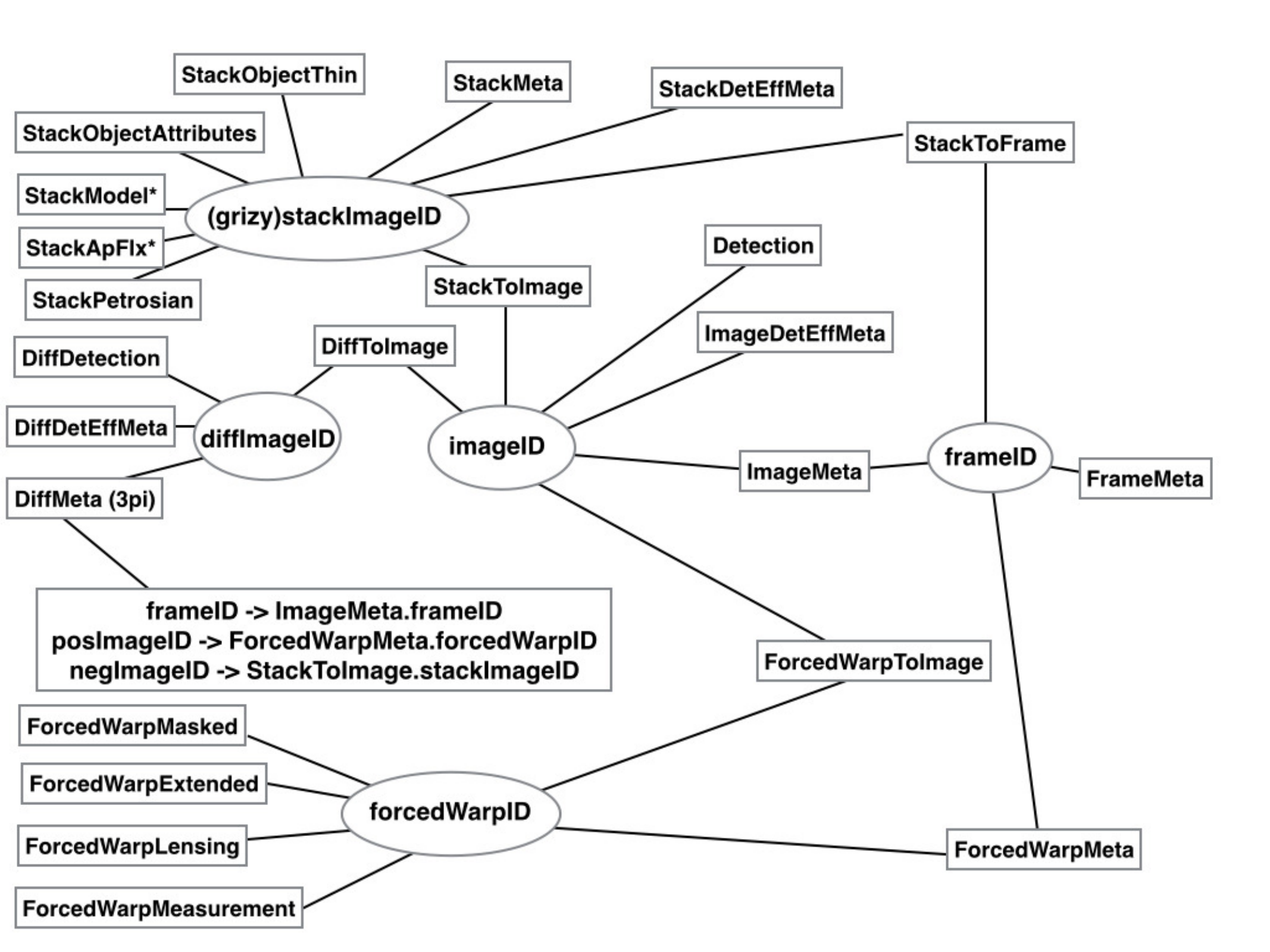}}
\caption{\noindent This diagram shows how to find the \texttt{imageID} and \texttt{frameID} from each of the different detection tables and metadata tables. The rectangular boxes with words inside represent the different table names.  The ovals with words inside represent the column names to use to join the tables.  Black lines connect table names to columns (i.e., {\em Detection} has a line to \texttt{imageID} which has a line to {\em ImageMeta}) - this shows that {\em Detection} and {\em ImageMeta} share the same index \texttt{imageID}). }
\label{fig:imageidmap}
\end{figure*}

\begin{table}
\caption{\noindent Summary of the different database tables, their types, and other comments. The column labeled `Release' specifies the first Data Release a specific product became available. Note that all of the DR1 tables were regenerated for DR2, in order to address minor bugs and inconsistencies discovered in DR1. }
\begin{center}
\begin{tabular}{llll}
\hline
\hline
PSPS Table Name & Table Type & Release\\
\hline
Filter  & System Metadata & DR1 \\
FitModel & System Metadata & DR1 \\
Survey & System Metadata & DR1 \\
PhotoCal & System Metadata & DR1 \\
StackType & System Metadata & DR1 \\
DiffType& System Metadata & DR1 \\
Tessellation Type& System Metadata & DR1 \\
ImageFlags& System Metadata & DR1 \\
DetectionFlags& System Metadata & DR1 \\
DetectionFlags2& System Metadata & DR1 \\
DetectionFlags3& System Metadata & DR1 \\
ObjectInfoFlags& System Metadata & DR1 \\
ObjectFilterFlags& System Metadata & DR1 \\
ObjectQualityFlags& System Metadata & DR1 \\
ForcedGalaxyShapeFlags& System Metadata & DR1 \\
FrameMeta& Observational Metadata & DR2 \\
ImageMeta& Observational Metadata & DR1/DR2 \\
Detection & Detection table & DR2 \\
ImageDetEffMeta& Observational Metadata &  DR2\\
ObjectThin  & Object / Mean properties  &  DR1\\ 
MeanObject & Object / Mean properties  &  DR1\\ 
StackMeta& Observational Metadata &  DR1 \\
StackObjectThin & Detection table &  DR1\\
StackObjectAttributes& Detection table &  DR1\\
StackApFlx& Detection table &  DR1\\
StackModelFitExp& Detection table &  DR1\\
StackModelFitDeV& Detection table &  DR1\\
StackModelFitSer& Detection table &  DR1\\
StackApFlxExGalUnc& Detection table &  DR1\\
StackApFlxExGalCon6& Detection table &  DR1\\
StackApFlxExGalCon8& Detection table &  DR1\\
StackPetrosian& Detection table &  DR1\\
StackToImage& Observational Metadata &  DR1 \\
StackToFrame& Observational Metadata &   DR1\\
StackDetEffMeta& Observational Metadata &   DR1\\
DiffMeta& Observational Metadata & DR3 \\
DiffDetection& Detection table &DR3 \\
DiffToImage& Observational Metadata & DR3 \\
DiffDetEffMeta& Observational Metadata & DR3 \\
DiffDetObject & Object / Mean properties  & DR3\\ 
ForcedWarpMeta& Observational Metadata & DR2 \\
ForcedWarpMeasurement& Detection table & DR2\\
ForcedWarpMasked& Detection table & DR2\\
ForcedWarpExtended& Detection table & DR2\\
ForcedWarpLensing& Detection table & DR2 \\
ForcedWarpToImage& Observational Metadata &  DR2 \\
ForcedGalaxyShape & Object / Mean properties  & DR2 \\ 
ForcedMeanObject & Object / Mean properties  &  DR1\\ 
ForcedMeanLensing & Object / Mean properties  & DR1 \\ 
GaiaFrameCoordinate & Object / Mean properties  & DR1 only \\ 
\hline
\end{tabular}
\end{center}
\label{table:pspstables}
\end{table}%

\begin{table*}
\caption{Fundamental IPP data product database tables}
\begin{center}
\begin{tabular}{lllll}
\hline
Table Class  & PSPS Table Name   & Source   & Note & Release \\
\hline
\hline
Detection    &  Detection           & dvo and cam smf   & 1     & DR2\\
Object       &  ObjectThin          &                   & 2     & DR1\\
             &  MeanObject          & dvo               & 3     & DR1\\
             &  GaiaFrameCoordinate & dvo              & 4     & DR1\\
Stack        & StackObjectThin      & dvo and skycal cmf & 5    & DR1\\
             & StackObjectAttributes & dvo and skycal cmf & 6    & DR1\\
             & StackApFlx & dvo and skycal cmf & 7    & DR1\\
             & StackApFlxExGalUnc & dvo and skycal cmf & 7    & DR1\\
             & StackApFlxExGalCon6 & dvo and skycal cmf & 7    & DR1\\
             & StackApFlxExGalCon8 & dvo and skycal cmf & 7    & DR1\\
             & StackPetrosian & dvo and skycal cmf & 7    & DR1\\
             & StackModelFitExp & dvo and skycal cmf & 8    & DR1\\
             & StackModelFitDeV & dvo and skycal cmf & 8    & DR1\\
             & StackModelFitSer & dvo and skycal cmf & 8    & DR1\\
Difference   & DiffDetection  & dvo and diff skycal cmf & 9  & DR2\\
Difference   & DiffDetObject  & dvo & 10  & DR2\\
Forced       & ForcedMeanObject & dvo &  11 & DR1\\
             & ForcedWarpMeasurement & dvo and forced warp cmf & 12 & DR2\\
             & ForcedMeanLensing & dvo & 13 & DR2\\
             & ForcedWarpLensing & dvo and forced warp cmf & 13 & DR2\\
             & ForcedGalaxyShape & dvo & 15 & DR2\\
             & ForcedWarpMasked & dvo and forced warp cmf & 16 & DR2\\
\hline            
\end{tabular}
\end{center}
\label{table:fundamentalipp}
\end{table*}


\begin{table}
\caption{Observational Metadata}
\begin{center}
\begin{tabular}{ll}
\hline
\hline
PSPS Table Name &  Rel.\\
\hline
FrameMeta&   \\
ImageMeta&   \\
ImageDetEffMeta&   \\
StackMeta&  DR1 \\
StackToImage&   DR1 \\
StackToFrame&    DR1\\
StackDetEffMeta&   DR1\\
DiffMeta&   \\
DiffToImage&  \\
DiffDetEffMeta&   \\
ForcedWarpMeta&   \\
ForcedWarpToImage&  \\
\hline
\end{tabular}
\end{center}
\label{table:observationalmetadata}
\end{table}%

\begin{table}
\caption{System Metadata tables.  Note that DR1 lacked flag information for various bits, this has been corrected for DR2.}
\begin{center}
\begin{tabular}{ll}
\hline
\hline
PSPS Table Name &  Release\\
\hline
Filter  & DR1 \\
FitModel & DR1 \\
Survey &  DR1 \\
PhotoCal &  DR1 \\
StackType &  DR1 \\
DiffType&  DR1 \\
TessellationType&  DR1 \\
ImageFlags& DR1 \\
DetectionFlags&  DR1 \\
DetectionFlags2&  DR1 \\
DetectionFlags3&  DR1 \\
ObjectInfoFlags&  DR1 \\
ObjectFilterFlags&  DR1 \\
ObjectQualityFlags& DR1 \\
ForcedGalaxyShapeFlags&  DR1 \\
\hline
\end{tabular}
\end{center}
\label{table:systemmetadata}
\end{table}%

\subsection{System Metadata Tables}
\label{sec:schemameta}


The system metadata tables primarily contain static information of flags, filters, surveys and other information that is specific to \PS.  

There are several tables that describe the different flag bits: {\em ImageFlags}, {\em DetectionFlags}, {\em DetectionFlags2}, {\em DetectionFlags3}, {\em ObjectInfoFlags}, {\em ObjectFilterFlags}, {\em ObjectQualityFlags}, and {\em ForcedGalaxyShapeFlags}. 

There are a handful of other system metadata tables: {\em Filter}, {\em Survey}, {\em FitModel}, {\em PhotoCal}, {\em StackType}, {\em DiffType}, {\em TessellationType}. The descriptions of each of those are provided below.

\vskip 0.2cm

\noindent {\bf \em Filter}: Contains a description of the optical filters used in the survey \citep{Tonry2012}. Filters (g,r,i,z,y) are assigned integer values from 1 to 5 (\texttt{filterID}).

\vskip 0.1cm
\noindent {\bf \em FitModel}: Contains descriptions of the models used in fitting detections in images, both PSF-like and extended galaxies \citep[see][]{deVaucouleurs1948,Sersic1963}. Describes the values for column \texttt{psfModelID} (located in various tables).

\vskip 0.1cm
\noindent {\bf \em Survey}: Contains descriptions of the various PS1 Science Consortium Surveys. The 3$\pi$ \texttt{SurveyID} is 0.

\vskip 0.1cm
\noindent {\bf \em PhotoCal}: Contains photometric calibration information for each filter and detector image combinations. Describes the values of \texttt{photoCalID} within {\em ImageMeta}, {\em StackMeta}, {\em ForcedWarpMeta}, and {\em DiffMeta}.

\vskip 0.1cm
\noindent {\bf \em StackType}: Contains descriptions of the types of stacked images constructed. For 3$\pi$, all {\em stacks} are \texttt{DEEP\_STACK}.  

\vskip 0.1cm
\noindent {\bf \em DiffType}: Contains descriptions of the types of difference images constructed. For 3$\pi$,  all {\em diffs} are \texttt{WARP\_STACK}, meaning they are constructed by subtracting warps from single exposures from the deep stacks for the corresponding part of sky. 

\vskip 0.1cm
\noindent {\bf \em TessellationType}: Contains descriptions of the types of image tessellations for the sky. For 3$\pi$, this is \texttt{RINGS.V3}.  Each MD field has its own {\em TessellationType} (\texttt{MD01.V3}, \texttt{MD02.V3}, etc.). The RINGS.V3 Tessellations are described in more detail in \citet{Magnier2017a}, and the MD tessellations in \citet{Huber2017}. 



\subsubsection{Flag Tables}
\label{sec:schemaflagsbitmasks}



\noindent There are 45 flag columns within the \PS\ database schema (example: Detection.infoFlag, Detection.infoFlag2, and others), and 8 different types of flags ({\em ObjectInfoFlags}/{\em ObjectQualityFlags}, {\em ObjectFilterFlags}, {\em ImageFlags}, {\em ForcedGalaxyShapeFlags}, {\em DetectionFlags}, {\em DetectionFlags2}, {\em DetectionFlags3}). The 8 different types of flags describes the bits in the flag columns (example: {\em ObjectInfoFlags} describes the flag bits for the column {\em ObjectThin}.objectInfoFlag) This section will give a brief overview of the 8 different types of flags. 
Table~\ref{table:flagtablestable} organizes the available flags and is intended to be used as a reference to select the appropriate flag information tables.

\vskip 0.2cm

\noindent {\bf \em ObjectInfoFlags} Contains information flag values for {\em ObjectThin}.objectInfoFlag, {\em ForcedMeanObject}.(grizy)Flags, {\em DiffDetObject}.objectInfoFlag, and {\em GaiaFrameCoordinate}.gaiaFlag.  These Flags are specifically useful for objects, several bits flag the object if it has been identified as QSO, variable, transient, or a known solar system object, if it has large proper motions, if it is extended, and the source of the average position information.  

\vskip 0.1cm
\noindent {\bf \em ObjectQualityFlags} Contains information flag values that denote if an object is real or a possible false positive. This is a subset of flags from {\em ObjectInfoFlags}, specifically the ones if the object is extended, has good measurements in individual exposures, and has good measurements in the stacks.  This describes the flags used in {\em ObjectThin}.qualityFlag and {\em DiffDetObject}.qualityFlag.  
 
\vskip 0.1cm
\noindent {\bf \em ObjectFilterFlags} Contains information flags for the photometric calibration of an object ({\em MeanObject}.(grizy)Flags). Bits include if it is ubercal'd, where photometry comes from (single exposures or stacks), information about synthetic photometry, and how the average magnitude was calculated. 

\vskip 0.1cm
\noindent {\bf \em ImageFlags} Contains information flag values for {\em ImageMeta}.qaFlags. Primarily flags of whether the image is bad or if there are too few measurements.  

\vskip 0.1cm
\noindent {\bf \em ForcedGalaxyShapeFlags} Contains information flag values that define ForcedGalaxyShape chi-squared surface fit failures, used in {\em ForcedGalaxyShape}.(grizy)GalFlags.

\vskip 0.1cm
\noindent {\bf \em DetectionFlags} Contains information flag values for  {\em Detection}.infoFlag, {\em StackObjectThin}.(grizy)infoFlag, {\em ForcedWarpMeasurements}.FinfoFlag and {\em DiffDetections}.DinfoFlag. Primarily information bits on sources, whether it is blended, used for PSF model, saturated, and many other types of defects, as well as information on types of magnitudes calculated, if it is extended, and fit information.

\vskip 0.1cm
\noindent {\bf \em DetectionFlags2} Contains information flag values for {\em Detection}.infoFlag2, {\em StackObjectThin}.(grizy)infoFlag2, {\em ForcedWarpMeasurements}.FinfoFlag2, and {\em  DiffDetections}.DinfoFlag2).  These flags contain bits specific to difference imaging, if source is near diffraction spikes, star core, burntool, flags relating to petrosians, and information on the fits (if they failed). 

\vskip 0.1cm
\noindent {\bf \em DetectionFlags3} Contains information flag values for {\em Detection}.infoFlag3, {\em StackObjectThin}.(grizy)infoFlag3, {\em ForcedWarpMeasurements}.infoFlag3, and {\em DiffDetections}.infoFlag3. Contains flags of whether detection was used for many different mean calculations (astrometric, photometric), and various associations with the DVO database. 



\newpage
\subsection{Object type tables}
\label{sec:schemaobject}

\noindent The object type tables originate from the DVO database, specifically, the DVO tables that have information about objects, their mean astrometric and photometric properties and information such as the number of detections per objects and other statistics and information.  The object type tables form the equivalent of a telephone book for all of the objects, with \texttt{ObjID} being the equivalent of the phone number or social security number.  A key defining feature is that \texttt{objID} is unique in these object type tables, there are no instances of 2 objects with the same \texttt{ObjID} in the same object type table. If an object is not in these tables, it has not been detected in any of the stages of processing.  The object type tables are {\em ObjectThin}, {\em MeanObject}, {\em DiffDetObject}, {\em ForcedMeanObject}, {\em ForcedMeanLensing}, {\em ForcedGalaxyShape} and {\em GaiaFrameCoordinate}.  For DR1, the available tables are {\em ObjectThin}, {\em MeanObject}, {\em ForcedMeanObject}, {\em ForcedMeanLensing}, and {\em GaiaFrameCoordinate}. What follows is a description of each of the object type tables, for DR1 and beyond.

\vskip 0.2cm
\noindent {\bf \em ObjectThin} Contains the positional information for objects in a number of coordinate systems. The objects associate single epoch detections and the stacked detections within a one arcsecond radius. The mean position from all the available single epoch data is used as the basis for coordinates when available, or the mean position over each filter of an object in the {\em stack} when it is not. The right ascension and declination for both the {\em stack} and single epoch mean is provided. The number of detections in each filter from single epoch data is listed, along with which filters the object has a {\em stack} detection \citep[see][]{Szalay2007}. Use \texttt{ObjID} to join to most tables; {\em uniquePspsOBid} to join to {\em MeanObject}.  For DR1 only, the Gaia-calibrated positions were calculated after {\em ObjectThin} was populated, they are provided in {\bf \em GaiaFrameCoordinate}.  For DR2 and beyond, there is no {\em GaiaFrameCoordinate table}, as ObjectThin has be re-calibrated and has the Gaia-calibrated positions.

\vskip 0.1cm
\noindent {\bf \em MeanObject} Contains the mean photometric information for objects based on the single epoch data, calculated as described in \citet{Magnier2013}. To be included in this table, an object must be bright enough to have been detected at least once in an individual exposure. PSF, \citet{Kron1980}, and total aperture-based \citep{Stoughton2002} magnitudes and statistics are listed for all filters. Use \texttt{ObjID} to join to most tables; {\em uniquePspsOBid} to join to {\em MeanObject}.


\vskip 0.1cms
\noindent {\bf \em DiffDetObject}	Contains the positional information for difference detection objects in a number of coordinate systems. The objects associate difference detections within a one arcsecond radius. The number of detections in each filter is listed, along with maximum coverage fractions \citep[see][]{Szalay2007}). Use \texttt{diffObjID} to join to most diff tables.  \texttt{diffObjID} and \texttt{uniquePspsDOid} are unique for {\em DiffDetObject}. Note that (\texttt{diffObjID} and  \texttt{objID} will be similar, but not identical, and it will not be easy to join to non-diff tables.  We recommend comparing the R.A. and Declination for objects between Diff* and non-diff tables.

\vskip 0.1cm
\noindent {\bf \em ForcedMeanObject} Contains the mean of single-epoch photometric information for sources detected in the stacked data, calculated as described in \citet{Magnier2013}. The mean is calculated for detections associated into objects within a one arcsecond correlation radius. PSF, \citet{Kron1980}, and SDSS aperture R5 ($r = 3.00$ arcsec), R6 ($r = 4.63$ arcsec), and R7 ($r = 7.43$ arcsec) total aperture-based \citep{Stoughton2002} magnitudes and statistics are listed for all filters. See also \citet{Kaiser1995,Magnier2013}.  Use \texttt{ObjID} to join to most tables, and use {\em uniquePspsFOid} to join to {\em ForcedMeanLensing}. \texttt{ObjID} is not unique, but \texttt{uniquePspsFOid} is.

\vskip 0.1cm
\noindent {\bf \em ForcedMeanLensing}  Contains the mean \citet{Kaiser1995} lensing parameters measured from the forced photometry of objects detected in stacked images on the individual single epoch data. Use \texttt{ObjID} to join to most tables; use \texttt{uniquePspsFOid} to join to {\em ForcedMeanObject}. \texttt{ObjID} is not unique, but \texttt{uniquePspsFOid} is.

\vskip 0.1cm
\noindent {\bf \em ForcedGalaxyShape} 	Contains the extended source galaxy shape parameters. All filters are matched into a single row. The positions, magnitudes, fluxes, and Sersic indices are inherited from their parent measurement in the {\em StackModelFit} tables, and are reproduced here for convenience. The major and minor axes and orientation are recalculated on a warp-by-warp basis from the best fit given these inherited properties \citep{Sersic1963}. Use \texttt{ObjID} to join to most tables. \texttt{ObjID} is not unique, but \texttt{uniquePspsFGid} is.

\vskip 0.1cm
\noindent {\bf \em GaiaFrameCoordinate} This table contains PSPS objects calibrated against Gaia astrometry, this is the best RA and Dec to use for an object.  Use \texttt{ObjID} to join to most tables. Note that this table is only present in the DR1 version of the database. The DR2 version of the database has the PSPS objects calibrated against Gaia astrometry in the ObjectThin tables. 

\subsection{Individual Exposure Detection Type Tables}
\label{sec:schemadetections}

\noindent The majority of the data in the database is in the form of detection type tables.  These are tables that are based on individual stages of processing from the IPP. Specifically, these are tables from the catalog outputs of {\em camera}, {\em stack}, {\em forced photometry} and {\em diff} stages of the IPP. Each of these categories of tables are described below.  

 Note that for these tables, if it has an objID column, objID will not unique, and there will be multiple entries with the same objID. If doing joins between tables from the same stage, do the join using the appropriate uniquePspsXXid. For example, to join between StackObjectThin and StackApFlx, use uniquePspsSTid.


\subsubsection{Tables based on the `camera' stage of IPP}
\label{sec:schemap2}

\noindent Images processed through the {\em camera} stage of the IPP have been detrended, and have had astrometry and photometry calculated.  Basic information from the images are then merged into the DVO database.  The core tables based on the {\em camera} stage are {\em FrameMeta}, {\em ImageMeta}, {\em Detection}, and {\em ImageDetEffMeta}. Each image ingested into the PSPS database has a unique \texttt{imageID}; this can be used to find out, via the {\em FrameMeta}, {\em ImageMeta}, and {\em ImageDetEffMeta} tables, information about each image such as the filter, RA and Dec, exposure time, etc.  All of the detections measured in the image are ingested into the Detection table, which also has the \texttt{imageID}, allowing for single detections to be traced back to the OTA that it was imaged on.  

\vskip 0.2cm

\noindent {\bf \em FrameMeta} Contains metadata related to an individual exposure. A {\em Frame} refers to the collection of all images obtained by the 60 OTA devices in the camera in a single exposure. The camera configuration, telescope pointing, observation time, and astrometric solution from the detector focal plane (L,M) to the sky (RA,Dec) is provided.

\vskip 0.1cm
\noindent {\bf \em ImageMeta} Contains metadata related to an individual OTA image that comprises a portion of the full exposure. The characterization of the image quality, the detrends applied, and the astrometric solution from the raw pixels (X,Y) to the detector focal plane (L,M) is provided.

\vskip 0.1cm
\noindent {\bf \em Detection} Contains single epoch photometry of individual detections from a single exposure. The identifiers connecting the detection back to the original image and to the object association are provided. PSF, aperture, and \citet{Kron1980} photometry are included, along with sky and detector coordinate positions. Use \texttt{ObjID} to join to other tables with photometry/astrometry parameters. \texttt{ObjID} is not unique.

\vskip 0.1cm
\noindent {\bf \em ImageDetEffMeta} Contains the detection efficiency information for a given individual OTA image. Provides the number of recovered sources out of 500 injected fake source and statistics about the magnitudes of the recovered sources for a range of magnitude offsets.

\subsubsection{Tables based on the `stack' stage of IPP}
\label{sec:schemast}

\noindent There are 15 tables based on the {\em stack} stage of processing.  The tables can be categorized into different groups, many of the tables are similar in shape (same column names), but based off of different stack Images (unconvolved vs convolved) or different extended models. The most important table is {\em StackObjectThin}.  This table contains the positional and photometric information for point-source photometry of stack detections. {\em StackObjectAttributes} extends information from {\em StackObjectThin} and it contains the PSF, Kron, and aperture fluxes for all objects.

StackApFlx, contains the unconvolved fluxes for 3 SDSS radii, for all sources.  ({\em StackApFLxExGalUnc}, {\em StackApFlxExGalCon6}, {\em StackApFlxExGalCon8}) are based on the (unconvolved, convolved to 6 pixels, convovolved to 8 pixels) fluxes for 9 SDSS radii, but only for sources not in the galactic plane. ({\em StackModelFitExp}, {\em StackModelFitDeV}, {\em StackModelFitSer}) all contain fit parameters for extended sources for different types of models (Exponential, DeVaucouleurs, Sersic). These are only measured outside the galactic plane and for objects with S/N$>20$. {\em StackPetrosian} contains the magnitudes and radii for extended sources.

There are several stack metadata tables, they are {\em StackMeta}, {\em StackToImage}, {\em StackToFrame} and {\em StackDetEffMeta}.  They provide general information about the stack and can be used to find out the exposures used in the stack.

Joins between any of the stack tables, except for the stack meta tables, should join using uniquePspsSTID.

\vskip 0.2cm
\noindent {\bf \em StackMeta} Contains the metadata describing the stacked image produced from the combination of a set of single epoch exposures.  The nature of the stack is given by the \texttt{StackTypeID}.  The astrometric and photometric calibration of the stacked image are listed.

\vskip 0.1cm
\noindent {\bf \em StackObjectThin}  Contains the positional and magnitude information for PSF, \citet{Kron1980} and total aperture-based measurements of stack detections.  The information for all filters are joined into a single row, with metadata indicating if this stack object represents the primary detection.  Due to overlaps in the stack tessellations, an object may appear in multiple stack images.  The primary detection is the detection that lies in the area of a stack image defined to provide the best sky coverage with minimal projection stretching.  As such areas are unique on the sky, all other detections of the object in that filter on other, overlapping stacks are secondary, regardless of their properties.  The detection flagged as best is the primary detection if that detection has a \texttt{psfQf} value greater than 0.98;  if that is not met, then any of the primary or secondary detections with the highest \texttt{psfQf} value is flagged as best \citep[see][]{Kron1980,Magnier2017a}. 

\vskip 0.1cm
\noindent {\bf \em StackObjectAttributes} Contains the PSF, \citet{Kron1980}, and total aperture-based fluxes for all filters in a single row, along with point-source object shape parameters. Negative fluxes are recorded. The magnitudes in {\em StackObjectThin} are derived from this table, where the flux is positive. See {\em StackObjectThin} table for discussion of primary, secondary, and best detections.

\vskip 0.1cm
\noindent {\bf \em StackApFlx} Contains the unconvolved fluxes within the SDSS R5 ($r = 3.00$ arcsec), R6 ($r = 4.63$ arcsec), and R7 ($r = 7.43$ arcsec) apertures \citep{Stoughton2002}.  Convolved fluxes within these same apertures are also provided for images convolved to 6 sky pixels (1.5 arcsec) and 8 sky pixels (2.0 arcsec).  All filters are matched into a single row.  See {\em StackObjectThin} table for discussion of primary, secondary, and best detections.


\vskip 0.1cm
\noindent {\bf \em StackModelFitExp} Contains the exponential fit parameters to extended sources.  See {\em StackObjectThin} table for discussion of primary, secondary, and best detections. 

\vskip 0.1cm
\noindent {\bf \em StackModelFitDeV} Contains the \citet{deVaucouleurs1948} fit parameters to extended sources.  See {\em StackObjectThin} table for discussion of primary, secondary, and best detections.  

\vskip 0.1cm
\noindent {\bf \em StackModelFitSer} Contains the \citet{Sersic1963} fit parameters to extended sources.  See {\em StackObjectThin} table for discussion of primary, secondary, and best detections.

\vskip 0.1cm
\noindent {\bf \em StackApFlxExGalUnc} Contains the unconvolved fluxes within the SDSS R3 ($r = 1.03$ arcsec), R4 ($r = 1.76$ arcsec), R5 ($r = 3.00$ arcsec), R6 ($r = 4.63$ arcsec), R7 ($r = 7.43$ arcsec), R8 ($r = 11.42$ arcsec), R9 ($r = 18.20$ arcsec), R10 ($r = 28.20$ arcsec), and R11 ($r = 44.21$ arcsec) apertures \citep{Stoughton2002} for extended sources.  These measurements are only provided for objects in the extragalactic sky, i.e., they are not provided for objects in the Galactic plane because they are not useful in crowded areas.  See {\em StackObjectThin} table for discussion of primary, secondary, and best detections. 

\vskip 0.1cm
\noindent {\bf \em StackApFlxExGalCon6} Contains the fluxes within the SDSS R3, R4, R5, R6, R7, R8, R9, R10, and R11 apertures \citep{Stoughton2002} for extended sources after the images have been convolved to a target of 6 sky pixels (1.5 arcsec).  These measurements are only provided for objects in the extragalactic sky, i.e., they are not provided for objects in the Galactic plane because they are not useful in crowded areas.  See {\em StackObjectThin} table for discussion of primary, secondary, and best detections.  

\vskip 0.1cm
\noindent {\bf \em StackApFlxExGalCon8} Contains the fluxes within the SDSS R3 , R4, R5, R6, R7, R8, R9, R10, and R11 apertures \citep{Stoughton2002} for extended sources after the images have been convolved to a target of 8 sky pixels (2.0 arcsec).  These measurements are only provided for objects in the extragalactic sky, i.e., they are not provided for objects in the Galactic plane because they are not useful in crowded areas.  See {\em StackObjectThin} table for discussion of primary, secondary, and best detections.  

\vskip 0.1cm
\noindent {\bf \em StackPetrosian} Contains the \citet{Petrosian1976} magnitudes and radii for extended sources.  See {\em StackObjectThin} table for discussion of primary, secondary, and best detections.  

\vskip 0.1cm
\noindent {\bf \em StackToImage} Contains the mapping of which input images were used to construct a particular stack.

\vskip 0.1cm
\noindent {\bf \em StackToFrame} Contains the mapping of input frames used to construct a particular stack along with processing stats.

\vskip 0.1cm
\noindent {\bf \em StackDetEffMeta} Contains the detection efficiency information for a given stacked image.  Provides the number of recovered sources out of 500 injected sources for each magnitude bin and statistics about the magnitudes of the recovered sources for a range of magnitude offsets.

\subsubsection{Tables from the `forced photometry' stage of IPP}
\label{sec:schemafw}
\noindent The {\em forced photometry} stage of the IPP uses the positions of detections found on the {\em stacks} as a list of positions to force calculations of photometry on the individual {\em warps} from images.  The {\em warp} images are essentially the detrended images rotated, rebinned, and chopped into skycells.  For a given {\em stack} skycell with a list of detected objects, the corresponding {\em warps} (from individual images) will have their photometry calculated, based on the list of detected objects in the {\em stack} skycell.  More extended photometric calculations are performed in the {\em forced photometry} stage compared to the {\em camera} stage, for example, lensing parameters and extended photometry is calculated for this stage and not for the {\em camera} stage.  Similar to the {\em stack} and {\em camera} stages, the end product of this stage of processing is a catalog file, basic information from the catalog file is ingested into the DVO, and then information from both the DVO and the catalog file are later ingested into the PSPS.  Basic information about each {\em forced warp} skycell can be found in {\em ForcedWarpMeta} and a mapping to the individual exposures can be found in {\em ForcedWarpToImage}, use \texttt{forcedWarpID} and \texttt{imageID} to make these joins. Tables with photometric information are in {\em ForcedWarpMeasurement} and {\em ForcedWarpExtended}.  {\em ForcedWarpMasked} allows one to know if there were no unmasked pixels for a specific {\em forced warp} measurement.  Joins between {\em ForcedWarpMeasurement}, {\em ForcedWarpMasked}, {\em ForcedWarpExtended}, {\em ForcedWarpLensing} should use uniquePspsFWid.

\vskip 0.2cm

\noindent {\bf \em ForcedWarpMeta} Contains the metadata related to a sky-aligned distortion corrected {\em warp} image, upon which forced photometry is performed. The astrometric and photometric calibration of the {\em warp} image are listed. 

\vskip 0.1cm
\noindent {\bf \em ForcedWarpMeasurement} Contains single epoch forced photometry of individual measurements of objects detected in the stacked images. The identifiers connecting the measurement back to the original image and to the object association are provided. PSF, total aperture-based, and \citet{Kron1980} photometry are included, along with sky and detector coordinate positions.

\vskip 0.1cm
\noindent {\bf \em ForcedWarpMasked} Contains an entry for objects detected in the stacked images which were in the footprint of a single epoch exposure, but for which there are no unmasked pixels at that epoch.

\vskip 0.1cm
\noindent {\bf \em ForcedWarpExtended} Contains the single epoch forced photometry fluxes within the SDSS R5 ($r = 3.00$ arcsec), R6 ($r = 4.63$ arcsec), and R7 ($r = 7.43$ arcsec) apertures \citep{Stoughton2002} for objects detected in the stacked images. 

\vskip 0.1cm
\noindent {\bf \em ForcedWarpLensing} Contains the \citet{Kaiser1995} lensing parameters measured from the forced photometry of objects detected in stacked images on the individual single epoch data. 

\vskip 0.1cm
\noindent {\bf \em ForcedWarpToImage} Contains the mapping of which input image comprises a particular {\em warp} image used for forced photometry.

\subsubsection{Tables based on the `diff' stage of IPP}
\label{sec:schemadiff}

\noindent The {\em diff} stage for 3$\pi$ processing subtracts {\em warp} images from the deep {\em stack} images.  Other types of {\em diffs} are also possible, but are not used for the 3$\pi$ database.  There is one {\em diff} image for each single exposure within the 3$\pi$ survey. The {\em diff} input images (the {\em stack} and {\em warp} images) are convolved to have similar PSFs \citep{Waters2017}.  Once the {\em diff} images are created, basic photometry is performed in order to find new sources, and {\em diff} catalog files are created.  The {\em diff} catalog files are then ingested into the {\em diff} DVO, and later ingested into the PSPS.  There are 4 database tables for this stage of processing.  Each {\em diff} has a unique {\em diffImageId}, and all 4 {\em diff} tables use this to join to each other. {\em DiffMeta}, {\em DiffToImage}, and {\em DiffDetEffMeta} are metadata tables that describe basic properties of the difference images as well as allows the {\em diffs} to be mapped to the {\em stacks} and single exposures that were used in the creation of the diffs.  {\em DiffDetection} contains the photometry from the detections measured from the difference images.  More details about these tables are described below.

\vskip 0.2cm

\vskip 0.1cm
\noindent {\bf \em DiffMeta} Contains metadata related to a difference image constructed by subtracting a stacked image from a single epoch image, or in the case of the MD Survey from a nightly {\em stack} (stack made from all exposures in a single filter in a single night). The astrometric calibration of the reference {\em stack} is listed.

\vskip 0.1cm
\noindent {\bf \em DiffDetection} Contains the photometry of individual detections from a difference image. The identifiers connecting the detection back to the difference image and to the object association are provided. PSF, aperture, and \citet{Kron1980} photometry are included, along with sky and detector coordinate positions. 

\vskip 0.1cm
\noindent {\bf \em DiffToImage} Contains the mapping of which input images were used to construct a particular difference image. 

\vskip 0.1cm
\noindent {\bf \em DiffDetEffMeta} Contains the detection efficiency information for a given individual difference image. Provides the number of recovered sources out of 500 injected sources and statistics about the magnitudes of the recovered sources for a range of magnitude offsets. 
\subsection{Description of \texttt{ObjID} and its relation to R.A. and Dec.} 
\label{sec:schemaobjid}

\noindent The indices \texttt{objID} and \texttt{diffObjID} are derived from right ascension and declination.  The derivation is the same for both. While it is possible to calculate the RA and Dec from the \texttt{objID}, it is not recommended to do this, because the \texttt{objID} is based on the astrometric solutions from individual exposures and {\em stacks} as they are ingested into the DVO database, and is not calibrated against 2MASS or Gaia. It is recommended to use \texttt{raMean}, \texttt{decMean} from {\em ObjectThin} or (for DR1 only) \texttt{ra}, \texttt{dec} from {\em GaiaFrameCoordinate}.  For DR1, {\em ObjectThin} is calibrated against 2MASS, and {\em GaiaFrameCoordinate} is calibrated against Gaia.  For DR2 and beyond, {\em ObjectThin} is calibrated against 2MASS and Gaia, and there is no {\em GaiaFrameCoordinate} as the gaia-calibrated coordinates are included in {\em ObjectThin}. Included below is the C code for the translation between R.A. and Dec., for users interested in the relationship.

\vskip 0.5cm

\texttt{
uint64\_t \\
CreatePSPSObjectID(double ra, double dec) \\
\{\\
    double zh = 0.0083333;\\
    double zid = (dec + 90.) / zh;             // 0 - 180*60*2 = 21600 < 15 bits\\
    int izone = (int) floor(zid);\\
    double zresid = zid -  ((float) izone);    // 0.0 - 1.0\\
\\
    uint64\_t part1, part2, part3;\\
    part1 = (uint64\_t)( izone  * 10000000000000LL) ; \\
    part2 = ((uint64\_t)(ra * 1000000.)) * 10000 ; // 0 - 360*1e6 = 3.6e8 (< 29 bits)\\
    part3 = (int) (zresid * 10000.0) ; // 0 - 10000 (1 bit == 30/10000 arcsec) (< 14 bits)\\
\\
    return part1 + part2 + part3;\\
\}\\
}

\begin{figure}

\centerline{\includegraphics[width=\columnwidth,angle=0]{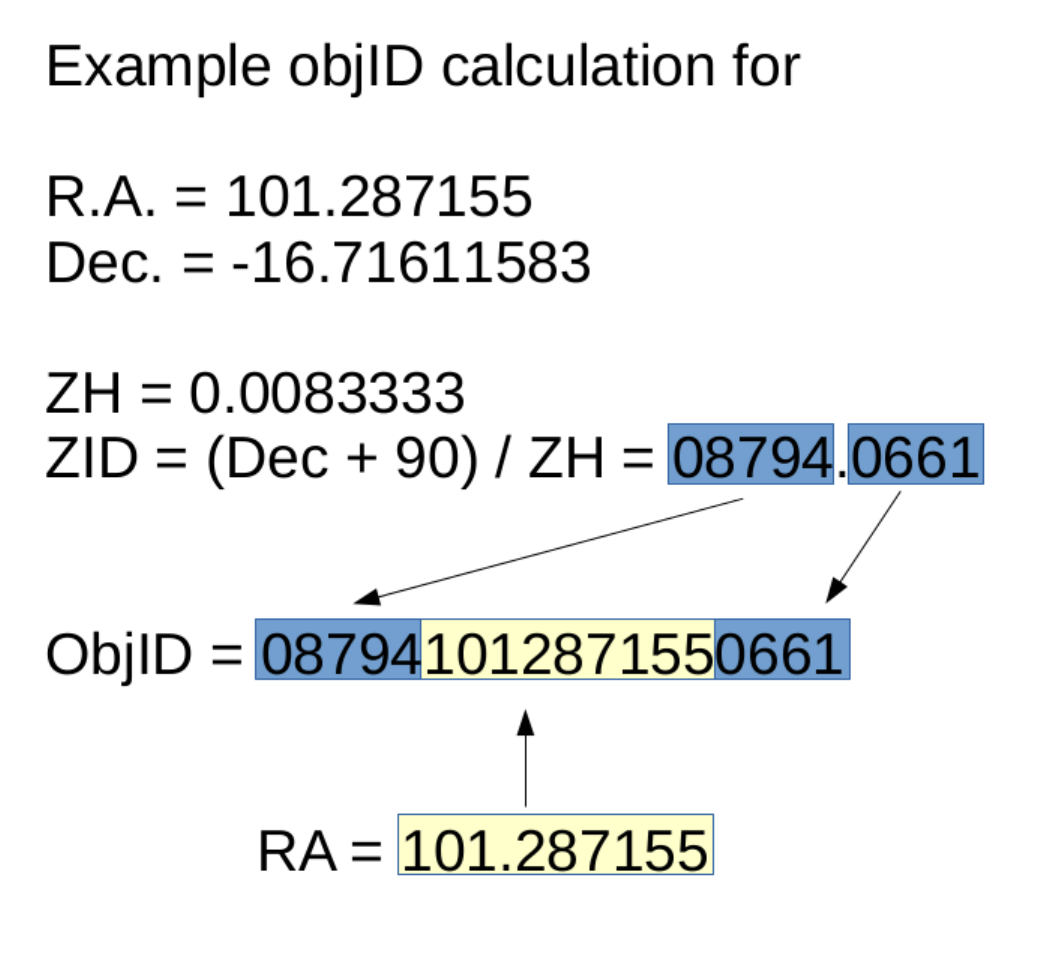}}

\vskip -0.5cm
\caption{Graphical description of how \texttt{ObjID} is calculated from RA and Dec. It is not recommended to derive the RA and Dec from the \texttt{objID} as this will result in an innaccurate RA and Dec, the \texttt{ObjID} is assigned when the stack/skycal cmfs are ingested into the database, and are not yet calibrated against 2MASS and Gaia. ObjID is primarily used for indexing the database. }
\vskip 0.5cm
\label{fig:objiddef}
\end{figure}

\subsection{Which RA and Dec to use?}
\label{sec:schemaradec}

\noindent There are multiple tables that contain columns that provide the R.A. and Dec.; this section gives information on each so that the user can choose the appropriate R.A. and Dec. A summary of these tables are provided in Table~\ref{table:radec} Generally, if the user is not interested in proper motion or moving objects, the best R.A. and Dec. to use is the R.A. and Dec. in {\em GaiaFrameCoordinate} if using DR1, as this is the weighted mean RA and Dec (similar to {\em ObjectThin}), but calibrated to Gaia.  The reason why this is in a separate table and not part of {\em ObjectThin} is because the mean properties were calculated and ingested into PSPS prior to Gaia's DR1.  {\em ObjectThin}'s \texttt{raMean} and \texttt{decMean} is calibrated against 2MASS, and is different from Gaia on average by {\color{red}X} \citep{Magnier2017c}. 

For DR2, there is no {\em GaiaFrameCoordinate} table, as DR2's {\em ObjectThin} table is already calibrated to Gaia. The best R.A. and Dec. to use for most cases is ObjectThin's \texttt{raMean} and \texttt{decMean}. If the proper motion is high, or the object is moving, and the user is interested in the single epoch photometry, they should use \texttt{ra} and \texttt{dec} in the {\em Detection} table. 

\begin{table*}
\caption{Various locations of RA and Dec columns}
\begin{center}
\begin{tabular}{cccc}
\hline
\hline

PSPS Table & RA column name & Dec column name & comments\\
\hline
FrameMeta & raBore & decBore & RA/Dec of telescope boresite \\
ObjectThin & raMean & decMean & mean RA and Dec from single exposure, calibrated against 2MASS \\
ObjectThin & raStack & decStack & mean RA and Dec calculated from {\em stack} skycells \\
Detection & RA & Dec & RA and Dec for single exposure detections \\
StackObjectThin & (grizy)ra & (grizy)dec & RA and Dec calculated from individual {\em stack} skycells \\
DiffDetection & RA & Dec & RA and Dec for single {\em diff} exposure detections  \\
DiffDetObject & RA & Dec & similar to raMean/decMean, calculated for {\em diff} objects \\
GaiaFrameCoordinate & RA & Dec & \textbf{Best RA and Dec, recalibrated to Gaia.}\\
\hline
\end{tabular}
\end{center}
\label{table:radec}
\end{table*}

\subsection{Indexes and Joins}
\label{sec:schemaindexjoins}

There are multiple columns within the schema that are indexed and designed to be used to join tables together. Generally, if a column name ends in ``\texttt{ID}", it is designed to be joined to other tables, either to system metadata tables (examples include \texttt{filterID}, \texttt{surveyID}, 
\texttt{ccdID}), or to fundamental data tables (for example, \texttt{}{objID}, \texttt{}{diffObjID}, \texttt{uniquePspsSTid}). There are a few exceptions, \texttt{randomID} and \texttt{random(stage)ID} should not be used for joins, the contents of each are random numbers to aid the user in selecting repeatable random subsets of data. Also, some of the \texttt{uniquePspsXXIds} are only present in one table, they are used to provide unique IDs for each stage of processing but some stages ({\em DiffDetObject}, for example) only have one corresponding table and therefore nothing to join to.

\noindent Figure~\ref{fig:objidmap} and Figure~\ref{fig:imageidmap} show graphical representations of how to join various tables.  On these figures, the table names are boxed, while the columns to be used to join are in ovals.  For the diagram with \texttt{objID} in the middle, it shows that any of the boxes connected to \texttt{objID} can be joined to each other using \texttt{objID}.  

\noindent All tables with photometric or astrometric information involving different sources or objects have an index, called \texttt{objID}.  \texttt{ObjID} is only unique for the object type of tables, and is loosely based on RA and Dec, see Section~\ref{sec:schemaobjid} for more information. It is possible to use the \texttt{ObjID} to get a rough estimate of the RA and Dec, but this should not be used for the definitive RA and Dec. Use {\em ObjectThin} to get the RA and Dec calibrated to 2MASS, and use {\em GaiaFrameCoordinate}(for DR1) or {\em ObjectThin}(for DR2)to get the RA and Dec calibrated to Gaia astrometry.  When available and possible, if joining 2 tables and they both have the same column name like \texttt{uniquePspsXXId}, join those 2 tables using the \texttt{uniquePspsXXId}. See Table~\ref{table:uniqueuepsps} to see the \texttt{uniquePsps} column name.  \texttt{UniquePspsXXId} is designed to be unique, specifically for the cases of when there are multiple detections that are sufficiently close by, they will have the same \texttt{objID} but different \texttt{uniquePspsXXId}s.  
 
\noindent It is possible to join every detection, no matter what stage of processing it is from, back to the original exposure(s) and to the OTAs.  Figure~\ref{fig:imageidmap} shows how to do this.  For each stage of processing, there is an associated (stage)imageID that is mapped back to the \texttt{imageID} via tables of the name \texttt{(stage)ToImage}.  For example, if one wanted to find out which exposures contributed to a detection in {\em StackObjectThin}, they would join to {\em StackToImage} using the \texttt{stackImageID}.  This allows the user to find data within the database as well as to find out the corresponding images to download from MAST.

\begin{table}
\caption{Unique indicies for various PSPS tables}
\begin{center}
\begin{tabular}{cccc}
\hline
\hline
PSPS Table & uniquePspsID name & Release\\
\hline
ObjectThin & uniquePspsOBid  & DR1 \\
MeanObject & uniquePspsOBid & DR1 \\
Detection & uniquePspsP2id & DR2\\
DiffDetection  & uniquePspsDFid & DR2\\
DiffDetObject& uniquePspsDOid & DR2\\
ForcedGalaxyShape& uniquePspsFGid & DR2\\
ForcedMeanObject& uniquePspsFOid & DR1\\
ForcedMeanLensing& uniquePspsFOid & DR1\\
ForcedWarpMeasurement& uniquePspsFWid & DR2\\
ForcedWarpExtended& uniquePspsFWid & DR2\\
ForcedWarpLensing& uniquePspsFWid & DR2\\
ForcedWarpMasked& uniquePspsFWid & DR2\\
GaiaFrameCoordinate& uniquePspsGOid & DR1\\
StackObject* (2 tables)& uniquePspsSTid & DR1\\
StackApFlx* (4 tables)& uniquePspsSTid & DR1\\
StackModelFit* (4 tables)& uniquePspsSTid & DR1\\
StackPetrosian& uniquePspsSTid & DR1\\
\hline
\end{tabular}
\end{center}
\label{table:uniqueuepsps}
\end{table}%
 
\subsection{NULLS as $-999$}
\label{sec:schemanulls}

The PSPS uses \texttt{-999} to denote \texttt{NULL} values, as PSPS is based off of CasJobs which also does not use NULL. The justification for this is  explained at the following url: \url{http://skyserver.sdss.org/edr/en/sdss/skyserver/}. Specifically, they state "We also insist that all fields are non-null. These integrity constraints are invaluable tools in detecting errors during loading and they aid tools that automatically navigate the database.", and since our own database design has in its roots many of the same parts as the SDSS database, we also adopt this convention of non-null fields.




\begin{table}
\caption{Flag types within \PS}
\begin{center}
\begin{tabular}{cccc}
\hline
\hline
Flag Type &  Size   &  Comments \\
\hline
ObjectInfoFlags & INT &  \\
ObjectQualityFlags & SMALLINT & \\
ObjectFilterFlags & INT  & \\
ImageFlags & INT & \\
ForcedGalaxyShapeFlags & SMALLINT & \\
Detection & BIGINT & \\
Detection2 & INT & \\
Detection3 & INT & \\
\hline
\end{tabular}
\end{center}
\label{table:flagtypetables}
\end{table}%

\begin{table*}
\caption{Flag columns within \PS}
\begin{center}
\begin{tabular}{cccc}
\hline
\hline
Table Name  &     Flag Column    &  Size  &   Flag Information Table \\
\hline
ObjectThin  & objectInfoFlag  &  INT          & ObjectInfoFlags \\
ObjectThin  &  qualityFlag     &   SMALLINT   &   ObjectQualityFlags \\
GaiaFrameCoordinate & gaiaFlag & INT & ObjectInfoFlags \\
MeanObject & (grizy)Flags          & INT  & ObjectFilterFlags \\
ForcedMeanObject & (grizy)Flags & INT & ObjectInfoFlags \\
DiffDetObject & objectInfoFlag & INT & ObjectInfoFlags \\
DiffDetObject & qualityFlag & SMALLINT & ObjectQualityFlags  \\
ImageMeta  & qaFlags       & INT & ImageFlags \\
ForcedGalaxyShape   & (grizy)GalFlags  & SMALLINT & ForcedGalaxyShapeFlags \\
Detection    & infoFlag  & BIGINT   & DetectionFlags \\
Detection    & infoFlag2  & INT   & DetectionFlags2 \\
Detection    & infoFlag3 & INT   & DetectionFlags3 \\
StackObjectThin    & (grizy)infoFlag  & BIGINT   & DetectionFlags \\
StackObjectThin    & (grizy)infoFlag2  & INT   & DetectionFlags2 \\
StackObjectThin    & (grizy)infoFlag3 & INT   & DetectionFlags3 \\
DiffDetection    & DinfoFlag  & BIGINT   & DetectionFlags \\
DiffDetection    & DinfoFlag2  & INT   & DetectionFlags2 \\
DiffDetection    & DinfoFlag3 & INT   & DetectionFlags3 \\
ForcedWarpMeasurement    & FinfoFlag  & BIGINT   & DetectionFlags \\
ForcedWarpMeasurement    & FinfoFlag2  & INT   & DetectionFlags2 \\
ForcedWarpMeasurement    & FinfoFlag3 & INT   & DetectionFlags3 \\
\hline
\end{tabular}
\end{center}
\label{table:flagtablestable}
\end{table*}%

\subsection{Tables and Views}
\label{sec:schemaviews}

\noindent The PSPS has defined several views to aid the user in making database queries.  A view is a virtual table that is based on the results from an SQL statement, and looks like a table to the user.  These views are constructed to aid the user, to alleviate the need for common joins, and produce query results faster than joins. Table~\ref{table:views} describes the views currently in PSPS. 


\begin{table*}
\caption{Currently defined views within PSPS}
\begin{center}
\resizebox{\textwidth}{!}{
\begin{tabular}{ccc}
\hline
\hline
View Name  &  Tables used to create view\\
\hline
DetectionObjectView & ObjectThin, MeanObject, Detection\\
MeanObjectView & ObjectThin, MeanObject\\
StackObjectView & ObjectThin, StackObjectThin, StackObjectAttributes\\ 
StackApFlxObjectView & ObjectThin, StackApFlx\\ 
StackApFlxExGalUncObjectView & ObjectThin, StackApFlxExGalUnc\\
StackApFlxExGalCon6ObjectView & ObjectThin, StackApFlxExGalCon6\\
StackApFlxExGalCon8ObjectView & ObjectThin, StackApFlxExGalCon8\\
StackModelObjectView & ObjectThin, StackModelFitExp, StackModelFitDeV, StackModelFitSer, StackPetrosian\\     
StackModelFitExpObjectView & ObjectThin, StackModelFitExp\\
StackModelFitDeVObjectView & ObjectThin, StackModelFitDeV\\
StackModelFitSerObjectView& ObjectThin, StackModelFitSer\\
StackModelFitPetObjectView & ObjectThin, StackPetrosian&\\
StackObjectPrimaryView& ObjectThin,StackObjectThin,StackObjectAttributes,StackApFlx\\
DiffDetObjectView& DiffDetObject,DiffDetection\\
ForcedDetObjectView&ObjectThin,ForcedWarpMeasurement\\
ForcedMeanObjectView&ObjectThin,ForcedMeanObject\\
ForcedGalaxyModelView &ObjectThin,ForcedGalaxyShape\\
\hline
\end{tabular}}
\end{center}
\label{table:views}
\end{table*}%

\section{The $3\pi$ Database}
\label{sec:threepidatabase}

The $3\pi$ Survey is described in detail in \citet{Chambers2017}, it covers 3/4 of the sky (everything with Dec $> -30$) in 5 bands (\grizy), with approximately 60 exposures per patch of sky. For various reasons, a very tiny fraction of the data does not make it into the PSPS database.  Some of the data will be ingested at a later time (DR2 and beyond), however some will not be ingested at all.  We present the different sources of missing data that we are aware of. This section describes the numbers of exposures that were processed in each stage, including counts for what were expected and as well as counts for known faults and quality issues. 





\subsubsection{IPP Processing stages}
\label{sec:3piprocessing}

All of the $3\pi$ data was reprocessed in a consistent way with the same version of IPP code, internally called Processing Version 3 (PV3).  Details of this are in \citet{Magnier2017a}. Small fractions of the raw image data were simply not processed at all, or failed at processing and were not recovered.  This can happen at various stages of processing, and we can report the expected numbers here. A more detailed accounting of the various causes that lead to data being filtered out of PV3 requires further documentation and is beyond the scope of this section.



There are 388,177 raw exposures taken between 2009-06-03 and 2015-02-25, that have metadata indicating that they were observed for the 3$\pi$ survey. However, only 381,279 ($98.2\%$) of these were queued for the first stage of processing, the {\em chip} stage. These exposures were excluded because they did not meet various requirements: they were flagged by the observers to not be used, or they failed to be processed by nightly science processing because of bad seeing or camera issues or other observational or telescope conditions that can ruin an exposure. 

Of the 381,279 exposures that were queued up for {\em chip} processing, 375,573 ($98.5\%$) completed PV3 processing with good quality: no issues detected while detrending or finding sources and performing photometry.  

The {\em camera} stage started with these 375,573 exposures, of which 374,521 ($99.7\%$) completed processing.  The ones that did not, failed because of bad quality: no stars for astrometry analysis, failed single chip astrometry, or failed mosaic astrometry. 

The {\em fake/warp} stages started off with a larger number of exposures than expected: 379,973 instead of 374,521.  This is due to the fact that the images were processed on different computing clusters, some with more computing power, which presented additional difficulties.  Of the 379,973 exposures processed, 374,339 ($98.5\%$) are unique, and 1,234 are duplicates. Of the 379,973, 379,551 ($99.9\%$) have good quality. The {\em warp} stage is the first stage that repartitions the exposures into the skycell tessellation, and since all later stages process on a skycell level rather than an exposure level, it makes sense to report that the {\em warp} stage has 206177 distinct skycells. 

The {\em stack} stage has 200,730 distinct skycells, 200,725 of which have good quality.  This is less than the above number 206177, because there is a declination cut on creating stacks, and {\em stacks} at declination less than the lower cut will have significantly less coverage and are not useful here.  The static sky stage stage has 200,720 distinct skycells that are processed and all have good quality. The {\em skycal} stage, the final stage of {\em stack} processing, started with 200,722 skycells, of which 2 were essentially duplicate {\em stacks} (same inputs, same skycells), 200,684 processed with good quality data.

The {\em stack} stages have a couple of additional inconsistencies.  First of all, of the 200,684 {\em stacks} that we expect, there are 409 {\em stacks} ($0.2\%$) that include duplicate exposures, i.e. the same exposure appears in a {\em stack} twice.  There are also 130 {\em stacks} with camera testing files included (short exposures, 1 second, which have been mislabeled as 3$\pi$ data and managed to get successfully processed).  These inconsistencies were discovered during the {\em IppToPsps} and PSPS phases, and will be present in the database.  Secondly, there are two {\em stack} duplicates present (i.e. two {\em stacks} for the same filter and skycell), these are also present in the PSPS. 

The {\em forced warp stages} Forced warps are queued by skycell and filter, and there are 994,890 of these, resulting in 19,266,450 cmf files (one per exposure per skycell) in 373,743 exposures and 199151 distinct skycells.   We see no inconsistencies in the numbers. There are slighltly less distinct skycells than in the stack stages. This is because there is a ragged edge at the survey limit of -30 (not consistently observed at that declination), and it made no sense to include those in forced photometry processing.

\subsection{Building the 3$\pi$ DVO Database}
\label{sec:building3pi}

The $3\pi$ database was built in stages, with many checks to verify all of the data was included. Full details of the construction and final calibration are in \citet{Magnier2017c}.  The $3\pi$ database contains the following counts of data for various processing stages, ordered by how they were ingested into the database:

\begin{description}
\item[stack/skycal cmf] The end product from the skycal stage produces 1 cmf file per stack skycell per filter. There are 1-5 cmfs per skycell, corresponding to different filters.  We expected 998,101 skycal cmfs, in 200,684 distinct skycells.  All were successfully ingested into the DVO.

\item[camera smfs] The camera stage produces 1 cmf file per exposure, with extensions for each of the 60 chips. We expected 374,521 camera stage smf, of which 374,446 of these were ingested with no faults.  There are 75 that repeatedly failed to be ingested in the DVO and will not be included in the PSPS. 

\item[forced warp cmfs]  The forced warp photometry stage produces many cmf files: each of the exposures that are within a given skycell will produce a cmf that is ingested into the database.  There are 19266450 of these cmf files, they come from 373,743 exposures, and are in 994,890 skycells. 

\item[forced warp summary cmfs] The forced warp photometry stage also produces a summary cmf of the mean properties for a given skycell and filter, based on the forced photometry results for all of the included exposures that are within a given skycell.  There are 994,890 of these cmfs, in 199,151 distinct skycells. There are 1-5 cmfs per skycell, corresponding to different filters.
\end{description}

Very tiny amounts of data were not ingested due to quality issues, and there are very minor duplicate issues with processing resulting in some duplicate files being ingested.  It is possible to remove those duplicates, but then the mean properties must be recalculated.  There are a small number of duplicates that were discovered in {\em ippToPsps} and PSPS, and it was not possible to remove and recalibrate them at this time.  


\subsection{IppToPsps Stage}
IppToPsps generates batches that includes all of the smf/cmf files categorized by stage of processing, and generates batches corresponding to each DVO file.  It is straightforward to verify all the data is accounted for and has been processed through ippToPsps, and easy to regenerate missing batches. This section describes the batch types and expected numbers.

\noindent {Stacks}: {\em IppToPsps} expected to generate 200,684 {\em stack} batches.  It created 200,681 batches (missing 3) for DR1, these can be recovered and they will be added shortly after DR1. For DR2, it generated 200,683 batches.  The 1 missing stack batch ended up being unsuitable data quality for the database.

\noindent {\em ObjectThin/MeanObject}: {\em IppToPsps} generated all of the expected 116,252 batches for DR1, and 111,505 for DR2.  There are less batches for DR2 because it avoids the ragged edge of the survey and excludes those with Dec $<-30$.

\noindent {\em GaiaFrameCoordinate}: {\em IppToPsps} generated all of the expected 116,252 batches for DR1. This batch type was not needed for DR2.

\noindent {\em Forced Mean Objects}: {\em IppToPsps} generated all of the expected 113,665 {\em Forced Object} batches for DR1, and 113,167 batches for DR2.  There are less batches than for ObjectThin because it avoids the ragged edge of the survey at Dec $= -30$. 

\noindent {\em Detection Table}: We expect 374,446 {\em Detection} batches, and we generated 374,344. The 102 missing batches were incorrectly marked as good exposures, are not science quality, and will not be added into the database.

\noindent {\em ForcedWarp Tables}: We expect 994,890 {\em Forced Warp} batches, and we generated 994,826. 

\noindent {\em Forced Galaxy Objects}: We expect 57,758 {\em Forced Galaxy Objects} and we generated 57758 {\em Forced Galaxy Object} batches. Not all areas of sky will have forced galaxy objects measured, so although the batches are subdivided in the sky the same way as for {\em ObjectThin/MeanObject} and {\em ForcedMeanObject}, we should expect a smaller number (57,758) than for the other Object types (typically > 110k).


\subsection{Loading into PSPS}

\begin{figure}
\vskip -0.2cm
\centerline{\includegraphics[width=1.1\columnwidth,angle=0]{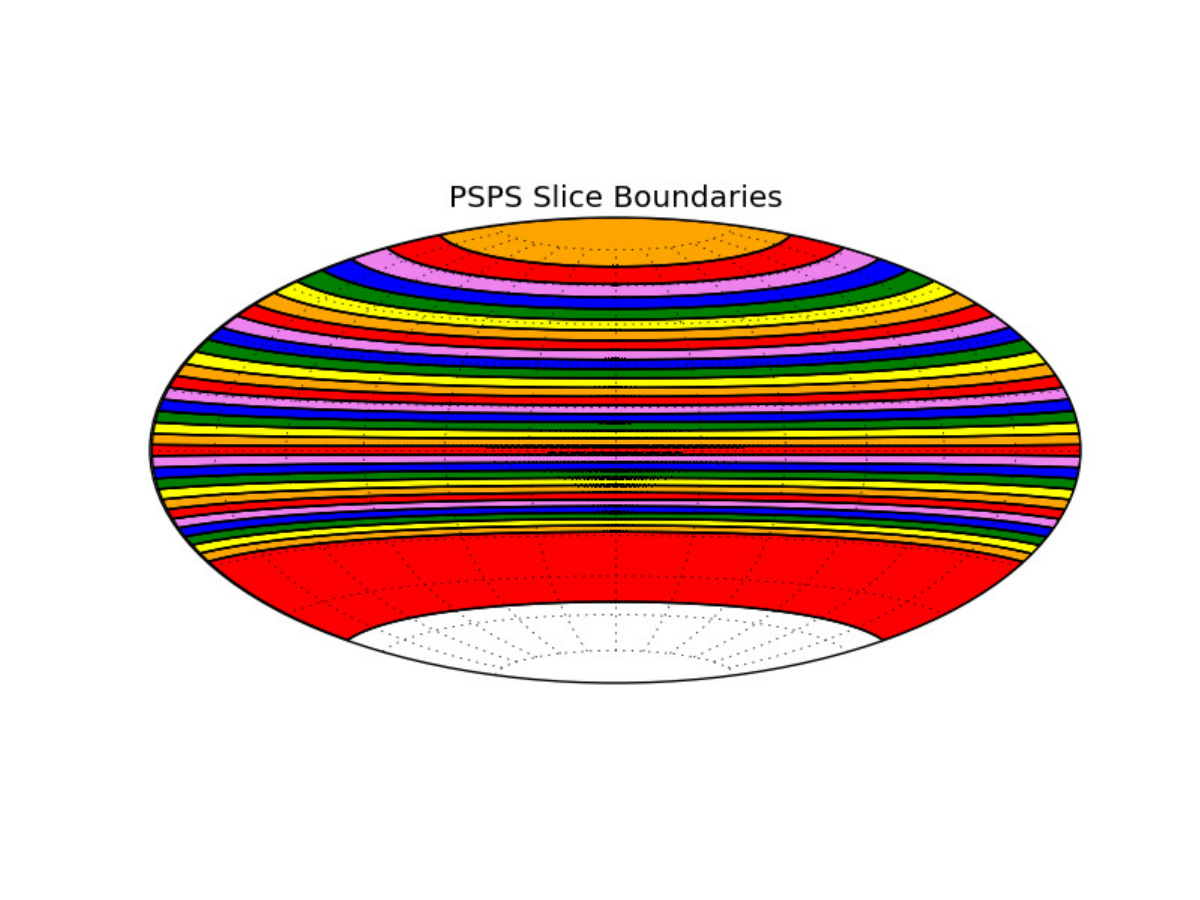}}
\vskip -1.5cm
\caption{\noindent The PSPS $3\pi$ database is subdivided into 32 slices, i.e, one slice per database machine. These 32 slices have Dec boundaries defined such that each slice contains a similar amount of data. The database tables ({\em Detectio}, {\em StackObjectThin}, etc.) are composed of views combining the 32 slices.  This aspect of the database is intended to be transparent to the user, however, power users may find it useful to know how the slices are subdivided.}
\label{fig:pspsslices}
\end{figure}

The PSPS database is partitioned into 32 slices, specifically chosen for the $3\pi$ database to have similar amounts of data in each slice.  Fig~\ref{fig:pspsslices} shows the different database slices for $3\pi$, and Table~\ref{table:pspsslices} describes the names of the slices and the declination ranges for each slice. 

The PSPS has loaded the tables for DR1, however there are a few inconsistencies discovered:

{\em IppToPsps} expected to generate 200,684 {\em stack} batches.  It created 200,681 batches (missing 3), of those that are expected, 1,940 partially loaded, and 14 failed to load into PSPS by the DR1 deadline.  These missing batches (1957) will not be included in DR1; they will be added shortly after DR1.

\noindent {\em Missing from ObjectThin/MeanObject}: {\em ippToPsps} generated 116,252 batches.  Of the 116,252 batches, 2,902 batches were only partially ingested into PSPS; this represents 2.5\% of the total number of batches, and 1.2\% of the total number of objects.  The missing batch data will be released shortly after DR1.

\noindent {\em Missing Forced Objects}: We expect 113,665 {\em Forced Object} batches.  There are 7,086 batches that have been partially loaded into PSPS, the rest of the batches are fully loaded.  These missing batches will not be included in DR1, they will be added shortly after DR1.


\begin{table}
\caption{PSPS Dec slice boundaries}
\begin{center}
\begin{tabular}{lll}
\hline
\hline
Slice Name & Min Dec & Max Dec\\
\hline
Slice 1 &$-54.82$ &  $-28.68$\\
Slice 2 &$-28.68$ &  $-26.41$\\
Slice 3& $-26.41$ &  $-24.12$\\
Slice 4& $-24.12$ &  $-21.88$\\
Slice 5& $-21.88$ &  $-19.55$\\
Slice 6& $-19.55$  & $-17.20$\\
Slice 7 &$-17.20$ &  $-14.78$\\
Slice 8 &$-14.78$ &  $-12.24$\\
Slice 9 &$-12.24$ &  $-9.64$\\
Slice 10 &$-9.64$ &   $-7.00$\\
Slice 11 &$-7.00$  &  $-4.29$\\
Slice 12 &$-4.29$  &  $-1.40$\\
Slice 13& $-1.40$  &  $+1.38$ \\
Slice 14 &$+1.38$   & $+4.13$\\
Slice 15 &$+4.13$  &  $+6.91$\\
Slice 16& $+6.91$ &   $+9.78$\\
Slice 17& $+9.78$  &  $+12.70$\\
Slice 18 &$+12.70$ &  $+15.72$\\
Slice 19 &$+15.72$ &  $+18.79$\\
Slice 20 &$+18.79$ &  $+21.93$\\
Slice 21 &$+21.93$ &   $+25.24$\\
Slice 22 &$+25.24$ &  $+28.59$\\
Slice 23 &$+28.59$  & $+31.95$\\
Slice 24 &$+31.95$  & $+35.44$ \\
Slice 25 &$+35.44$  & $+38.98$ \\
Slice 26 &$+38.98$ &  $+42.73$\\
Slice 27 &$+42.73$ &  $+46.73$\\
Slice 28  &$+46.73$&   $+50.90$\\
Slice 29 &$+50.90$ &  $+55.41$\\
Slice 30 &$+55.41$&  $+60.62$\\
Slice 31 &$+60.62$ &  $+67.83$\\
Slice 32 &$+67.83$ &  $+89.99$\\
\hline
\end{tabular}
\end{center}
\label{table:pspsslices}
\end{table}%

\subsubsection{Queries to find holes}
\label{sec:pspsholes}

\noindent It is possible to use the PSPS database to verify data and to check there is no missing data.  The table {\em GaiaFrameCoordinate} is complete and has no missing data. It can be used to do quick integrity checks on other tables within PSPS. Finding missing objects from {\em ObjectThin} is straightforward: Do a \texttt{FULL OUTER JOIN} for {\em GaiaFrameCoordinate} and {\em ObjectThin} using \texttt{ObjID}.  Objects with a \texttt{NULL} {\em ObjectThin}.\texttt{batchID} are missing. Missing objects from {\em ForcedMeanObject} can be found in a similar way as those in {\em ObjectThin}, do a \texttt{FULL OUTER JOIN} between {\em GaiaFrameCoordinate} and {\em ForcedMeanObject}.  Objects with a NULL {\em ForcedMeanObject}.\texttt{batchID} are missing. Finding missing objects in {\em StackObjectThin} is trickier, because not all objects have {\em stack} photometry.  This requires a \texttt{FULL OUTER JOIN} on {\em StackObjectThin} to {\em ObjectThin} to {\em GaiaFrameCoordinate}.  Missing {\em stacks} will have valid {\em ObjectThin}.\texttt{raStack} and {\em ObjectThin}.\texttt{decStack} and \texttt{NULL} {\em StackObjectThin}.\texttt{objID}.





\begin{table}
\caption{\noindent Expected exposure numbers in various stages of processing}
\begin{center}
\begin{tabular}{lll}
\hline
\hline
IPP stage & number queued & successful \\
\hline
raw exposures & 388177   & ...  \\
chip stage& 381279  & 375573  \\
cam stage& 375573   & 374521\\
fake/warp (exp)& 379973  & 379551 \\
fake/warp (skycells) &   & 206177 \\ 
stack & 200730     &  200725    \\ 
staticsky& 200720     & 200720      \\
skycal & 200722   & 200684 \\
fullforce&  200684      &  200684    \\
\hline
\end{tabular}
\end{center}
\label{table:ippcounts}
\end{table}%

\begin{table}
\caption{Expected numbers in DVO and PSPS}
\begin{center}
\begin{tabular}{lllll}
\hline
\hline
IPP stage & successfull IPP & in DVO &{\em IppToPsps}& PSPS \\
\hline
cam stage& 374521 & 374446 &  &  \\
skycal& 200684 & 200684 & 200681 & 198727 \\
fullforce& 200684 & 200684  &  not yet      &     \\
forced galaxy& 113665  &    113665  & not yet      &     \\
object & 116252 & 116252  & 116252 & 113350\\
gaia  &  116252 & 116252 & 116252 & 116252\\
forced object & 113665 & 113665 & 106579   \\
\hline
\end{tabular}
\end{center}
\label{table:ipptopspscounts}
\end{table}%

\section{Conclusion}
\label{sec:conclusion}
The Pan-STARRS database contains 10,723,304,629 objects. It is the largest data release from the largest digital sky survey to date, distilling the information from 1.6 petabytes of images and tables into a form that is accessible to the astronomical community through MAST. Nevertheless, sifting through such a large database can prove daunting, and this work is intended describe the primary tables and quantities within the database, together with example queries. Data from Pan-STARRS has been used for myriad purposes including detecting moving objects within the solar system (and in the case of ‘Oumuamua, from outside it!), the analysis of tens of thousands of high-energy transient events, mapping the 3D structure of dust within our Galaxy, and studies of the large scale structure of our Universe. Yet these only scratch the surface, and it is likely that mining the database will lead to discoveries that were missed and correlations that were overlooked. As we enter the era of multi-messenger astrophysics, the Pan-STARRS data products will be essential to identifying the host galaxies and electromagnetic counterparts of events detected by gravitational wave, high-energy particle, neutrino and radio observatories. While we have provided various tools to work with this data release, we anticipate that it will spur the development of new interfaces and ways of working with high-dimensional datasets. This work will be critical to science with future surveys such as LSST. Combining this Pan-STARRS data release with other large catalogs such as \emph{GALEX}, 2MASS and \emph{Gaia} will provide a rich, high-dimensional dataset that will enable new scientific studies, and may yield astronomical treasures that we have not even begun to imagine.

{\color{red} }

\section{Acknowledgments}

{\it Facilities:} \facility{PS1 (GPC1)}

The Pan-STARRS1 Surveys (PS1) have been made possible through contributions of the Institute for Astronomy, the University of Hawaii, the Pan-STARRS Project Office, the Max-Planck Society and its participating institutes, the Max Planck Institute for Astronomy, Heidelberg, and the Max Planck Institute for Extraterrestrial Physics, Garching, The Johns Hopkins University, Durham University, the University of Edinburgh, Queen's University Belfast, the Harvard-Smithsonian Center for Astrophysics, the Las Cumbres Observatory Global Telescope Network Incorporated, the National Central University of Taiwan, the Space Telescope Science Institute, the National Aeronautics and Space Administration under Grant No. NNX08AR22G issued through the Planetary Science Division of the NASA Science Mission Directorate, the National Science Foundation under Grant No. AST-1238877, the University of Maryland,  Eotvos Lorand University (ELTE), the Los Alamos National Laboratory and the Gordon and Betty Moore foundation. 

This work has made use of data from the European Space Agency (ESA)
mission {\it Gaia} (\url{http://www.cosmos.esa.int/gaia}), processed by
the {\it Gaia} Data Processing and Analysis Consortium (DPAC,
\url{http://www.cosmos.esa.int/web/gaia/dpac/consortium}). Funding
for the DPAC has been provided by national institutions, in particular
the institutions participating in the {\it Gaia} Multilateral Agreement.

HF has some more acknowledgments to add but for now would especially like to thank G.~Hasinger for extensive \LaTeX{} help, G.~Narayan for magical \LaTeX{} ninja skills, and S.~Isani (Ministry of Fonts) for checking \LaTeX{} for typos (he disagrees with my font choice).  

\bibliographystyle{apj}
\bibliography{main}{}

\appendix

\section{Query Examples}
\label{sec:query}

\noindent This section shows example queries for the \PS\ database. The progression will be from simple queries to more complicated queries, with a focus on queries for DR1.  Queries for DR2 will be added later.
 

SQL has no requirements on case.  We adopt the standard convention of using \texttt{CAPITAL LETTERS} for SQL reserved words and functions, and \texttt{CamelCase} for the tables and columns within the PSPS database schema.

\begin{enumerate}
\item \textbf{Counting the number of rows in a large table}

This is an example of a simple query, it needs to be run in the slow queue. The difference between \texttt{COUNT\_BIG()} and \texttt{COUNT()} is that \texttt{COUNT\_BIG()} returns a \texttt{BIGINT}, while \texttt{COUNT()} returns an \texttt{INT}.  The PSPS tables are so large that \texttt{COUNT()}, which goes up to 2.14 billion, is too small of a number. Users should choose the method of counting rows that is appropriate for their data ranges. Unless it involves large tables and large areas of sky, \texttt{COUNT()} is recommended.

\texttt{SELECT COUNT\_BIG(objID) FROM ObjectThin}

\item \textbf{Return mean PSF magnitudes and errors for all filters (grizy) for a rectangular patch of sky}

\texttt{SELECT ObjectThin.objID, nDetections, raMean, decMean, \\
gMeanPSFMag, gMeanPSFMagErr, rMeanPSFMag, rMeanPSFMagErr, \\
iMeanPSFMag, iMeanPSFMagErr, zMeanPSFMag, zMeanPSFMagErr, \\
yMeanPSFMag, yMeanPSFMagErr \\
FROM ObjectThin \\
INNER JOIN MeanObject ON ObjectThin.objID = MeanObject.objID \\
WHERE \\
raMean > 100.0 \\ 
AND raMean < 100.1 \\
AND decMean > 0.0 \\
AND decMean <  0.1 \\
}

This returns 3869 objects. The majority of these objects have only been detected once.  

\item \textbf{Make a simple text histogram of ObjectThin.nDetections for a rectangular patch of sky}

It is possible to save queries into your own personal MyDB, as well as to make queries on your MyDB. Do the query from above, but save it to your MyDB as 'MyDBtest'. 
Run the following query on your MyDB to make a histogram of \texttt{nDetections}.

\texttt{SELECT nDetections, COUNT(nDetections) FROM MyDBtest GROUP BY nDetections ORDER BY nDetections}

The results are ordered by \texttt{ndetections}, and it is apparent that most of the objects have 0-2 detections.  The \texttt{ndetections} column refers to the number of times something is detected from the the individual exposures.  Objects that have 0 detections are so faint that they are not visible in the individual exposures but are detected in the stacks.  Objects with 1-2 (or a few detections), might be spurious detections, moving objects, or faint objects.  If the user is interested in objects that are more likely to be well measured in several epochs and also of astrophysical nature, it is best to add a restriction on \texttt{ndetections}.  If the user is interested in the static sky, and in {\em stack} photometry, it is best to do a \texttt{JOIN} on \texttt{StackObjectThin}. See the next 2 queries for examples of each of those types of queries.

\item \textbf{Select mean PSF magnitudes and errors for filters griyz and for a rectangular patch of sky, with a restriction of $>$ 10 \texttt{nDetections}}

This is an example of a query to get mean PSF magnitudes and errors for all filters in a rectangular patch of sky, for objects with $>$ 10 \texttt{nDetections}. The reason we chose 10 detections is somewhat arbitrary, and can be adjusted, but is used primarily to cut out objects for which there are only a few detections. Objects with very few detections might not be astrophysical, or they might be too faint to be seen multiple times.

\texttt{SELECT ObjectThin.objID, raMean,decMean, \\
gMeanPSFMag, gMeanPSFMagErr, rMeanPSFMag, rMeanPSFMagErr, \\
iMeanPSFMag, iMeanPSFMagErr, zMeanPSFMag, zMeanPSFMagErr, \\
yMeanPSFMag, yMeanPSFMagErr \\
FROM ObjectThin \\
INNER JOIN MeanObject ON ObjectThin.objID = MeanObject.objID \\
WHERE\\
nDetections > 10 \\
AND raMean >100.0 \\ 
AND raMean < 100.1 \\
AND decMean > 0.0 \\
AND decMean <  0.1 
}

This returns 748 objects, a significant reduction from the 3869 on the previous query. 

\item \textbf{Select {\em stack} PSF magnitudes for all filters for a rectangular patch of sky}

This is an example of a query to get {\em stack} PSF magnitudes for the same rectangular patch of sky. No restrictions on \texttt{nDetections} is necessary, the expectation is that sources on {\em stacks} are more likely to be astrophysical.

\texttt{SELECT ObjectThin.objID, raStack, decStack, \\
gPSFMag, gPSFMagErr, rPSFMag, rPSFMagErr,iPSFMag, iPSFMagErr, \\ 
zPSFMag, zPSFMagErr, yPSFMag, yPSFMagErr \\
FROM ObjectThin \\
INNER JOIN StackObjectThin ON ObjectThin.objID = StackObjectThin.objID \\
WHERE raMean >100.0 \\
AND raMean < 100.1 \\
AND decMean > 0.0 \\
AND decMean <  0.1 \\
}

This returns 1808 objects. 

\item \textbf{An example of finding rows with \texttt{NULL} values, using \texttt{TOP} to limit results}

The PSPS uses \texttt{-999} to denote \texttt{NULL} values. The following query returns some objects that are detected in single exposures but not in the stacks. The numbers are limited by TOP to return the first 10 rows.

\texttt{
SELECT TOP 10 \\
objectThin.objID, raMean, decMean, raStack, decStack, \\
nDetections, ng, nr, ni, nz, ny, imeanpsfmag \\
FROM objectThin \\
JOIN MeanObject on objectThin.objID = meanObject.objid \\
WHERE raStack = -999\\
}

\item \textbf{Basic search using \texttt{BETWEEN} to limit ranges}
 
Similar to the query above, except uses \texttt{BETWEEN} to limit RA and Dec ranges as well as iPSFMag ranges.

\texttt{
SELECT ObjectThin.objID, raStack, decStack, \\
gPSFMag, gPSFMagErr, rPSFMag, rPSFMagErr,iPSFMag, iPSFMagErr, \\ 
zPSFMag, zPSFMagErr, yPSFMag, yPSFMagErr \\
FROM ObjectThin \\
INNER JOIN StackObjectThin ON ObjectThin.objID = StackObjectThin.objID \\
WHERE raMean BETWEEN 100.0 AND 100.1\\
AND decMean  BETWEEN 0.0 AND 0.1  \\
AND iPSFMag  BETWEEN 18.0 AND 21.0  \\
}

\item \textbf{Using built in functions to do a box search}

ObjectThin contains Hierarchical triangular mesh information, making it possible to use the built in function dbo.fGetObjFromRectEq(minra, mindec, maxra, maxdec) to do a rectangular search. Tables which have htm, cx,cy, cz can use this built in function.

\texttt{
SELECT o.objID, o.raMean, o.decMean\\
FROM ObjectThin o, dbo.fGetObjFromRectEq(56.65, 23.92, 57.05, 24.32) r\\
WHERE p.objID = r.objID \\
}

\item \textbf{Using built in functions to do a cone search}

ObjectThin contains Hierarchical triangular mesh information, making it possible to use the built in function dbo.fGetNearbyObjEq(ra, dec, conesize(deg)) to do a radial search for objects near a given ra and dec (cone search). Tables which have htm, cx,cy, cz can use this built in function.

\texttt{
SELECT o.objID, raMean, decMean, gPSFMag, gPSFMagErr\\
FROM ObjectThin o, dbo.fGetNearbyObjEq(56.85 , 24.12, 0.2) n\\
WHERE o.objID = n.objID \\
}

\item \textbf{Cone search of high fidelity stellar-like objects}

We want to get all objects with R degrees of a given position that are high fidelity stellar-like objects.
We get all objects within 0.2 degree of RA=334.0 and Dec=0.0 which have mean magnitudes in griz (i.e. at least 1 detection in each band that can be used for the mean mag). In addition, we require QfPerfect>0.85 in all bands. We select stars with small ($<0.05$) difference between Kron and PSF magnitudes.

\texttt{
SELECT o.objID,  
o.raMean, o.decMean, o.raMeanErr, o.decMeanErr, 
o.qualityFlag,
o.gMeanPSFMag, o.gMeanPSFMagErr, o.gMeanPSFMagNpt,
o.rMeanPSFMag, o.rMeanPSFMagErr, o.rMeanPSFMagNpt,
o.iMeanPSFMag, o.iMeanPSFMagErr, o.iMeanPSFMagNpt,
o.zMeanPSFMag, o.zMeanPSFMagErr, o.zMeanPSFMagNpt,
o.yMeanPSFMag, o.yMeanPSFMagErr, o.yMeanPSFMagNpt, 
o.rMeanKronMag, o.rMeanKronMagErr,
o.nDetections, o.ng, o.nr, o.ni, o.nz,o.ny,
o.gFlags, o.gQfPerfect,
o.rFlags, o.rQfPerfect,
o.iFlags, o.iQfPerfect,
o.zFlags, o.zQfPerfect,
o.yFlags, o.yQfPerfect,
soa.primaryDetection, soa.bestDetection
INTO mydb.[HighFidelityStarsDR2]
FROM dbo.fGetNearbyObjEq(334, 0.0, 0.2*60.0) as x
JOIN MeanObjectView o on o.ObjID=x.ObjId
LEFT JOIN StackObjectAttributes AS soa ON soa.objID = x.objID
WHERE o.nDetections>5 
AND soa.primaryDetection>0 
AND o.gQfPerfect>0.85 and o.rQfPerfect>0.85 and o.iQfPerfect>0.85 and o.zQfPerfect>0.85 
AND (o.rmeanpsfmag - o.rmeankronmag < 0.05)
}

\item \textbf{Galaxy Candidates for K2 SN Search}

The Kepler Extra-Galactic Survey (KEGS) is a program using the Kepler telescope to search for supernovae, active galactic nuclei, and other transients in galaxies. We have to identify galaxies in a suitable redshift range (z<=0.12) a priori, which will be monitored by K2. Here is an example to get galaxies for Campaign 14. We only select objects with r<=19.5, and we make a cut on (rmeanpsfmag - rmeankronmag)>=0.05 in order to remove stars. We only want to use objects for which the majority of pixels were not masked, thus the cut on QFperfect>=0.95. We also obtain the petrosian radii in order to be able to select galaxies by size.

\texttt{
SELECT  o.objID,
ot.raStack, ot.decStack, ot.raMean, ot.decMean, 
ot.ng,  o.gMeanPSFMag,o.gMeanPSFMagErr,o.gMeanKronMag,o.gMeanKronMagErr,
ot.nr,  o.rMeanPSFMag,o.rMeanPSFMagErr,o.rMeanKronMag,o.rMeanKronMagErr,
ot.ni,  o.iMeanPSFMag,o.iMeanPSFMagErr,o.iMeanKronMag,o.iMeanKronMagErr,
ot.nz,  o.zMeanPSFMag,o.zMeanPSFMagErr,o.zMeanKronMag,o.zMeanKronMagErr,
ot.ny,  o.yMeanPSFMag,o.yMeanPSFMagErr,o.yMeanKronMag,o.yMeanKronMagErr,
o.gQfPerfect, o.rQfPerfect, o.iQfPerfect, o.zQfPerfect, o.yQfPerfect,
ot.qualityFlag, ot.objInfoFlag,
sp.gpetRadius, sp.rpetRadius, sp.ipetRadius, sp.zpetRadius, sp.ypetRadius,
sp.gpetR50,sp.rpetR50,sp.ipetR50,sp.zpetR50,sp.ypetR50,
soa.primaryDetection, soa.bestDetection
       INTO mydb.[C14]
FROM MeanObject AS o
JOIN fgetNearbyObjEq(160.68333, 6.85167 , 8.5*60.0) cone ON cone.objid = o.objID 
JOIN ObjectThin AS ot ON ot.objID = o.objID
LEFT JOIN StackPetrosian AS sp ON sp.objID = o.objID
LEFT JOIN StackObjectAttributes AS soa ON soa.objID = o.objID
 WHERE ot.ni >= 3
     AND ot.ng >= 3
     AND ot.nr >= 3
     AND soa.primaryDetection>0 
     AND (o.rMeanKronMag > 0 AND o.rMeanKronMag <= 19.5 )
     AND (o.gQfPerfect >= 0.95)
     AND (o.rQfPerfect >= 0.95)
     AND (o.iQfPerfect >= 0.95)
     AND (o.zQfPerfect >= 0.95)
     AND (o.rmeanpsfmag - o.rmeankronmag > 0.05)}

\item \textbf{Find the objID of a single object}

Star CSS J030521.9+013231 (Catalina Sky Survey), 584630948352256 (GAIA) is an RR Lyrae with period = 0.55547 days and coordinates RA = 46.341468915923 and DEC = 1.54199810825252 (ref. GAIA DR2,  2018yCat.1345....0G). In the following, we obtain the PSF and aperture photometry light-curves, both forced and unforced, for this star.

\texttt{
SELECT d.ID\_GAIA,d.RA\_GAIA as GAIARA, d.DEC\_GAIA as GAIADec, d.Gmag, d.period,
o.objID,  
o.raMean, o.decMean, o.raMeanErr, o.decMeanErr, 
o.qualityFlag,
o.gMeanPSFMag, o.gMeanPSFMagErr, o.gMeanPSFMagNpt,
o.rMeanPSFMag, o.rMeanPSFMagErr, o.rMeanPSFMagNpt,
o.iMeanPSFMag, o.iMeanPSFMagErr, o.iMeanPSFMagNpt,
o.zMeanPSFMag, o.zMeanPSFMagErr, o.zMeanPSFMagNpt,
o.yMeanPSFMag, o.yMeanPSFMagErr, o.yMeanPSFMagNpt, 
o.rMeanKronMag, o.rMeanKronMagErr,
o.nDetections, o.ng, o.nr, o.ni, o.nz, o.ny,
o.gFlags, o.gQfPerfect,
o.rFlags, o.rQfPerfect,
o.iFlags, o.iQfPerfect,
o.zFlags, o.zQfPerfect,
o.yFlags, o.yQfPerfect,
soa.primaryDetection, soa.bestDetection
 INTO mydb.[RRL\_584630948352256\_PS1]
 FROM mydb.[rrl\_584630948352256] d
CROSS APPLY dbo.fGetNearbyObjEq(46.341468915923, 1.54199810825252, 1.0/60.0) as x
JOIN MeanObjectView o on o.ObjID=x.ObjId
LEFT JOIN StackObjectAttributes AS soa ON soa.objID = x.objID
WHERE o.nDetections>8 
AND soa.primaryDetection>0 
AND o.gQfPerfect>0.85 and o.rQfPerfect>0.85 and o.iQfPerfect>0.85 and o.zQfPerfect>0.85 
AND (o.rmeanpsfmag - o.rmeankronmag < 0.05)}

\item \textbf{Obtain lightcurve for a given object (Detections)}

The query above gives an ObjID of 109850463414820867 for that RRLyrae star. Knowing the objID it is possible to query to get the lightcurve.  

\texttt{
SELECT o.ID\_GAIA,o.GAIARA, o.GAIADec, o.Gmag, o.period,
o.objID, o.raMean, o.decMean,
d.ra, d.dec, d.raErr, d.decErr, 
d.detectID, d.obstime, d.exptime, d.airmass, d.psfflux, d.psffluxErr, 
d.psfQf, d.psfQfPerfect, d.psfLikelihood, d.psfChiSq, d.extNSigma, d.zp, d.apFlux, d.apFluxErr,
d.imageID, d.filterID,
d.sky, d.skyerr, d.infoflag, d.infoflag2, d.infoflag3,
o.qualityFlag,
o.gMeanPSFMag, o.gMeanPSFMagErr, o.gMeanPSFMagNpt,
o.rMeanPSFMag, o.rMeanPSFMagErr, o.rMeanPSFMagNpt,
o.iMeanPSFMag, o.iMeanPSFMagErr, o.iMeanPSFMagNpt,
o.zMeanPSFMag, o.zMeanPSFMagErr, o.zMeanPSFMagNpt,
o.yMeanPSFMag, o.yMeanPSFMagErr, o.yMeanPSFMagNpt, 
o.rMeanKronMag, o.rMeanKronMagErr,
o.nDetections, o.ng, o.nr, o.ni, o.nz,o.ny,
o.gFlags, o.gQfPerfect,
o.rFlags, o.rQfPerfect,
o.iFlags, o.iQfPerfect,
o.zFlags, o.zQfPerfect,
o.yFlags, o.yQfPerfect,
o.primaryDetection, o.bestDetection 
INTO mydb.[RRL\_584630948352256\_PS1det]
 FROM mydb.[RRL\_584630948352256\_PS1] o
JOIN Detection d on d.ObjID = o.ObjID
}

\begin{figure}[htpb]
    \centering
    \includegraphics[width=0.8\textwidth]{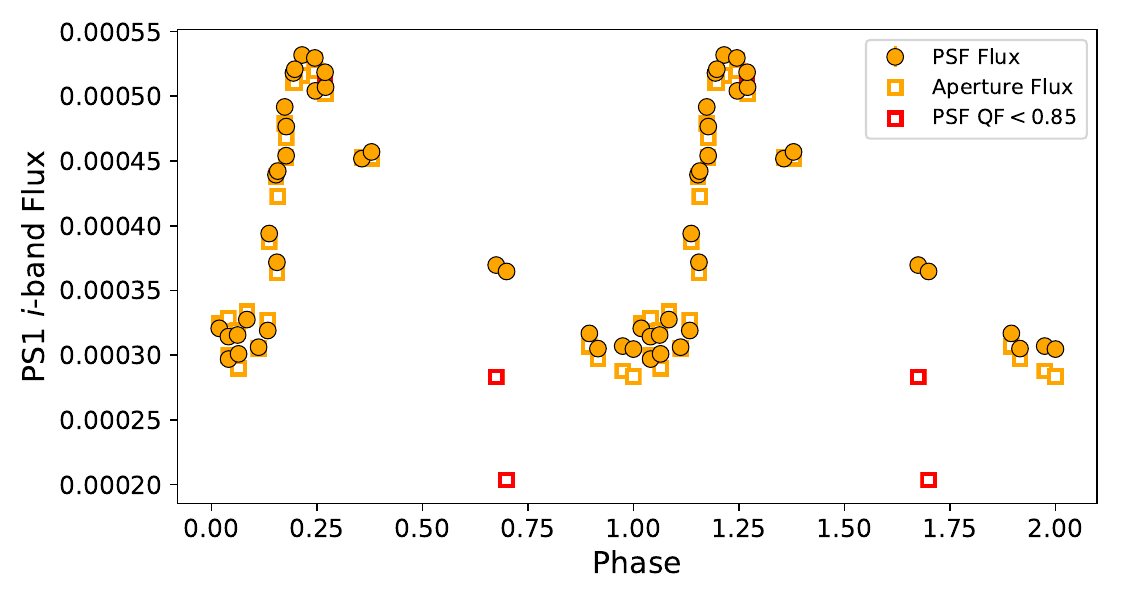}
    \caption{A key feature of this data release is the inclusion of the individual detections, allowing time-resolved studies, illustrated by the \textit{i}-band phase curve of one of the PS1 RR Lyrae. The IPP aperture and PSF fluxes are in good agreement, and either can be used for studies of the hundreds of thousands of variable and transient sources within PS1. Users are encouraged to check the provided flags for various bits documented in this work, as these may indicate sub-optimal photometry. We highlight two such measurements in red. Uncertainties are typically smaller than the marker size.}
    \label{fig:rrlyr}
\end{figure}

\item \textbf{Obtain lightcurve for a given object (Forced photometry)}

Similar to the query above, except this one provides the forced photometry for the star with ObjID = 109850463414820867.

\texttt{
SELECT o.ID\_GAIA,o.GAIARA, o.GAIADec, o.Gmag, o.period,
o.objID, o.raMean, o.decMean,
fwm.detectID, fwm.obstime, fwm.exptime, fwm.airmass, fwm.Fpsfflux, fwm.FpsffluxErr, fwm.FpsfQf, fwm.FpsfQfPerfect, fwm.FpsfChiSq, fwm.zp, fwm.FapFlux, fwm.FapFluxErr,
fwm.forcedWarpID, fwm.filterID,
fwm.Fsky, fwm.Fskyerr, fwm.Finfoflag, fwm.Finfoflag2, fwm.Finfoflag3,
o.qualityFlag,
o.gMeanPSFMag, o.gMeanPSFMagErr, o.gMeanPSFMagNpt,
o.rMeanPSFMag, o.rMeanPSFMagErr, o.rMeanPSFMagNpt,
o.iMeanPSFMag, o.iMeanPSFMagErr, o.iMeanPSFMagNpt,
o.zMeanPSFMag, o.zMeanPSFMagErr, o.zMeanPSFMagNpt,
o.yMeanPSFMag, o.yMeanPSFMagErr, o.yMeanPSFMagNpt,
o.rMeanKronMag, o.rMeanKronMagErr,
o.nDetections, o.ng, o.nr, o.ni, o.nz,o.ny,
o.gFlags, o.gQfPerfect,
o.rFlags, o.rQfPerfect,
o.iFlags, o.iQfPerfect,
o.zFlags, o.zQfPerfect,
o.yFlags, o.yQfPerfect,
o.primaryDetection, o.bestDetection
FROM mydb.[RRL\_584630948352256\_PS1] o
JOIN ForcedWarpMeasurement fwm on fwm.ObjID = o.ObjID}

\end{enumerate}















\section{Abbreviations and Acronyms}
\label{sec:acronyms}

\begin{description}

\item[3pi or 3$\pi$] Three Pi Survey. This survey covers 3 pi steradians of the sky (3/4 of the sky), everything with declination greater than -30 degrees.

\item[DR1]  Data Release 1.  Covers 3$\pi$ data release for ObjectThin, MeanObject, Stack*, ForcedMean* and related metadata tables. Covers the static sky.

\item[DR2]  Data Release 2. 3$\pi$ data release of time domain tables, including Detection, ForcedWarp*, Diff* and related metadata.

\item[DVO]  Desktop Virtual Observatory - written by Eugene Mangier, IPP uses this to store and manipulate catalog data. 

\item[GPC1]  Giga Pixel Camera 1.  This is the name of the camera that is part of the \PS\ telescope.  1.4 gigapixels, and sees 7 square degrees per exposure.  It is made up of 60 orthogonal transfer arrays (OTA), each with 64 cells per OTA.  

\item[IPP]  Image Processing Pipeline.  Code developed to process and manage all aspects of \PS processing, starting from downloading the images to the summit to generating data in the final database schema form.

\item[MD]  Short for Medium Deep fields. A set of 10 fields, each of which is 7 square degrees, observed at a high cadence primarily to be used for searches for transient objects.  These will be released at a later time.

\item[MOPS]  Short for Medium Deep fields. A set of 10 fields, each of which is 7 square degrees, observed at a high cadence primarily to be used for searches for transient objects.  These will be released at a later time.

\item[OTA]  orthogonal transfer array, the name of the devices that make up the GPC1 camera

\item[PSI]  Pan-STARRS1 Science interface. The interface used by consortium members to access earlier versions of the data.

\item[PSPS]  Published Science Products Subsystem - the \PS\ databases


\item[PV3]  Processing Version 3, refers to the processing / code iteration, this is the processing version of the first public \PS data release (covers DR1-DR2).

\end{description}

\section{Flag Tables}
\label{sec:FlagTables}

There are 8 different classes of Flag tables for the database schema.  This section lists the different flags as well as their descriptions.  See \ref{sec:schemaflagsbitmasks} for more details on flags and bitmasks, and \ref{sec:query} for some example queries. 

\begin{description}
\item[ObjectInfoFlags] Describes the flag bits used for: ObjectThin.objectInfoFlag, GaiaFrameCoordinate.gaiaFlag, DiffDetObject.objectInfoFlag, ForcedMeanObject.gFlag, ForcedMeanObject.rFlag, ForcedMeanObject.iFlag, ForcedMeanObject.zFlag, ForcedMeanObject.yFlag.
See Table~\ref{table:objectinfoflags}

\item[ObjectQualityFlags] Describes the flags used for: ObjectThin.qualityFlag, DiffDetObject.qualityFlag.
See Table~\ref{table:objectqualityflags}

\item[ObjectFilterFlags] Describes the flags used for: MeanObject.gFlags, MeanObject.rFlags, MeanObject.iFlags, MeanObject.zFlags, MeanObject.yFlags, StackObjectThin.gInfoFlag4, StackObjectThin.rInfoFlag4, StackObjectThin.iInfoFlag4, StackObjectThin.zInfoFlag4, StackObjectThin.yInfoFlag4,.
See Table~\ref{table:objectfilterflags}

\item[ImageFlags] Describes the flags used for: ImageMeta.qaFlags.
See Table~\ref{table:imageflags}

\item[DetectionFlags] Describes the flags used for: Detection.infoFlag, StackObjectThin.ginfoFlag, StackObjectThin.rinfoFlag , StackObjectThin.iinfoFlag, StackObjectThin.zinfoFlag, StackObjectThin.yinfoFlag, DiffDetection.DinfoFlag, ForcedWarpMeasurement.FinfoFlag.
See Table~\ref{table:detectionflags}

\item[DetectionFlags2] Describes the flags used for: Detection.infoFlag2, StackObjectThin.ginfoFlag2, StackObjectThin.rinfoFlag2 , StackObjectThin.iinfoFlag2, StackObjectThin.zinfoFlag2, StackObjectThin.yinfoFlag2,  DiffDetection.DinfoFlag2,ForcedWarpMeasurement.FinfoFlag2.
See Table~\ref{table:detectionflags2}

\item[DetectionFlags3] Describes the flags used for: Detection.infoFlag3, StackObjectThin.ginfoFlag3, StackObjectThin.rinfoFlag3 , StackObjectThin.iinfoFlag3, StackObjectThin.zinfoFlag3, StackObjectThin.yinfoFlag3,  DiffDetection.DinfoFlag3,ForcedWarpMeasurement.FinfoFlag3.
See Table~\ref{table:detectionflags3}

\item[ForcedGalaxyShapeFlags] Describes the flags used for: ForcedGalaxyShape.gGalFlags, ForcedGalaxyShape.rGalFlags, ForcedGalaxyShape.iGalFlags, ForcedGalaxyShape.zGalFlags, ForcedGalaxyShape.yGalFlags. 
See Table~\ref{table:forcedgalaxyshapeflags}

\end{description}

\clearpage
\begin{table}
\caption{ObjectInfoFlags}
\begin{center}
\begin{tabular}{llll}
\hline
\hline
Name & Hexadecimal & Value & Description \\
\hline
DEFAULT & 0x00000000 & 0 & Initial value; resets all bits. \\
FEW&0x00000001&1&Used within relphot; skip star.\\
POOR&0x00000002&2&Used within relphot; skip star.\\
ICRF\_QSO&0x00000004&4&Object IDed with known ICRF quasar \\
       &    &     & (may have ICRF position measurement)\\
HERN\_QSO\_P60&0x00000008&8&Identified as likely QSO (Hernitschek et al 2015);  \\
      &     &      & P\_QSO $>=$ 0.60\\
HERN\_QSO\_P05&	0x00000010&	16& 	Identified as possible QSO (Hernitschek et al 2015), P\_QSO $>=$ 0.05\\ 
HERN\_RRL\_P60&	0x00000020&	32 &	Identified as likely RR Lyra (Hernitschek et al 2015), P\_RRLyra $>=$ 0.60\\
HERN\_RRL\_P05&	0x00000040&	64 &	Identified as possible RR Lyra (Hernitschek et al 2015), P\_RRLyra $>=$ 0.05\\
HERN\_VARIABLE&	0x00000080&	128 &	Identified as a variable based on ChiSq? (Hernitschek et al 2015)\\
TRANSIENT&	0x00000100&	256& 	Identified as a non-periodic (stationary) transient\\
HAS\_SOLSYS\_DET&	0x00000200&	512& 	At least one detection identified with a known solar-system object (asteroid or other).\\
MOST\_SOLSYS\_DET&	0x00000400&	1024 &	Most detections identified with a known solar-system object (asteroid or other). \\
LARGE\_PM&0x00000800&2048&Star with large proper motion\\
RAW\_AVE&0x00001000&4096&Simple weighted average position was used (no IRLS fitting)\\
FIT\_AVE&0x00002000&8192&Average position was fitted\\
FIT\_PM&0x00004000&16384&Proper motion model was fitted\\
FIT\_PAR&0x00008000&32768&Parallax model was fitted\\
USE\_AVE&0x00010000&65536&Average position used (not PM or PAR)\\
USE\_PM&0x00020000&131072&Proper motion used (not AVE or PAR)\\
USE\_PAR&0x00040000&262144&Parallax used (not AVE or PM)\\
NO\_MEAN\_ASTROM&0x00080000&524288&Mean astrometry could not be measured\\
STACK\_FOR\_MEAN&0x00100000&1048576&Stack position used for mean astrometry\\
MEAN\_FOR\_STACK&0x00200000&2097152&Mean astrometry used for {\em stack} position\\
BAD\_PM&0x00400000&4194304&Failure to measure proper-motion model\\
EXT&0x00800000&8388608&Extended in our data (eg; PS)\\
EXT\_ALT&0x01000000&16777216&Extended in external data (eg; 2MASS)\\
GOOD&0x02000000&33554432&Good-quality measurement in our data (eg; PS)\\
GOOD\_ALT&0x04000000&67108864&Good-quality measurement in  external data (eg; 2MASS)\\
GOOD\_STACK&0x08000000&134217728&Good-quality object in the {\em stack} ($>$ 1 good stack measurement)\\
BEST\_STACK&0x10000000&268435456&The primary {\em stack} measurements are the best measurements\\
SUSPECT\_STACK&0x20000000&536870912&Suspect object in the {\em stack}\\
     &   &    &  (no more than 1 good measurement, 2 or more suspect or good stack measurement)\\
BAD\_STACK&0x40000000&1073741824&Poor-quality {\em stack} object (no more than 1 good or suspect measurement)\\
\hline
\end{tabular}
\end{center}
\label{table:objectinfoflags}
\end{table}%

\begin{table}
\caption{ObjectQualityFlags}
\begin{center}
\begin{tabular}{llll}
\hline
\hline
Name & Hexadecimal & Value & Description \\
\hline
DEFAULT&0x00000000&0&Initial value; resets all bits. \\
QF\_OBJ\_EXT&0x00000001&1&Extended in our data (eg; PS). \\
QF\_OBJ\_EXT\_ALT&0x00000002&2&Extended in external data (eg; 2MASS). \\
QF\_OBJ\_GOOD&0x00000004&4&Good-quality measurement in our data (eg; PS). \\
QF\_OBJ\_GOOD\_ALT&0x00000008&8&Good-quality measurement in  external data (eg; 2MASS). \\
QF\_OBJ\_GOOD\_STACK&0x00000010&16&Good-quality object in the {\em stack} ($>$ 1 good stack measurement). \\
QF\_OBJ\_BEST\_STACK&0x00000020&32&The primary stack measurements are the best measurements.\\
QF\_OBJ\_SUSPECT\_STACK&0x00000040&64&Suspect object in the {\em stack}\\
& & & (no more than 1 good measurement, 2 or more suspect or good stack measurement). \\
QF\_OBJ\_BAD\_STACK&0x00000080&128&Poor-quality {\em stack} object (no more than 1 good or suspect measurement). \\

\hline
\end{tabular}
\end{center}
\label{table:objectqualityflags}
\end{table}%

\begin{table}
\caption{ObjectFilterFlags}
\begin{center}
\begin{tabular}{llll}
\hline
\hline
Name & Hexadecimal & Value & Description \\
\hline
DEFAULT&0x00000000&0&Initial value; resets all bits.\\
SECF\_STAR\_FEW&0x00000001&1&Used within relphot: skip star.\\
SECF\_STAR\_POOR&0x00000002&2&Used within relphot: skip star.\\
SECF\_USE\_SYNTH&0x00000004&4&Synthetic photometry used in average measurement.\\
SECF\_USE\_UBERCAL&0x00000008&8&Ubercal photometry used in average measurement.\\
SECF\_HAS\_PS1&0x00000010&16&PS1 photometry used in average measurement.\\
SECF\_HAS\_PS1\_STACK&0x00000020&32&PS1 {\em stack} photometry exists.\\
SECF\_HAS\_TYCHO&0x00000040&64&Tycho photometry used for synthetic magnitudes.\\
SECF\_FIX\_SYNTH&0x00000080&128&Synthetic magnitudes repaired with zeropoint map.\\
SECF\_RANK\_0&0x00000100&256&Average magnitude calculated in 0th pass.\\
SECF\_RANK\_1&0x00000200&512&Average magnitude calculated in 1st pass.\\
SECF\_RANK\_2&0x00000400&1024&Average magnitude calculated in 2nd pass.\\
SECF\_RANK\_3&0x00000800&2048&Average magnitude calculated in 3rd pass.\\
SECF\_RANK\_4&0x00001000&4096&Average magnitude calculated in 4th pass.\\
SECF\_STACK\_PRIMARY&0x00004000&16384&PS1 {\em stack} photometry comes from primary skycell.\\
SECF\_STACK\_BESTDET&0x00008000&32768&PS1 {\em stack} best measurement is a detection (not forced).\\
SECF\_STACK\_PRIMDET&0x00010000&65536&PS1 stack primary measurement is a detection (not forced).\\
SECF\_HAS\_SDSS&0x00100000&1048576&This photcode has SDSS photometry.\\
SECF\_HAS\_HSC& 0x00200000&2097152&This photcode has HSC photometry.\\
SECF\_HAS\_CFH& 0x00400000&4194304&This photcode has CFH photometry (mostly megacam).\\
SECF\_HAS\_DES& 0x00800000&8388608&This photcode has DES photometry.\\
SECF\_OBJ\_EXT& 0x01000000&16777216&Extended in this band.\\
\hline
\end{tabular}
\end{center}
\label{table:objectfilterflags}
\end{table}%

\begin{table}
\caption{ImageFlags}
\begin{center}
\begin{tabular}{llll}
\hline
\hline
Name & Hexadecimal & Value & Description \\
\hline
NEW&0x00000000&0&No relphot / relastro attempted. \\
PHOTOM\_NOCAL&0x00000001&1&Used within relphot to mean 'don't apply fit'. \\
PHOTOM\_POOR&0x00000002&2&Relphot says image is bad (dMcal $>$ limit). \\
PHOTOM\_SKIP&0x00000004&4&External information image has bad photometry. \\
PHOTOM\_FEW&0x00000008&8&Currently too few measurements for good value for photometry. \\
ASTROM\_NOCAL&0x00000010&16&User-set value used within relastro: ignore. \\
ASTROM\_POOR &0x00000020&32&Relastro says image is bad (dR,dD $>$ limit). \\
ASTROM\_FAIL &0x00000040&64&Relastro fit diverged, fit not applied.\\
ASTROM\_SKIP &0x00000080&128&External information image has bad astrometry.\\
ASTROM\_FEW& 0x00000100&256&Currently too few measurements for good value for astrometry.\\
PHOTOM\_UBERCAL&0x00000200&512&Externally-supplied photometry zero point from ubercal analysis.\\ 
ASTROM\_GMM &0x00000400&1024&Image was fitted to positions corrected by the galaxy motion model.\\
\hline
\end{tabular}
\end{center}
\label{table:imageflags}
\end{table}%

\begin{table}
\caption{ForcedGalaxyShapeFlags}
\begin{center}
\begin{tabular}{llll}
\hline
\hline
Name & Hexadecimal & Value & Description \\
\hline
NO\_ERROR&0x00000000&0&No error condition raised. \\
FAIL\_FIT&0x00000001&1&Fit failed to converge or was degenerate \\
TOO\_FEW&0x00000002&2&Not enough points to fit the model \\
OUT\_OF\_RANGE&0x00000004&4&Fit minimum too far outside data range \\
BAD\_ERROR&0x00000008&8&Invalid size error (nan or inf) \\
\hline
\end{tabular}
\end{center}
\label{table:forcedgalaxyshapeflags}
\end{table}%

\begin{table}
\caption{DetectionFlags}
\begin{center}
\begin{tabular}{llll}
\hline
\hline
Name & Hexadecimal & Value & Description \\
\hline
DEFAULT&0x00000000&0&Initial value; resets all bits.\\
PSFMODEL&0x00000001&1&Source fitted with a psf model (linear or non-linear).\\
EXTMODEL&0x00000002&2&Source fitted with an extended-source model.\\
FITTED&0x00000004&4&Source fitted with non-linear model (PSF or EXT; good or bad).\\
FAIL&0x00000008&8&Fit (non-linear) failed (non-converge; off-edge; run to zero).\\
POOR&0x00000010&16&Fit succeeds; but low-S/N; high-Chisq; or large (for PSF -- drop?).\\
PAIR&0x00000020&32&Source fitted with a double PSF.\\
PSFSTAR&0x00000040&64&Source used to define PSF model.\\
SATSTAR&0x00000080&128&Source model peak is above saturation.\\
BLEND&0x00000100&256&Source is a blend with other sources.\\
EXTERNAL&0x00000200&512&Source based on supplied input position.\\
BADPSF&0x00000400&1024&Failed to get good estimate of object's PSF.\\
DEFECT&0x00000800&2048&Source is thought to be a defect.\\
SATURATED&0x00001000&4096&Source is thought to be saturated pixels (bleed trail).\\
CR\_LIMIT&0x00002000&8192&Source has crNsigma above limit.\\
EXT\_LIMIT&0x00004000&16384&Source has extNsigma above limit.\\
MOMENTS\_FAILURE&0x00008000&32768&Could not measure the moments.\\
SKY\_FAILURE&0x00010000&65536&Could not measure the local sky.\\
SKYVAR\_FAILURE&0x00020000&131072&Could not measure the local sky variance.\\
BELOW\_MOMENTS\_SN&0x00040000&262144&Moments not measured due to low S/N.\\
UNDEF\_1&0x00080000&524288&Unused bit value.\\
BIG\_RADIUS&0x00100000&1048576&Poor moments for small radius; try large radius.\\
AP\_MAGS&0x00200000&2097152&Source has an aperture magnitude.\\
BLEND\_FIT&0x00400000&4194304&Source was fitted as a blend.\\
EXTENDED\_FIT&0x00800000&8388608&Full extended fit was used.\\
EXTENDED\_STATS&0x01000000&16777216&Extended aperture stats calculated.\\
LINEAR\_FIT&0x02000000&33554432&Source fitted with the linear fit.\\
NONLINEAR\_FIT&0x04000000&67108864&Source fitted with the non-linear fit.\\
RADIAL\_FLUX&0x08000000&134217728&Radial flux measurements calculated.\\
SIZE\_SKIPPED&0x10000000&268435456&Size could not be determined.\\
PEAK\_ON\_SPIKE&0x20000000&536870912&Peak lands on diffraction spike.\\
PEAK\_ON\_GHOST&0x40000000&1073741824&Peak lands on ghost or glint.\\
PEAK\_OFF\_CHIP&0x80000000&2147483648&Peak lands off edge of chip.\\
\hline
\end{tabular}
\end{center}
\label{table:detectionflags}
\end{table}%

\begin{table}
\caption{DetectionFlags2}
\begin{center}
\begin{tabular}{llll}
\hline
\hline
Name & Hexadecimal & Value & Description \\
\hline
DEFAULT&0x00000000&0&Initial value; resets all bits.\\
DIFF\_WITH\_SINGLE&0x00000001&1&Difference source matched to a single positive detection.\\
DIFF\_WITH\_DOUBLE&0x00000002&2&Difference source matched to positive detections in both images.\\
MATCHED&0x00000004&4&Source generated based on another image (forced photometry \\
&&&at source location).\\
ON\_SPIKE&0x00000008&8&More than 25\% of (PSF-weighted) pixels land on diffraction spike.\\
ON\_STARCORE&0x00000010&16&More than 25\% of (PSF-weighted) pixels land on starcore.\\
ON\_BURNTOOL&0x00000020&32&More than 25\% of (PSF-weighted) pixels land on burntool.\\
ON\_CONVPOOR&0x00000040&64&More than 25\% of (PSF-weighted) pixels land on convpoor.\\
PASS1\_SRC&0x00000080&128&Source detected in first pass analysis\\
HAS\_BRIGHTER\_NEIGHBOR&0x00000100&256&Peak is not the brightest in its footprint\\
BRIGHT\_NEIGHBOR\_1&0x00000200&512& Flux\_negative $/$ ($r^2$ flux\_positive) $>$  1.\\
BRIGHT\_NEIGHBOR\_10&0x00000400&1024&Flux\_negative $/$ ($r^2$ flux\_positive) $ >$ 10.\\
DIFF\_SELF\_MATCH&0x00000800&2048&Positive detection match is probably this source.\\
SATSTAR\_PROFILE&0x00001000&4096&Saturated source is modeled with a radial profile.\\
ECONTOUR\_FEW\_PTS&0x00002000&8192&Too few points to measure the elliptical contour.\\
RADBIN\_NAN\_CENTER&0x00004000&16384&Radial bins failed with too many NaN center bin.\\
PETRO\_NAN\_CENTER&0x00008000&32768&Petrosian (1976) radial bins failed with too many NaN center bin.\\
PETRO\_NO\_PROFILE&0x00010000&65536&Petrosian (1976) not built becaues radial bins missing.\\
PETRO\_INSIG\_RATIO&0x00020000&131072&Insignificant measurement of Petrosian (1976) ratio.\\
PETRO\_RATIO\_ZEROBIN&0x00040000&262144&Petrosian (1976) ratio in the 0th bin (likely bad).\\
EXT\_FITS\_RUN&0x00080000&524288&Attempted to run extended fits on this source.\\
EXT\_FITS\_FAIL&0x00100000&1048576&At least one of the model fits failed.\\
EXT\_FITS\_RETRY&0x00200000&2097152&One of the model fits was retried with new window.\\
EXT\_FITS\_NONE&0x00400000&4194304&All of the model fits failed. \\
\hline
\end{tabular}
\end{center}
\label{table:detectionflags2}
\end{table}%

\begin{table}
\caption{DetectionFlags3}
\begin{center}
\begin{tabular}{llll}
\hline
\hline
Name & Hexadecimal & Value & Description \\
\hline
DEFAULT&0x00000000&0&Initial value; resets all bits.\\
NOCAL&0x00000001&1&Detection ignored for this analysis (photcode; time range) -- internal only.\\
POOR\_PHOTOM&0x00000002&2&Detection is photometry outlier.\\
SKIP\_PHOTOM&0x00000004&4&Detection was ignored for photometry measurement.\\
AREA&0x00000008&8&Detection near image edge.\\
POOR\_ASTROM&0x00000010&16&Detection is astrometry outlier.\\
SKIP\_ASTROM&0x00000020&32&Detection was ignored for astrometry measurement.\\
USED\_OBJ&0x00000040&64&Detection was used during update objects\\
USED\_CHIP&0x00000080&128&Detection was used during update chips to measure astrometry with IRLS fit.\\
BLEND\_MEAS&0x00000100&256&Detection is within radius of multiple objects.\\
BLEND\_OBJ&0x00000200&512&Multiple detections within radius of object.\\
WARP\_USED&0x00000400&1024&Measurement used to find mean {\em warp} photometry.\\
UNMASKED\_ASTRO&0x00000800&2048&Detection was unmasked in update chips to determine astrometry parameter errors.\\
BLEND\_MEAS\_X&0x00001000&4096&Detection is within radius of multiple objects across catalogs.\\
ARTIFACT&0x00002000&8192&Detection is thought to be non-astronomical.\\
SYNTH\_MAG&0x00004000&16384&Magnitude is synthetic.\\
PHOTOM\_UBERCAL&0x00008000&32768&Externally-supplied zero point from ubercal analysis.\\
STACK\_PRIMARY&0x00010000&65536&This {\em stack} measurement is in the primary skycell.\\
STACK\_PHOT\_SRC&0x00020000&131072&This measurement supplied the {\em stack} photometry.\\
ICRF\_QSO&0x00040000&262144&This measurement is an ICRF reference position.\\
IMAGE\_EPOCH&0x00080000&524288&This measurement is registered to the image epoch \\
&  &   & (not tied to the reference catalog epoch).\\
PHOTOM\_PSF&0x00100000&1048576&This measurement is used for the mean PSF magnitude.\\
PHOTOM\_APER&0x00200000&2097152&This measurement is used for the mean aperture magnitude.\\
PHOTOM\_KRON&0x00400000&4194304&This measurement is used for the mean Kron (1980) magnitude.\\
MASKED\_PSF& 0x01000000&16777216&This measurement is masked based on IRLS weights for the mean PSF magnitude.\\
MASKED\_APER&0x02000000&33554432&This measurement is masked based on IRLS weights for the mean aperture magnitude.\\
MASKED\_KRON&     0x04000000&67108864&This measurement is masked based on IRLS weights for the mean Kron (1980) magnitude.\\
OBJECT\_HAS\_2MASS&0x10000000&268435456&This measurement comes from an object with 2MASS data.\\
OBJECT\_HAS\_GAIA& 0x20000000&536870912&This measurement comes from an object with Gaia data.\\
OBJECT\_HAS\_TYCHO&0x40000000&1073741824&This measurement comes from an object with Tycho data.\\
\hline
\end{tabular}
\end{center}
\label{table:detectionflags3}
\end{table}%




\begin{table}
\caption{Filter}
\begin{center}
\begin{tabular}{ll}
\hline
\hline
FilterID & Filter \\
\hline
1 & g\\
2 & r\\
3 &  i\\
4 & z\\
5 & y\\
\hline
\end{tabular}
\end{center}
\label{table:filters}
\end{table}%

\section{Schema}
\label{sec:schema}

%

\subsection{Object / Mean Object Tables}

\clearpage
\begin{table}
\caption{ObjectThin: Contains the positional information for objects in a number of coordinate systems.  The objects associate single epoch detections and the stacked detections within a one arcsecond radius.  The mean position from the single epoch data is used as the basis for coordinates when available, or the position of an object in the {\em stack} when it is not.  The right ascension and declination for both the {\em stack} and single epoch mean is provided.  The number of detections in each filter from single epoch data is listed, along with which filters the object has a {\em stack} detection \citep[see]{Szalay2007}.}
\begin{center}
\begin{tabular}{lllll}
\hline
\hline
column name & units & data type & default & description\\
\hline
objName & - & VARCHAR(32) & NA  &IAU name for this object.\\
objNameHMS & - & VARCHAR(32) & NA  & Alternate sexigesimal name for this object (DR2 only).\\
objPSOName & - & VARCHAR(32) & NA  &Alternate Pan-STARRS name for this object (DR1 only).\\
objAltName1 & - & VARCHAR(32) & NA  &Alternate name for this object.\\
objAltName2 & - & VARCHAR(32) &   &Altername name for this object.\\
objAltName3 & - & VARCHAR(32) &   &Altername name for this object.\\
objPopularName & - & VARCHAR(140) &   &Well known name for this object.\\
objID & - & BIGINT & NA  &Unique object identifier.\\
uniquePspsOBid & - & BIGINT & NA  &Unique internal PSPS object identifier.\\
ippObjID & - & BIGINT & NA  &IPP internal object identifier.\\
surveyID & - & TINYINT & NA  &Survey identifier.  Details in the Survey table.\\
htmID & - & BIGINT & NA  &Hierarchical triangular mesh (Szalay 2007) index.\\
zoneID & - & INT & NA  &Local zone index, found by dividing the sky into bands of declination 1/2\\  
&   &     &     & arcminute in height: zoneID=floor((90+Dec)/0.0083333).\\
tessID & - & TINYINT & 0  &Tessellation identifier.  Details in the TessellationType table.\\
projectionID & - & SMALLINT & -1  &Projection cell identifier.\\
skyCellID & - & TINYINT & 255  &Skycell region identifier.\\
randomID & - & FLOAT & NA  &Random value drawn from the interval between zero and one.\\
batchID & - & BIGINT & NA  &Internal database batch identifier.\\
dvoRegionID & - & INT & -1  &Internal DVO region identifier.\\
processingVersion & - & TINYINT & NA  &Data release version.\\
objInfoFlag & - & INT & 0  &Information flag bitmask indicating details of the photometry.  Values\\ 
& & & & listed in ObjectInfoFlags.\\
qualityFlag & - & TINYINT & 0  &Subset of objInfoFlag denoting whether this object is real or a likely\\
& & & &false positive.  Values listed in ObjectQualityFlags.\\
raStack & degrees & FLOAT & -999  &Right ascension from {\em stack} detections, weighted mean value across \\
& & & & filters, in equinox J2000.  See StackObjectThin for {\em stack} epoch information.\\
decStack & degrees & FLOAT & -999  &Declination from {\em stack} detections, weighted mean value across\\
& & & & filters, in equinox J2000.  See StackObjectThin for {\em stack} epoch information.\\
raStackErr & arcsec & REAL & -999  &Right ascension standard deviation from {\em stack} detections.\\
decStackErr & arcsec & REAL & -999  &Declination standard deviation from {\em stack} detections.\\
raMean & degrees & FLOAT & -999  &Right ascension from single epoch detections (weighted mean) in \\
& & & & equinox J2000 at the mean epoch given by epochMean.\\
decMean & degrees & FLOAT & -999  &Declination from single epoch detections (weighted mean) in equinox\\
& & & & J2000 at the mean epoch given by epochMean.\\
raMeanErr & arcsec & REAL & -999  &Right ascension standard deviation from single epoch detections.\\
decMeanErr & arcsec & REAL & -999  &Declination standard deviation from single epoch detections.\\
epochMean & days & FLOAT & -999  &Modified Julian Date of the mean epoch corresponding to raMean,\\
& & & & decMean (equinox J2000).\\
posMeanChisq & - & REAL & -999  &Reduced chi squared value of mean position.\\
cx & - & FLOAT & NA  &Cartesian x on a unit sphere. \\
cy & - & FLOAT & NA  &Cartesian y on a unit sphere. \\
cz & - & FLOAT & NA  &Cartesian z on a unit sphere. \\
lambda & degrees & FLOAT & -999  &Ecliptic longitude.\\
beta & degrees & FLOAT & -999  &Ecliptic latitude.\\
l & degrees & FLOAT & -999  &Galactic longitude.\\
b & degrees & FLOAT & -999  &Galactic latitude.\\
nStackObjectRows & - & SMALLINT & -999  &Nr. of independent StackObjectThin rows associated with this object.\\
nStackDetections & - & SMALLINT & -999  &Number of {\em stack} detections.\\
nDetections & - & SMALLINT & -999  &Number of single epoch detections in all filters.\\
ng & - & SMALLINT & -999  &Number of single epoch detections in g filter.\\
nr & - & SMALLINT & -999  &Number of single epoch detections in r filter.\\
ni & - & SMALLINT & -999  &Number of single epoch detections in i filter.\\
nz & - & SMALLINT & -999  &Number of single epoch detections in z filter.\\
ny & - & SMALLINT & -999  &Number of single epoch detections in y filter.\\
\hline
\end{tabular}
\end{center}
\label{table:ObjectThin}
\end{table}%

\clearpage
\begin{table}
\caption{MeanObject: Contains the mean photometric information for objects based on the single epoch data, calculated as described in \citet{Magnier2013}.  To be included in this table, an object must be bright enough to have been detected at least once in an individual exposure.  PSF, Kron (1980), and aperture magnitudes and statistics are listed for all filters.}
\begin{center}
\begin{tabular}{lllll}
\hline
\hline
column name & units & data type & default & description\\
\hline
objID & - & BIGINT & NA  &Unique object identifier.\\
uniquePspsOBid & - & BIGINT & NA  &Unique internal PSPS object identifier.\\
gQfPerfect & - & REAL & -999  &Maximum PSF weighted fraction of pixels totally unmasked\\
& & & & from g filter detections.\\
gMeanPSFMag & AB & REAL & -999  &Mean PSF magnitude from g filter detections.\\
gMeanPSFMagErr & AB & REAL & -999  &Error in mean PSF magnitude from g filter \\
& & & & detections.\\
gMeanPSFMagStd & AB & REAL & -999  &Standard deviation of PSF magnitudes from g filter\\
& & & & detections.\\
gMeanPSFMagNpt & - & SMALLINT & -999  &Number of measurements included in mean PSF\\
& & & & magnitude from g filter detections.\\
gMeanPSFMagMin & AB & REAL & -999  &Minimum PSF magnitude from g filter detections.\\
gMeanPSFMagMax & AB & REAL & -999  &Maximum PSF magnitude from g filter detections.\\
gMeanKronMag & AB & REAL & -999  &Mean Kron (1980) magnitude from g filter detections.\\
gMeanKronMagErr & AB & REAL & -999  &Error in mean Kron (1980) magnitude from g filter\\
& & & & detections.\\
gMeanKronMagStd & AB & REAL & -999  &Standard deviation of Kron (1980) magnitudes from\\
& & & & g filter detections.\\
gMeanKronMagNpt & - & SMALLINT & -999  &Number of measurements included in mean Kron\\
& & & & (1980) magnitude from g filter detections.\\
gMeanApMag & AB & REAL & -999  &Mean aperture magnitude from g filter detections.\\
gMeanApMagErr & AB & REAL & -999  &Error in mean aperture magnitude from g filter \\
& & & & detections.\\
gMeanApMagStd & AB & REAL & -999  &Standard deviation of aperture magnitudes from g\\
& & & & filter detections.\\
gMeanApMagNpt & - & SMALLINT & -999  &Number of measurements included in mean aperture\\
& & & & magnitude from g filter detections.\\
gFlags & - & INT & 0  &Information flag bitmask for mean object from g filter\\
& & & & detections.  Values listed in ObjectFilterFlags.\\
rQfPerfect \\
... & & & & same entries repeated for r, i, z, and y filters \\
yFlags \\
\hline
\end{tabular}
\end{center}
\label{table:MeanObject}
\end{table}%


\subsection{GaiaFrameCoordinate}

\begin{table}
\caption{GaiaFrameCoordinate: PSPS objects calibrated against Gaia astrometry}
\begin{center}
\begin{tabular}{lllll}
\hline
\hline
column name & units & data type & default & description\\
\hline
objID & - & BIGINT & NA  &Unique object identifier.\\
uniquePspsGOid & - & BIGINT & NA  &Unique internal PSPS object identifier.\\
ippObjID & - & BIGINT & NA  &IPP internal object identifier.\\
batchID & - & BIGINT & NA  &Internal database batch identifier.\\
gaiaFlag & - & INT &   &Information flag bitmask.\\
ra & degrees & FLOAT & -999  &Right ascension from single epoch detections (weighted mean) in\\
& & & & equinox J2000 at the mean epoch given by epochMean and calibrated against Gaia.\\
dec & degrees & FLOAT & -999  &Declination from single epoch detections (weighted mean) in equinox\\
& & & & J2000 at the mean epoch given by epochMean and calibrated against Gaia.\\
raErr & arcsec & REAL & -999  &Right ascension standard deviation from single epoch detections.\\
decErr & arcsec & REAL & -999  &Declination standard deviation from single epoch detections.\\
\hline
\end{tabular}
\end{center}
\label{table:GaiaFrameCoordinate}
\end{table}%



\subsection{Stack Tables}
\clearpage
\begin{table}
\caption{StackMeta: Contains the metadata describing the stacked image produced from the combination of a set of single epoch exposures.  The nature of the {\em stack} is given by the StackTypeID.  The astrometric and photometric calibration of the stacked image are listed.}
\begin{center}
\begin{tabular}{lllll}
\hline
\hline
column name & units & data type & default & description\\
\hline
stackImageID & - & BIGINT & NA  &Unique {\em stack} identifier.\\
batchID & - & BIGINT & NA  &Internal database batch identifier.\\
surveyID & - & TINYINT & NA  &Survey identifier.  Details in the Survey table.\\
filterID & - & TINYINT & NA  &Filter identifier.  Details in the Filter table.\\
stackTypeID & - & TINYINT & 0  &Stack type identifier.  Details in the StackType table.\\
tessID & - & TINYINT & 0  &Tessellation identifier.  Details in the TessellationType table.\\
projectionID & - & SMALLINT & -1  &Projection cell identifier.\\
skyCellID & - & TINYINT & 255  &Skycell region identifier.\\
photoCalID & - & INT & NA  &Photometric calibration identifier.  Details in the PhotoCal table.\\
analysisVer & - & VARCHAR(100) &   &IPP software analysis release version.\\
md5sum & - & VARCHAR(100) &   &IPP MD5 Checksum.\\
expTime & seconds & REAL & -999  &Exposure time of the stack.  Necessary for converting listed\\
& & & & fluxes and magnitudes back to measured ADU counts.\\
nP2Images & - & SMALLINT & -999  &Number of input exposures/frames contributing to this stack.\\
detectionThreshold & magnitudes & REAL & -999  &Reference magnitude for detection efficiency calculation.\\
astroScat & - & REAL & -999  &Measurement of the calibration (not astrometric error) defined to\\
& & & & be the sum in quadrature of the standard deviations in the X and\\
& & & & Y directions.\\
photoScat & - & REAL & -999  &Photometric scatter relative to reference catalog.\\
nAstroRef & - & INT & -999  &Number of astrometric reference sources.\\
nPhotoRef & - & INT & -999  &Number of photometric reference sources.\\
recalAstroScatX & arcsec & REAL & -999  &Measurement of the re-calibration (not astrometric error)\\
& & & & in the X direction.\\
recalAstroScatY & arcsec & REAL & -999  &Measurement of the re-calibration (not astrometric error)\\
& & & & in the Y direction.\\
recalNAstroStars & - & INT & -999  &Number of astrometric reference sources used in\\
& & & & recalibration.\\
recalphotoScat & magnitudes & REAL & -999  &Photometric scatter relative to reference catalog.\\
recalNPhotoStars & - & INT & -999  &Number of astrometric reference sources used in\\
& & & & recalibration.\\
psfModelID & - & INT & -999  &PSF model identifier.\\
psfFWHM & arcsec & REAL & -999  &Mean PSF full width at half maximum at image center.\\
psfWidMajor & arcsec & REAL & -999  &PSF major axis FWHM at image center.\\
psfWidMinor & arcsec & REAL & -999  &PSF minor axis FWHM at image center.\\
psfTheta & degrees & REAL & -999  &PSF major axis orientation at image center.\\
photoZero & magnitudes & REAL & -999  &Locally derived photometric zero point for this stack.\\
photoZeroAperture & magnitudes & REAL & -999 & Locally derived photometric zero point for this stack\\
& & & & (Aperture-like measurements only) (DR2).\\
ctype1 & - & VARCHAR(100) &   &Name of astrometric projection in right ascension.\\
ctype2 & - & VARCHAR(100) &   &Name of astrometric projection in declination.\\
crval1 & degrees & FLOAT & -999  &Right ascension corresponding to reference pixel.\\
crval2 & degrees & FLOAT & -999  &Declination corresponding to reference pixel.\\
crpix1 & sky pixels & FLOAT & -999  &Reference pixel for right ascension.\\
crpix2 & sky pixels & FLOAT & -999  &Reference pixel for declination.\\
cdelt1 & degrees/pixel & FLOAT & -999  &Pixel scale in right ascension.\\
cdelt2 & degrees/pixel & FLOAT & -999  &Pixel scale in declination.\\
pc001001 & - & FLOAT & -999  &Linear transformation matrix element between image pixel x and \\
& & & & right ascension.\\
pc001002 & - & FLOAT & -999  &Linear transformation matrix element between image pixel y and \\
& & & & right ascension.\\
pc002001 & - & FLOAT & -999  &Linear transformation matrix element between image pixel x and \\
& & & & declination.\\
pc002002 & - & FLOAT & -999  &Linear transformation matrix element between image pixel y and \\
& & & & declination.\\
processingVersion & - & TINYINT & NA  &Data release version.\\
\hline
\end{tabular}
\end{center}
\label{table:StackMeta}
\end{table}%

\clearpage
\begin{table}
\caption{StackObjectThin: Contains the positional and photometric information for point-source photometry of {\em stack} detections.  The information for all filters are joined into a single row, with metadata indicating if this {\em stack} object represents the primary detection.  Due to overlaps in the {\em stack} tessellations, an object may appear in multiple {\em stack} images.  The primary detection is the unique detection from the {\em stack} image that provides the best coverage with minimal projection stretching.  All other detections of the object in that filter are secondary, regardless of their properties.  The detection flagged as best is the primary detection if that detection has a psfQf value greater than 0.98;  if that is not met, then any of the primary or secondary detections with the highest psfQf value is flagged as best.}
\begin{center}
\begin{tabular}{lllll}
\hline
\hline
column name & units & data type & default & description\\
\hline
objID & - & BIGINT & NA  &Unique object identifier.\\
uniquePspsSTid & - & BIGINT & NA  &Unique internal PSPS {\em stack} identifier.\\
ippObjID & - & BIGINT & NA  &IPP internal object identifier.\\
surveyID & - & TINYINT & NA  &Survey identifier.  Details in the Survey table.\\
tessID & - & TINYINT & 0  &Tessellation identifier.  Details in the TessellationType table.\\
projectionID & - & SMALLINT & -1  &Projection cell identifier.\\
skyCellID & - & TINYINT & 255  &Skycell region identifier.\\
randomStackObjID & - & FLOAT & NA  &Random value drawn from the interval between zero and one.\\
primaryDetection & - & TINYINT & 255  &Identifies if this row is the primary {\em stack} detection.\\
bestDetection & - & TINYINT & 255  &Identifies if this row is the best detection.\\
dvoRegionID & - & INT & -1  &Internal DVO region identifier.\\
processingVersion & - & TINYINT & NA  &Data release version.\\
gippDetectID & - & BIGINT & NA  &IPP internal detection identifier.\\
gstackDetectID & - & BIGINT & NA  &Unique {\em stack} detection identifier.\\
gstackImageID & - & BIGINT & NA  &Unique {\em stack} identifier for g filter detection.\\
gra & degrees & FLOAT & -999  &Right ascension from g filter {\em stack} detection.\\
gdec & degrees & FLOAT & -999  &Declination from g filter {\em stack} detection.\\
graErr & arcsec & REAL & -999  &Right ascension error from g filter {\em stack} detection.\\
gdecErr & arcsec & REAL & -999  &Declination error from g filter {\em stack} detection.\\
gEpoch & days & FLOAT & -999  &Modified Julian Date of the mean epoch of images contributing to the\\
& & & & g-band {\em stack} (equinox J2000).\\
gPSFMag & AB & REAL & -999  &PSF magnitude from g filter {\em stack} detection.\\
gPSFMagErr & AB & REAL & -999  &Error in PSF magnitude from g filter {\em stack} detection.\\
gApMag & AB & REAL & -999  &Aperture magnitude from g filter {\em stack} detection.\\
gApMagErr & AB & REAL & -999  &Error in aperture magnitude from g filter {\em stack} detection.\\
gKronMag & AB & REAL & -999  &Kron (1980) magnitude from g filter {\em stack} detection.\\
gKronMagErr & AB & REAL & -999  &Error in Kron (1980) magnitude from g filter {\em stack} detection.\\
ginfoFlag & - & BIGINT & 0  &Information flag bitmask indicating details of the g filter {\em stack}\\
& & & & photometry.  Values listed in DetectionFlags.\\
ginfoFlag2 & - & INT & 0  &Information flag bitmask indicating details of the g filter {\em stack}\\
& & & & photometry.  Values listed in DetectionFlags2.\\
ginfoFlag3 & - & INT & 0  &Information flag bitmask indicating details of the g filter {\em stack}\\
& & & & photometry.  Values listed in DetectionFlags3.\\
ginfoFlag4 & - & INT & 0 & Information flag bitmask indicating details of the g filter stack photometry.\\
& & & &  Values listed in ObjectFilterFlags. (DR2)\\
gnFrames & - & INT & -999  &Number of input frames/exposures contributing to the g filter {\em stack}\\
& & & & detection.\\
rippDetectID \\
... & & & & same entries repeated for r, i, z, and y filters \\
ynFrames \\
\hline
\end{tabular}
\end{center}
\label{table:StackObjectThin}
\end{table}%

\clearpage
\begin{table}
\caption{StackObjectAttributes: Contains the PSF, \citet{Kron1980}, and aperture fluxes for all filters in a single row, along with point-source object shape parameters.  See {\em StackObjectThin} table for discussion of primary, secondary, and best detections.
}
\begin{center}
\begin{tabular}{lllll}
\hline
\hline
column name & units & data type & default & description\\
\hline
objID & - & BIGINT & NA  &Unique object identifier.\\
uniquePspsSTid & - & BIGINT & NA  &Unique internal PSPS {\em stack} identifier.\\
ippObjID & - & BIGINT & NA  &IPP internal object identifier.\\
randomStackObjID & - & FLOAT & NA  &Random value drawn from the interval between zero and one.\\
primaryDetection & - & TINYINT & 255  &Identifies if this row is the primary {\em stack} detection.\\
bestDetection & - & TINYINT & 255  &Identifies if this row is the best detection.\\
gippDetectID & - & BIGINT & NA  &IPP internal detection identifier.\\
gstackDetectID & - & BIGINT & NA  &Unique {\em stack} detection identifier.\\
gstackImageID & - & BIGINT & NA  &Unique {\em stack} identifier for g filter detection.\\
gxPos & sky pixels & REAL & -999  &PSF x center location from g filter {\em stack} detection.\\
gyPos & sky pixels & REAL & -999  &PSF y center location from g filter {\em stack} detection.\\
gxPosErr & sky pixels & REAL & -999  &Error in PSF x center location from g filter {\em stack} detection.\\
gyPosErr & sky pixels & REAL & -999  &Error in PSF y center location from g filter {\em stack} detection.\\
gpsfMajorFWHM & arcsec & REAL & -999  &PSF major axis FWHM from g filter {\em stack} detection.\\
gpsfMinorFWHM & arcsec & REAL & -999  &PSF minor axis FWHM from g filter {\em stack} detection.\\
gpsfTheta & degrees & REAL & -999  &PSF major axis orientation from g filter {\em stack} detection.\\
gpsfCore & - & REAL & -999  &PSF core parameter k from g filter {\em stack} detection, where \\
& & & & $F = F0 / (1 + k r^2 + r^{3.33})$.\\
gpsfLikelihood & - & REAL & -999  &Likelihood that this g filter {\em stack} detection is best fit\\
& & & & by a PSF.\\
gpsfQf & - & REAL & -999  &PSF coverage factor for g filter {\em stack} detection.\\
gpsfQfPerfect & - & REAL & -999  &PSF-weighted fraction of pixels totally unmasked for g filter {\em stack} detection.\\
gpsfChiSq & - & REAL & -999  &Reduced chi squared value of the PSF model fit for g filter {\em stack} detection.\\
gmomentXX & $arcsec^2$ & REAL & -999  &Second moment $M_{xx}$ for g filter {\em stack} detection.\\
gmomentXY & $arcsec^2$ & REAL & -999  &Second moment $M_{xy}$ for g filter {\em stack} detection.\\
gmomentYY & $arcsec^2$ & REAL & -999  &Second moment $M_{yy}$ for g filter {\em stack} detection.\\
gmomentR1 & arcsec & REAL & -999  &First radial moment for g filter {\em stack} detection.\\
gmomentRH & $arcsec^{0.5}$ & REAL & -999  &Half radial moment ($r^{0.5}$ weighting) for g filter {\em stack} detection.\\
gPSFFlux & Jy & REAL & -999  &PSF flux from g filter {\em stack} detection.\\
gPSFFluxErr & Jy & REAL & -999  &Error in PSF flux from g filter {\em stack} detection.\\
gApFlux & Jy & REAL & -999  &Aperture flux from g filter {\em stack} detection.\\
gApFluxErr & Jy & REAL & -999  &Error in aperture flux from g filter {\em stack} detection.\\
gApFillFac & - & REAL & -999  &Aperture fill factor from g filter {\em stack} detection.\\
gApRadius & arcsec & REAL & -999  &Aperture radius for g filter {\em stack} detection.\\
gKronFlux & Jy & REAL & -999  &Kron (1980) flux from g filter {\em stack} detection.\\
gKronFluxErr & Jy & REAL & -999  &Error in Kron (1980) flux from g filter {\em stack} detection.\\
gKronRad & arcsec & REAL & -999  &Kron (1980) radius from g filter {\em stack} detection.\\
gexpTime & seconds & REAL & -999  &Exposure time of the g filter stack.  Necessary for converting \\
& & & & listed fluxes and magnitudes back to measured ADU counts.\\
gExtNSigma & - & REAL & -999  &An extendedness measure for the g filter {\em stack} detection based on\\
& & & & the deviation between PSF and Kron (1980) magnitudes, normalized \\
& & & & by the PSF magnitude uncertainty.\\
gsky & $Jy/arcsec^2$ & REAL & -999  &Residual background sky level at the g filter {\em stack} detection.\\
gskyErr & $Jy/arcsec^2$ & REAL & -999  &Error in residual background sky level at the g filter {\em stack} detection.\\
gzp & magnitudes & REAL & 0  &Photometric zeropoint for the g filter stack.  Necessary for converting\\
& & & & listed fluxes and magnitudes back to measured ADU counts.\\
gzpAPER & magnitudes & REAL & 0 & Photometric zeropoint for the g filter stack (APERTURE-like magnitudes only).\\  
& & & & Needed to convert fluxes or magnitudes back to measured ADU counts. (DR2)\\
gPlateScale & arcsec/pixel & REAL & 0  &Local plate scale for the g filter stack.\\
rippDetectID \\
... & & & & same entries repeated for r, i, z, and y filters \\
yPlateScale \\
\hline
\end{tabular}
\end{center}
\label{table:StackObjectAttributes}
\end{table}%

\clearpage
\begin{table}
\caption{StackApFlx: Contains the unconvolved fluxes within the SDSS R5 (r = 3.00 arcsec), R6 (r = 4.63 arcsec), and R7 (r = 7.43 arcsec) apertures \citep{Stoughton2002}.  Convolved fluxes within these same apertures are also provided for images convolved to 6 sky pixels (1.5 arcsec) and 8 sky pixels (2.0 arcsec).  All filters are matched into a single row.  See {\em StackObjectThin} table for discussion of primary, secondary, and best detections.}
\begin{center}
\begin{tabular}{lllll}
\hline
\hline
column name & units & data type & default & description\\
\hline
objID & - & BIGINT & NA  &Unique object identifier.\\
uniquePspsSTid & - & BIGINT & NA  &Unique internal PSPS {\em stack} identifier.\\
ippObjID & - & BIGINT & NA  &IPP internal object identifier.\\
randomStackObjID & - & FLOAT & NA  &Random value drawn from the interval between zero and one.\\
primaryDetection & - & TINYINT & 255  &Identifies if this row is the primary {\em stack} detection.\\
bestDetection & - & TINYINT & 255  &Identifies if this row is the best detection.\\
gstackDetectID & - & BIGINT & NA  &Unique {\em stack} detection identifier.\\
gstackImageID & - & BIGINT & NA  &Unique {\em stack} identifier for g filter detection.\\
gippDetectID & - & BIGINT & NA  &IPP internal detection identifier.\\
gflxR5 & Jy & REAL & -999  &Flux from g filter detection within an aperture of radius r = 3.00 arcsec.\\
gflxR5Err & Jy & REAL & -999  &Error in flux from g filter detection within an aperture of radius r = 3.00 arcsec.\\
gflxR5Std & Jy & REAL & -999  &Standard deviation of g filter flux within an aperture of radius r = 3.00 arcsec.\\
gflxR5Fill & - & REAL & -999  &Aperture fill factor for g filter detection within an aperture of radius r = 3.00 arcsec.\\
gflxR6 & Jy & REAL & -999  &Flux from g filter detection within an aperture of radius r = 4.63 arcsec.\\
gflxR6Err & Jy & REAL & -999  &Error in flux from g filter detection within an aperture of radius r = 4.63 arcsec.\\
gflxR6Std & Jy & REAL & -999  &Standard deviation of g filter flux within an aperture of radius r = 4.63 arcsec.\\
gflxR6Fill & - & REAL & -999  &Aperture fill factor for g filter detection within an aperture of radius r = 4.63 arcsec.\\
gflxR7 & Jy & REAL & -999  &Flux from g filter detection within an aperture of radius r = 7.43 arcsec.\\
gflxR7Err & Jy & REAL & -999  &Error in flux from g filter detection within an aperture of radius r = 7.43 arcsec.\\
gflxR7Std & Jy & REAL & -999  &Standard deviation of g filter flux within an aperture of radius r = 7.43 arcsec.\\
gflxR7Fill & - & REAL & -999  &Aperture fill factor for g filter detection within an aperture of radius r = 7.43 arcsec.\\
gc6flxR5 & Jy & REAL & -999  &Flux from g filter detection convolved to a target of 6 sky pixels\\
& & & & (1.5 arcsec) within an aperture of radius r = 3.00 arcsec.\\
gc6flxR5Err & Jy & REAL & -999  &Error in flux from g filter detection convolved to a target of \\
& & & & 6 sky pixels (1.5 arcsec) within an aperture of radius r = 3.00 arcsec.\\
gc6flxR5Std & Jy & REAL & -999  &Standard deviation of flux from g filter detection convolved to \\
& & & & a target of 6 sky pixels (1.5 arcsec) within an aperture of radius r = 3.00 arcsec.\\
gc6flxR5Fill & - & REAL & -999  &Aperture fill factor for g filter detection convolved to a \\
& & & & target of 6 sky pixels (1.5 arcsec) within an aperture of radius r = 3.00 arcsec.\\
gc6flxR6 & Jy & REAL & -999  &Flux from g filter detection convolved to a target of 6 sky pixels\\
& & & & (1.5 arcsec) within an aperture of radius r = 4.63 arcsec.\\
gc6flxR6Err & Jy & REAL & -999  &Error in flux from g filter detection convolved to a target of \\
& & & & 6 sky pixels (1.5 arcsec) within an aperture of radius r = 4.63 arcsec.\\
gc6flxR6Std & Jy & REAL & -999  &Standard deviation of flux from g filter detection convolved to \\
& & & & a target of 6 sky pixels (1.5 arcsec) within an aperture of radius r = 4.63 arcsec.\\
gc6flxR6Fill & - & REAL & -999  &Aperture fill factor for g filter detection convolved to a target of \\
& & & & 6 sky pixels (1.5 arcsec) within an aperture of radius r = 4.63 arcsec.\\
gc6flxR7 & Jy & REAL & -999  &Flux from g filter detection convolved to a target of 6 sky pixels\\
& & & & (1.5 arcsec) within an aperture of radius r = 7.43 arcsec.\\
gc6flxR7Err & Jy & REAL & -999  &Error in flux from g filter detection convolved to a target of \\
& & & & 6 sky pixels (1.5 arcsec) within an aperture of radius r = 7.43 arcsec.\\
gc6flxR7Std & Jy & REAL & -999  &Standard deviation of flux from g filter detection convolved to a target of \\
& & & & 6 sky pixels (1.5 arcsec) within an aperture of radius r = 7.43 arcsec.\\
gc6flxR7Fill & - & REAL & -999  &Aperture fill factor for g filter detection convolved to a target of 6 sky \\
& & & & pixels (1.5 arcsec) within an aperture of radius r = 7.43 arcsec.\\
gc8flxR5 & Jy & REAL & -999  &Flux from g filter detection convolved to a target of 8 sky pixels\\
& & & & (2.0 arcsec) within an aperture of radius r = 3.00 arcsec.\\
gc8flxR5Err & Jy & REAL & -999  &Error in flux from g filter detection convolved to a target of 8 sky pixels \\
& & & & (2.0 arcsec) within an aperture of radius r = 3.00 arcsec.\\
gc8flxR5Std & Jy & REAL & -999  &Standard deviation of flux from g filter detection convolved to a target of 8 sky \\
& & & & pixels (2.0 arcsec) within an aperture of radius r = 3.00 arcsec.\\
gc8flxR5Fill & - & REAL & -999  &Aperture fill factor for g filter detection convolved to a target of 8 sky \\
& & & & pixels (2.0 arcsec) within an aperture of radius r = 3.00 arcsec.\\
gc8flxR6 & Jy & REAL & -999  &Flux from g filter detection convolved to a target of 8 sky pixels\\
& & & & (2.0 arcsec) within an aperture of radius r = 4.63 arcsec.\\
gc8flxR6Err & Jy & REAL & -999  &Error in flux from g filter detection convolved to a target of 8 sky pixels\\
& & & & (2.0 arcsec) within an aperture of radius r = 4.63 arcsec.\\
gc8flxR6Std & Jy & REAL & -999  &Standard deviation of flux from g filter detection convolved to a target of \\
& & & & 8 sky pixels (2.0 arcsec) within an aperture of radius r = 4.63 arcsec.\\
gc8flxR6Fill & - & REAL & -999  &Aperture fill factor for g filter detection convolved to a target of 8 sky pixels \\
& & & & (2.0 arcsec) within an aperture of radius r = 4.63 arcsec.\\
gc8flxR7 & Jy & REAL & -999  &Flux from g filter detection convolved to a target of 8 sky pixels (2.0 arcsec) \\
& & & & within an aperture of radius r = 7.43 arcsec.\\
gc8flxR7Err & Jy & REAL & -999  &Error in flux from g filter detection convolved to a target 8 sky pixels \\
& & & & (2.0 arcsec) within an aperture of radius r = 7.43 arcsec.\\
gc8flxR7Std & Jy & REAL & -999  &Standard deviation of flux from g filter detection convolved to a target of \\
& & & & 8 sky pixels (2.0 arcsec) within an aperture of radius r = 7.43 arcsec.\\
gc8flxR7Fill & - & REAL & -999  &Aperture fill factor for g filter detection convolved to a target of 8 sky pixels \\
& & & & (2.0 arcsec) within an aperture of radius r = 7.43 arcsec.\\
rstackDetectID \\
... & & & & same entries repeated for r, i, z, and y filters \\
yc8flxR7Fill\\
\hline
\end{tabular}
\end{center}
\label{table:StackApFlx}
\end{table}%

\clearpage
\begin{table}
\caption{StackModelFitExp: Contains the exponential fit parameters to extended sources.  See {\em StackObjectThin} table for discussion of primary, secondary, and best detections. }
\begin{center}
\begin{tabular}{lllll}
\hline
\hline
column name & units & data type & default & description\\
\hline
objID & - & BIGINT & NA  &Unique object identifier.\\
uniquePspsSTid & - & BIGINT & NA  &Unique internal PSPS {\em stack} identifier.\\
ippObjID & - & BIGINT & NA  &IPP internal object identifier.\\
randomStackObjID & - & FLOAT & NA  &Random value drawn from the interval between zero and one.\\
primaryDetection & - & TINYINT & 255  &Identifies if this row is the primary {\em stack} detection.\\
bestDetection & - & TINYINT & 255  &Identifies if this row is the best detection.\\
gippDetectID & - & BIGINT & NA  &IPP internal detection identifier.\\
gstackDetectID & - & BIGINT & NA  &Unique {\em stack} detection identifier.\\
gstackImageID & - & BIGINT & NA  &Unique {\em stack} identifier for g filter detection.\\
gExpRadius & arcsec & REAL & -999  &Exponential fit radius for g filter {\em stack} detection.\\
gExpRadiusErr & arcsec & REAL & -999  &Error in exponential fit radius for g filter {\em stack} detection.\\
gExpMag & AB & REAL & -999  &Exponential fit magnitude for g filter {\em stack} detection.\\
gExpMagErr & AB & REAL & -999  &Error in exponential fit magnitude for g filter {\em stack} detection.\\
gExpAb & - & REAL & -999  &Exponential fit axis ratio for g filter {\em stack} detection.\\
gExpAbErr & - & REAL & -999  &Error in exponential fit axis ratio for g filter {\em stack} detection.\\
gExpPhi & degrees & REAL & -999  &Major axis position angle, phi, of exponential fit for g \\
& & & & filter {\em stack} detection.\\
gExpPhiErr & degrees & REAL & -999  &Error in major axis position angle of exponential fit for g\\
& & & & filter {\em stack} detection.\\
gExpRa & degrees & FLOAT & -999  &Right ascension of exponential fit center for g filter \\
& & & & {\em stack} detection.\\
gExpDec & degrees & FLOAT & -999  &Declination of exponential fit center for g filter \\
& & & & {\em stack} detection.\\
gExpRaErr & arcsec & REAL & -999  &Error in right ascension of exponential fit center for g \\
& & & & filter {\em stack} detection.\\
gExpDecErr & arcsec & REAL & -999  &Error in declination of exponential fit center for g \\
& & & & filter {\em stack} detection.\\
gExpChisq & - & REAL & -999  &Exponential fit reduced chi squared for g filter {\em stack}
detection.\\
rippDetectID \\
... & & & & same entries repeated for r, i, z, and y filters \\
yExpChisq \\
\hline
\end{tabular}
\end{center}
\label{table:StackModelFitExp}
\end{table}%

\begin{table}
\caption{StackModelFitDeV: Contains the \citet{deVaucouleurs1948} fit parameters to extended sources.  See {\em StackObjectThin} table for discussion of primary, secondary, and best detections.}
\begin{center}
\begin{tabular}{lllll}
\hline
\hline
column name & units & data type & default & description\\
\hline
objID & - & BIGINT & NA  &Unique object identifier.\\
uniquePspsSTid & - & BIGINT & NA  &Unique internal PSPS {\em stack} identifier.\\
ippObjID & - & BIGINT & NA  &IPP internal object identifier.\\
randomStackObjID & - & FLOAT & NA  &Random value drawn from the interval between zero and one.\\
primaryDetection & - & TINYINT & 255  &Identifies if this row is the primary {\em stack} detection.\\
bestDetection & - & TINYINT & 255  &Identifies if this row is the best detection.\\
gippDetectID & - & BIGINT & NA  &IPP internal detection identifier.\\
gstackDetectID & - & BIGINT & NA  &Unique {\em stack} detection identifier.\\
gstackImageID & - & BIGINT & NA  &Unique {\em stack} identifier for g filter detection.\\
gDeVRadius & arcsec & REAL & -999  &\citet{deVaucouleurs1948} fit radius for g filter {\em stack} detection.\\
gDeVRadiusErr & arcsec & REAL & -999  &Error in \citet{deVaucouleurs1948} fit radius for g filter {\em stack} detection.\\
gDeVMag & AB & REAL & -999  &\citet{deVaucouleurs1948} fit magnitude for g filter {\em stack} detection.\\
gDeVMagErr & AB & REAL & -999  &Error in \citet{deVaucouleurs1948} fit magnitude for g filter {\em stack} detection.\\
gDeVAb & - & REAL & -999  &\citet{deVaucouleurs1948} fit axis ratio for g filter {\em stack} detection.\\
gDeVAbErr & - & REAL & -999  &Error in \citet{deVaucouleurs1948} fit axis ratio for g filter {\em stack} detection.\\
gDeVPhi & degrees & REAL & -999  &Major axis position angle, phi, of \citet{deVaucouleurs1948} fit for g filter \\
& & & & {\em stack} detection.\\
gDeVPhiErr & degrees & REAL & -999  &Error in major axis position angle of \citet{deVaucouleurs1948} fit for g filter\\
& & & & {\em stack} detection.\\
gDeVRa & degrees & FLOAT & -999  &Right ascension of \citet{deVaucouleurs1948} fit center for g filter {\em stack} \\
& & & & detection.\\
gDeVDec & degrees & FLOAT & -999  &Declination of \citet{deVaucouleurs1948} fit center for g filter \\
& & & & {\em stack} detection.\\
gDeVRaErr & arcsec & REAL & -999  &Error in right ascension of \citet{deVaucouleurs1948} fit center for g filter {\em stack} \\
& & & & detection.\\
gDeVDecErr & arcsec & REAL & -999  &Error in declination of \citet{deVaucouleurs1948} fit center for g filter {\em stack} detection.\\
gDeVChisq & - & REAL & -999  &\citet{deVaucouleurs1948} fit reduced chi squared for g filter {\em stack} detection.\\
rippDetectID \\
... & & & & same entries repeated for r, i, z, and y filters \\
yDeVChisq \\
\hline
\end{tabular}
\end{center}
\label{table:StackModelFitDeV}
\end{table}%

\begin{table}
\caption{StackModelFitSer: Contains the \citet{Sersic1963} fit parameters to extended sources.  See {\em StackObjectThin} table for discussion of primary, secondary, and best detections \citep{Sersic1963}.}
\begin{center}
\begin{tabular}{lllll}
\hline
\hline
column name & units & data type & default & description\\
\hline
objID & - & BIGINT & NA  &Unique object identifier.\\
uniquePspsSTid & - & BIGINT & NA  &Unique internal PSPS {\em stack} identifier.\\
ippObjID & - & BIGINT & NA  &IPP internal object identifier.\\
randomStackObjID & - & FLOAT & NA  &Random value drawn from the interval between zero and one.\\
primaryDetection & - & TINYINT & 255  &Identifies if this row is the primary {\em stack} detection.\\
bestDetection & - & TINYINT & 255  &Identifies if this row is the best detection.\\
gippDetectID & - & BIGINT & NA  &IPP internal detection identifier.\\
gstackDetectID & - & BIGINT & NA  &Unique {\em stack} detection identifier.\\
gstackImageID & - & BIGINT & NA  &Unique {\em stack} identifier for g filter detection.\\
gSerRadius & arcsec & REAL & -999  &\citet{Sersic1963} fit radius for g filter {\em stack} detection.\\
gSerRadiusErr & arcsec & REAL & -999  &Error in \citet{Sersic1963} fit radius for g filter {\em stack} detection.\\
gSerMag & AB & REAL & -999  &\citet{Sersic1963} fit magnitude for g filter {\em stack} detection.\\
gSerMagErr & AB & REAL & -999  &Error in \citet{Sersic1963} fit magnitude for g filter {\em stack} detection.\\
gSerAb & - & REAL & -999  &\citet{Sersic1963} fit axis ratio for g filter {\em stack} detection.\\
gSerAbErr & - & REAL & -999  &Error in \citet{Sersic1963} fit axis ratio for g filter {\em stack} detection.\\
gSerNu & - & REAL & -999  &\citet{Sersic1963} fit index for g filter {\em stack} detection.\\
gSerNuErr & - & REAL & -999  &Error in \citet{Sersic1963} fit index for g filter {\em stack} detection.\\
gSerPhi & degrees & REAL & -999  &Major axis position angle, phi, of \citet{Sersic1963} fit for g filter {\em stack} detection.\\
gSerPhiErr & degrees & REAL & -999  &Error in major axis position angle of \citet{Sersic1963} fit for g filter {\em stack} detection.\\
gSerRa & degrees & FLOAT & -999  &Right ascension of \citet{Sersic1963} fit center for g filter {\em stack} detection.\\
gSerDec & degrees & FLOAT & -999  &Declination of \citet{Sersic1963} fit center for g filter {\em stack} detection.\\
gSerRaErr & arcsec & REAL & -999  &Error in right ascension of \citet{Sersic1963} fit center for g filter {\em stack} detection.\\
gSerDecErr & arcsec & REAL & -999  &Error in declination of \citet{Sersic1963} fit center for g filter {\em stack} detection.\\
gSerChisq & - & REAL & -999  &\citet{Sersic1963} fit reduced chi squared for g filter {\em stack} detection.\\
rippDetectID \\
... & & & & same entries repeated for r, i, z, and y filters \\
ySerChisq \\
\hline
\end{tabular}
\end{center}
\label{table:StackModelFitSer}
\end{table}%

\clearpage

\begin{table}
\caption{StackApFlxExGalUnc: Contains the unconvolved fluxes within the SDSS R3 (r = 1.03 arcsec), R4 (r = 1.76 arcsec), R5 (r = 3.00 arcsec), R6 (r = 4.63 arcsec), R7 (r = 7.43 arcsec), R8 (r = 11.42 arcsec), R9 (r = 18.20 arcsec), R10 (r = 28.20 arcsec), and R11 (r = 44.21 arcsec) apertures \citep{Stoughton2002} for extended sources.  These measurements are only provided for objects in the extragalactic sky, i.e., they are not provided for objects in the Galactic plane because they are not useful in crowded areas.  See {\em StackObjectThin} table for discussion of primary, secondary, and best detections.  }
\begin{center}
\begin{tabular}{lllll}
\hline
\hline
column name & units & data type & default & description\\
\hline
objID & - & BIGINT & NA  &Unique object identifier.\\
uniquePspsSTid & - & BIGINT & NA  &Unique internal PSPS {\em stack} identifier.\\
ippObjID & - & BIGINT & NA  &IPP internal object identifier.\\
randomStackObjID & - & FLOAT & NA  &Random value drawn from the interval between zero and one.\\
primaryDetection & - & TINYINT & 255  &Identifies if this row is the primary {\em stack} detection.\\
bestDetection & - & TINYINT & 255  &Identifies if this row is the best detection.\\
gippDetectID & - & BIGINT & NA  &IPP internal detection identifier.\\
gstackDetectID & - & BIGINT & NA  &Unique {\em stack} detection identifier.\\
gstackImageID & - & BIGINT & NA  &Unique {\em stack} identifier for g filter detection.\\
gflxR3 & Jy & REAL & -999  &Flux from g filter detection within an aperture of radius r = 1.03 arcsec.\\
gflxR3Err & Jy & REAL & -999  &Error in flux from g filter detection within an aperture of radius r = 1.03 arcsec.\\
gflxR3Std & Jy & REAL & -999  &Standard deviation of flux from g filter detection within an aperture of radius \\
& & & & r = 1.03 arcsec.\\
gflxR3Fill & - & REAL & -999  &Aperture fill factor for g filter detection within an aperture of radius r = 1.03 arcsec.\\
gflxR4 & Jy & REAL & -999  &Flux from g filter detection within an aperture of radius r = 1.76 arcsec.\\
gflxR4Err & Jy & REAL & -999  &Error in flux from g filter detection within an aperture of radius r = 1.76 arcsec.\\
gflxR4Std & Jy & REAL & -999  &Standard deviation of flux from g filter detection within an aperture of radius \\
& & & & r = 1.76 arcsec.\\
gflxR4Fill & - & REAL & -999  &Aperture fill factor for g filter detection within an aperture of radius r = 1.76 arcsec.\\
gflxR5 & Jy & REAL & -999  &Flux from g filter detection within an aperture of radius r = 3.00 arcsec.\\
gflxR5Err & Jy & REAL & -999  &Error in flux from g filter detection within an aperture of radius r = 3.00 arcsec.\\
gflxR5Std & Jy & REAL & -999  &Standard deviation of flux from g filter detection within an aperture of radius \\
& & & & r = 3.00 arcsec.\\
gflxR5Fill & - & REAL & -999  &Aperture fill factor for g filter detection within an aperture of radius r = 3.00 arcsec.\\
gflxR6 & Jy & REAL & -999  &Flux from g filter detection within an aperture of radius r = 4.63 arcsec.\\
gflxR6Err & Jy & REAL & -999  &Error in flux from g filter detection within an aperture of radius r = 4.63 arcsec.\\
gflxR6Std & Jy & REAL & -999  &Standard deviation of flux from g filter detection within an aperture of radius \\
& & & & r = 4.63 arcsec.\\
gflxR6Fill & - & REAL & -999  &Aperture fill factor for g filter detection within an aperture of radius r = 4.63 arcsec.\\
gflxR7 & Jy & REAL & -999  &Flux from g filter detection within an aperture of radius r = 7.43 arcsec.\\
gflxR7Err & Jy & REAL & -999  &Error in flux from g filter detection within an aperture of radius r = 7.43 arcsec.\\
gflxR7Std & Jy & REAL & -999  &Standard deviation of flux from g filter detection within an aperture of radius \\
& & & & r = 7.43 arcsec.\\
gflxR7Fill & - & REAL & -999  &Aperture fill factor for g filter detection within an aperture of radius r = 7.43 arcsec.\\
gflxR8 & Jy & REAL & -999  &Flux from g filter detection within an aperture of radius r = 11.42 arcsec.\\
gflxR8Err & Jy & REAL & -999  &Error in flux from g filter detection within an aperture of radius r = 11.42 arcsec.\\
gflxR8Std & Jy & REAL & -999  &Standard deviation of flux from g filter detection within an aperture of radius \\
& & & & r = 11.42 arcsec.\\
gflxR8Fill & - & REAL & -999  &Aperture fill factor for g filter detection within an aperture of radius r = 11.42 arcsec.\\
gflxR9 & Jy & REAL & -999  &Flux from g filter detection within an aperture of radius r = 18.20 arcsec.\\
gflxR9Err & Jy & REAL & -999  &Error in flux from g filter detection within an aperture of radius r = 18.20 arcsec.\\
gflxR9Std & Jy & REAL & -999  &Standard deviation of flux from g filter detection within an aperture of radius \\
& & & & r = 18.20 arcsec.\\
gflxR9Fill & - & REAL & -999  &Aperture fill factor for g filter detection within an aperture of radius r = 18.20 arcsec.\\
gflxR10 & Jy & REAL & -999  &Flux from g filter detection within an aperture of radius r = 28.20 arcsec.\\
gflxR10Err & Jy & REAL & -999  &Error in flux from g filter detection within an aperture of radius r = 28.20 arcsec.\\
gflxR10Std & Jy & REAL & -999  &Standard deviation of flux from g filter detection within an aperture of radius \\
& & & & r = 28.20 arcsec.\\
gflxR10Fill & - & REAL & -999  &Aperture fill factor for g filter detection within an aperture of radius r = 28.20 arcsec.\\
gflxR11 & Jy & REAL & -999  &Flux from g filter detection within an aperture of radius r = 44.21 arcsec.\\
gflxR11Err & Jy & REAL & -999  &Error in flux from g filter detection within an aperture of radius r = 44.21 arcsec.\\
gflxR11Std & Jy & REAL & -999  &Standard deviation of flux from g filter detection within an aperture of radius \\
& & & & r = 44.21 arcsec.\\
gflxR11Fill & - & REAL & -999  &Aperture fill factor for g filter detection within an aperture of radius r = 44.21 arcsec.\\
rippDetectID\\
... & & & & same entries repeated for r, i, z, and y filters \\
yflxR11Fill\\
\hline
\end{tabular}
\end{center}
\label{table:StackApFlxExGalUnc}
\end{table}%

\begin{table}
\caption{StackApFlxExGalCon6: Contains the fluxes within the SDSS R3 (r = 1.03 arcsec), R4 (r = 1.76 arcsec), R5 (r = 3.00 arcsec), R6 (r = 4.63 arcsec), R7 (r = 7.43 arcsec), R8 (r = 11.42 arcsec), R9 (r = 18.20 arcsec), R10 (r = 28.20 arcsec), and R11 (r = 44.21 arcsec) apertures (\citep{Stoughton2002} for extended sources after the images have been convolved to a target of 6 sky pixels (1.5 arcsec).  These measurements are only provided for objects in the extragalactic sky, i.e., they are not provided for objects in the Galactic plane because they are not useful in crowded areas.  See {\em StackObjectThin} table for discussion of primary, secondary, and best detections.}
\begin{center}
\begin{tabular}{lllll}
\hline
\hline
column name & units & data type & default & description\\
\hline
objID & - & BIGINT & NA  &Unique object identifier.\\
uniquePspsSTid & - & BIGINT & NA  &Unique internal PSPS {\em stack} identifier.\\
ippObjID & - & BIGINT & NA  &IPP internal object identifier.\\
randomStackObjID & - & FLOAT & NA  &Random value drawn from the interval between zero and one.\\
primaryDetection & - & TINYINT & 255  &Identifies if this row is the primary {\em stack} detection.\\
bestDetection & - & TINYINT & 255  &Identifies if this row is the best detection.\\
gippDetectID & - & BIGINT & NA  &IPP internal detection identifier.\\
gstackDetectID & - & BIGINT & NA  &Unique {\em stack} detection identifier.\\
gstackImageID & - & BIGINT & NA  &Unique {\em stack} identifier for g filter detection.\\
gc6flxR3 & Jy & REAL & -999  &Flux from g filter detection convolved to a target of 6 sky pixels (1.5 arcsec) \\
& & & & within an aperture of radius r = 1.03 arcsec.\\
gc6flxR3Err & Jy & REAL & -999  &Error in flux from g filter detection convolved to a target of 6 sky pixels \\
& & & & (1.5 arcsec) within an aperture of radius r = 1.03 arcsec.\\
gc6flxR3Std & Jy & REAL & -999  &Standard deviation of flux from g filter detection convolved to a target of 6 \\
& & & & sky pixels (1.5 arcsec) within an aperture of radius r = 1.03 arcsec.\\
gc6flxR3Fill & - & REAL & -999  &Aperture fill factor for g filter detection convolved to a target of 6 sky pixels \\
& & & & (1.5 arcsec) within an aperture of radius r = 1.03 arcsec.\\
... &  & & & gc6flxR3 ... gc6flxR3Fill columns repeated for R4 (r = 1.76 arcsec).\\
... &  & & & repeated for R5 (r = 3.00 arcsec).\\
... &  & & & repeated for R6 (r = 4.63 arcsec).\\
... &  & & & repeated for R7 (r = 7.43 arcsec).\\
... &  & & & repeated for R8 (r = 11.42 arcsec).\\
... &  & & & repeated for R9 (r = 18.20 arcsec).\\
... &  & & & repeated for R10 (r = 28.20 arcsec).\\
gc6flxR11 & Jy & REAL & -999  &Flux from g filter detection convolved to a target of 6 sky pixels (1.5 arcsec) \\
& & & & within an aperture of radius r = 44.21 arcsec.\\
gc6flxR11Err & Jy & REAL & -999  &Error in flux from g filter detection convolved to a target of 6 sky pixels \\
& & & & (1.5 arcsec) within an aperture of radius r = 44.21 arcsec.\\
gc6flxR11Std & Jy & REAL & -999  &Standard deviation of flux from g filter detection convolved to a target of 6 \\
& & & & sky pixels (1.5 arcsec) within an aperture of radius r = 44.21 arcsec.\\
gc6flxR11Fill & - & REAL & -999  &Aperture fill factor for g filter detection convolved to a target of 6 sky pixels \\
& & & & (1.5 arcsec) within an aperture of radius r = 44.21 arcsec.\\
rippDetectID\\
... & & & & same entries repeated for r, i, z, and y filters \\
yc6flxR11Fill \\
\hline
\end{tabular}
\end{center}
\label{table:StackApFlxExGalCon6}
\end{table}%

\begin{table}
\caption{StackApFlxExGalCon8: Contains the fluxes within the SDSS R3 (r = 1.03 arcsec), R4 (r = 1.76 arcsec), R5 (r = 3.00 arcsec), R6 (r = 4.63 arcsec), R7 (r = 7.43 arcsec), R8 (r = 11.42 arcsec), R9 (r = 18.20 arcsec), R10 (r = 28.20 arcsec), and R11 (r = 44.21 arcsec) apertures \citep{Stoughton2002} for extended sources after the images have been convolved to a target of 8 sky pixels (2.0 arcsec).  These measurements are only provided for objects in the extragalactic sky, i.e., they are not provided for objects in the Galactic plane because they are not useful in crowded areas.  See {\em StackObjectThin} table for discussion of primary, secondary, and best detections.}
\begin{center}
\begin{tabular}{lllll}
\hline
\hline
column name & units & data type & default & description\\
\hline
objID & - & BIGINT & NA  &Unique object identifier.\\
uniquePspsSTid & - & BIGINT & NA  &Unique internal PSPS {\em stack} identifier.\\
ippObjID & - & BIGINT & NA  &IPP internal object identifier.\\
randomStackObjID & - & FLOAT & NA  &Random value drawn from the interval between zero and one.\\
primaryDetection & - & TINYINT & 255  &Identifies if this row is the primary {\em stack} detection.\\
bestDetection & - & TINYINT & 255  &Identifies if this row is the best detection.\\
gippDetectID & - & BIGINT & NA  &IPP internal detection identifier.\\
gstackDetectID & - & BIGINT & NA  &Unique {\em stack} detection identifier.\\
gstackImageID & - & BIGINT & NA  &Unique {\em stack} identifier for g filter detection.\\
gc8flxR3 & Jy & REAL & -999  &Flux from g filter detection convolved to a target of 8 sky pixels (2.0 arcsec) \\
& & & & within an aperture of radius r = 1.03 arcsec.\\
gc8flxR3Err & Jy & REAL & -999  &Error in flux from g filter detection convolved to a target of 8 sky pixels \\
& & & & (2.0 arcsec) within an aperture of radius r = 1.03 arcsec.\\
gc8flxR3Std & Jy & REAL & -999  &Standard deviation of flux from g filter detection convolved to a target of 8 \\
& & & & sky pixels (2.0 arcsec) within an aperture of radius r = 1.03 arcsec.\\
gc8flxR3Fill & - & REAL & -999  &Aperture fill factor for g filter detection convolved to a target of 8 sky pixels \\
& & & & (2.0 arcsec) within an aperture of radius r = 1.03 arcsec.\\
... &  & & & gc8flxR3 ... gc8flxR3Fill columns repeated for R4 (r = 1.76 arcsec).\\
... &  & & & repeated for R5 (r = 3.00 arcsec).\\
... &  & & & repeated for R6 (r = 4.63 arcsec).\\
... &  & & & repeated for R7 (r = 7.43 arcsec).\\
... &  & & & repeated for R8 (r = 11.42 arcsec).\\
... &  & & & repeated for R9 (r = 18.20 arcsec).\\
... &  & & & repeated for R10 (r = 28.20 arcsec).\\
gc8flxR11 & Jy & REAL & -999  &Flux from g filter detection convolved to a target of 8 sky pixels (2.0 arcsec) \\
& & & & within an aperture of radius r = 44.21 arcsec.\\
gc8flxR11Err & Jy & REAL & -999  &Error in flux from g filter detection convolved to a target of 8 sky pixels \\
& & & & (2.0 arcsec) within an aperture of radius r = 44.21 arcsec.\\
gc8flxR11Std & Jy & REAL & -999  &Standard deviation of flux from g filter detection convolved to a target of 8 \\
& & & & sky pixels (2.0 arcsec) within an aperture of radius r = 44.21 arcsec.\\
gc8flxR11Fill & - & REAL & -999  &Aperture fill factor for g filter detection convolved to a target of 8 sky pixels \\
& & & & (2.0 arcsec) within an aperture of radius r = 44.21 arcsec.\\
rippDetectID \\
... & & & & same entries repeated for r, i, z, and y filters \\
yc8flxR11Fill \\
\hline
\end{tabular}
\end{center}
\label{table:StackApFlxExGalCon8}
\end{table}%

\clearpage


\begin{table}
\caption{StackPetrosian: Contains the \citet{Petrosian1976} magnitudes and radii for extended sources.  See {\em StackObjectThin} table for discussion of primary, secondary, and best detections.}
\begin{center}
\begin{tabular}{lllll}
\hline
\hline
column name & units & data type & default & description\\
\hline
objID & - & BIGINT & NA  &Unique object identifier.\\
uniquePspsSTid & - & BIGINT & NA  &Unique internal PSPS {\em stack} identifier.\\
ippObjID & - & BIGINT & NA  &IPP internal object identifier.\\
randomStackObjID & - & FLOAT & NA  &Random value drawn from the interval between zero and one.\\
primaryDetection & - & TINYINT & 255  &Identifies if this row is the primary {\em stack} detection.\\
bestDetection & - & TINYINT & 255  &Identifies if this row is the best detection.\\
gippDetectID & - & BIGINT & NA  &IPP internal detection identifier.\\
gstackDetectID & - & BIGINT & NA  &Unique {\em stack} detection identifier.\\
gstackImageID & - & BIGINT & NA  &Unique {\em stack} identifier for g filter detection.\\
gpetRadius & arcsec & REAL & -999  &Petrosian (1976) fit radius for g filter {\em stack} detection.\\
gpetRadiusErr & arcsec & REAL & -999  &Error in Petrosian (1976) fit radius for g filter {\em stack} detection.\\
gpetMag & AB & REAL & -999  &Petrosian (1976) magnitude from g filter {\em stack} detection.\\
gpetMagErr & AB & REAL & -999  &Error in Petrosian (1976) magnitude from g filter {\em stack} detection.\\
gpetR50 & arcsec & REAL & -999  &Petrosian (1976) fit radius for g filter {\em stack} detection. at 50\% light\\
gpetR50Err & arcsec & REAL & -999  &Error in Petrosian (1976) fit radius for g filter {\em stack} detection. at 50\% light\\
gpetR90 & arcsec & REAL & -999  &Petrosian (1976) fit radius for g filter {\em stack} detection. at 90\% light\\
gpetR90Err & arcsec & REAL & -999  &Error in Petrosian (1976) fit radius for g filter {\em stack} detection. at 90\% light\\
gpetCf & - & REAL & -999  &Petrosian (1976) fit coverage factor for g filter {\em stack} detection.\\
rippDetectID \\
... & & & & same entries repeated for r, i, z, and y filters \\
ypetCf \\
\hline
\end{tabular}
\end{center}
\label{table:StackPetrosian}
\end{table}%

\begin{table}
\caption{StackToImage: Contains the mapping of which input images were used to construct a particular stack.}
\begin{center}
\begin{tabular}{lllll}
\hline
\hline
column name & units & data type & default & description\\
\hline
stackImageID & - & BIGINT & NA  &Unique {\em stack} identifier.\\
imageID & - & BIGINT & NA  &Unique image identifier.  Constructed as (100 * frameID + ccdID).\\
\hline
\end{tabular}
\end{center}
\label{table:StackToImage}
\end{table}%


\begin{table}
\caption{StackToFrame: Contains the mapping of input frames used to construct a particular {\em stack} along with processing stats.}
\begin{center}
\begin{tabular}{lllll}
\hline
\hline
column name & units & data type & default & description\\
\hline
stackImageID & - & BIGINT & NA  &Unique {\em stack} identifier.\\
frameID & - & INT & NA  &Unique frame/exposure identifier.\\
scaleFactor & - & REAL & 0  &normalization factor applied to input image before stacking.\\
zp & magnitudes & REAL & 0  &Photometric zeropoint.  Necessary for converting listed fluxes and \\ 
& & & & magnitudes back to measured ADU counts.\\
expTime & seconds & REAL & -999  &Exposure time of the frame/exposure.  Necessary for converting \\
& & & & listed fluxes and magnitudes back to measured ADU counts.\\
airMass & - & REAL & 0  &Airmass at midpoint of the exposure.  Necessary for converting \\ 
& & & & listed fluxes and magnitudes back to measured ADU counts.\\
\hline
\end{tabular}
\end{center}
\label{table:StackToFrame}
\end{table}%


\begin{table}
\caption{StackDetEffMeta: Contains the detection efficiency information for a given stacked image.  Provides the number of recovered sources out of 500 injected sources for each magnitude bin and statistics about the magnitudes of the recovered sources for a range of magnitude offsets.}
\begin{center}
\begin{tabular}{lllll}
\hline
\hline
column name  & units    & data type & default & description\\
\hline
stackImageID & -         & BIGINT& NA  &Unique  stack identifier.\\
magref      & magnitudes & REAL & NA  &Detection efficiency reference magnitude.\\
nInjected   & -          & INT  & NA  &Number of fake sources injected in each magnitude bin.\\
offset01    & magnitudes & REAL & NA  &Detection efficiency magnitude offset for bin 1.\\
counts01    & -          & REAL & NA  &Detection efficiency count of recovered sources in bin 1.\\
diffMean01  & magnitudes & REAL & NA  &Detection efficiency mean magnitude difference in bin 1.\\
diffStdev01 & magnitudes & REAL & NA  &Detection efficiency standard deviation of magnitude differences in bin 1.\\
errMean01   & magnitudes & REAL & NA  &Detection efficiency mean magnitude error in bin 1.\\
offset02    & magnitudes & REAL & NA  &Detection efficiency magnitude offset for bin 2.\\
counts02    & -          & REAL & NA  &Detection efficiency count of recovered sources in bin 2.\\
diffMean02  & magnitudes & REAL & NA  &Detection efficiency mean magnitude difference in bin 2.\\
diffStdev02 & magnitudes & REAL & NA  &Detection efficiency standard deviation of magnitude differences in bin 2.\\
errMean02   & magnitudes & REAL & NA  &Detection efficiency mean magnitude error in bin 2.\\
offset03    & magnitudes & REAL & NA  &Detection efficiency magnitude offset for bin 3.\\
counts03    & -          & REAL & NA  &Detection efficiency count of recovered sources in bin 3.\\
diffMean03  & magnitudes & REAL & NA  &Detection efficiency mean magnitude difference in bin 3.\\
diffStdev03 & magnitudes & REAL & NA  &Detection efficiency standard deviation of magnitude differences in bin 3.\\
errMean03   & magnitudes & REAL & NA  &Detection efficiency mean magnitude error in bin 3.\\
...\\
offset13    & magnitudes & REAL & NA  &Detection efficiency magnitude offset for bin 13.\\
counts13    & -          & REAL & NA  &Detection efficiency count of recovered sources in bin 13.\\
diffMean13  & magnitudes & REAL & NA  &Detection efficiency mean magnitude difference in bin 13.\\
diffStdev13 & magnitudes & REAL & NA  &Detection efficiency standard deviation of magnitude differences in bin 13.\\
errMean13   & magnitudes & REAL & NA  &Detection efficiency mean magnitude error in bin 13.\\
\hline
\end{tabular}
\end{center}
\label{table:StackDetEffMeta}
\end{table}%

\clearpage

\subsection{Single Exposure Detection Tables}


\begin{table}
\caption{FrameMeta: Contains metadata related to an individual exposure.  A "Frame" refers to the collection of all images obtained by the 60 OTA devices in the camera in a single exposure. The camera configuration, telescope pointing, observation time, and astrometric solution from the detector focal plane (L,M) to the sky (RA,Dec) is provided.}
\begin{center}
\resizebox{\textwidth}{!}{%
\begin{tabular}{lllll}
\hline
\hline
column name & units & data type & default & description\\
\hline
frameID & - & INT & NA  &Unique frame/exposure identifier.\\
frameName & - & VARCHAR(32) & NA  &Frame/exposure name provided by the camera software.\\
surveyID & - & TINYINT & NA  &Survey identifier.  Details in the Survey table.\\
filterID & - & TINYINT & NA  &Filter identifier.  Details in the Filter table.\\
ippChipID & - & INT & NA  &IPP chipRun identifier.\\
ippCamID & - & INT & NA  &IPP camRun identifier.\\
ippWarpID & - & INT & NA  &IPP warpRun identifier.\\
cameraID & - & SMALLINT & NA  &Camera identifier.  Details in the CameraConfig table.\\
cameraConfigID & - & SMALLINT & NA  &Camera configuration identifier.  Details in the CameraConfig table.\\
telescopeID & - & SMALLINT & NA  &Telescope identifier.\\
analysisVer & - & VARCHAR(100) &   &IPP software analysis release version.\\
md5sum & - & VARCHAR(100) &   &IPP MD5 Checksum.\\
nOTA & - & SMALLINT & -999  &Number of valid OTA images in this frame/exposure.\\
photoScat & magnitudes & REAL & -999  &Photometric scatter relative to reference catalog across the full FOV.\\
nPhotoRef & - & INT & -999  &Number of photometric reference sources.\\
expStart & days & FLOAT & -999  &Modified Julian Date at the start of the exposure.\\
expTime & seconds & REAL & -999  &Exposure time of the frame/exposure.  Necessary for converting listed\\
& & & & fluxes and magnitudes back to measured ADU counts.\\
airmass & - & REAL & 0  &Airmass at midpoint of the exposure.  Necessary for converting listed \\
& & & & fluxes and magnitudes back to measured ADU counts.\\
raBore & degrees & FLOAT & -999  &Right ascension of telescope boresight.\\
decBore & degrees & FLOAT & -999  &Declination of telescope boresight.\\
ctype1 & - & VARCHAR(100) &   &Name of astrometric projection in RA.\\
ctype2 & - & VARCHAR(100) &   &Name of astrometric projection in Dec.\\
crval1 & degrees & FLOAT & -999  &Right ascension corresponding to reference pixel.\\
crval2 & degrees & FLOAT & -999  &Declination corresponding to reference pixel.\\
crpix1 & pixels & FLOAT & -999  &Reference pixel for RA.\\
crpix2 & pixels & FLOAT & -999  &Reference pixel for Dec.\\
cdelt1 & degrees/pixel & FLOAT & -999  &Pixel scale in RA.\\
cdelt2 & degrees/pixel & FLOAT & -999  &Pixel scale in Dec.\\
pc001001 & - & FLOAT & -999  &Linear transformation matrix element between focal plane pixel L and RA.\\
pc001002 & - & FLOAT & -999  &Linear transformation matrix element between focal plane pixel M and RA.\\
pc002001 & - & FLOAT & -999  &Linear transformation matrix element between focal plane pixel L and Dec.\\
pc002002 & - & FLOAT & -999  &Linear transformation matrix element between focal plane pixel M and Dec.\\
polyOrder & - & TINYINT & 255  &Polynomial order of astrometric fit between detector focal plane and sky.\\
pca1x3y0 & - & FLOAT & -999  &Polynomial coefficient for the astrometric fit component ($x^3$ $y^0$) for RA.\\
pca1x2y1 & - & FLOAT & -999  &Polynomial coefficient for the astrometric fit component ($x^2$ $y^1$) for RA.\\
pca1x1y2 & - & FLOAT & -999  &Polynomial coefficient for the astrometric fit component ($x^1$ $y^2$) for RA.\\
pca1x0y3 & - & FLOAT & -999  &Polynomial coefficient for the astrometric fit component ($x^0$ $y^3$) for RA.\\
pca1x2y0 & - & FLOAT & -999  &Polynomial coefficient for the astrometric fit component ($x^2$ $y^0$) for RA.\\
pca1x1y1 & - & FLOAT & -999  &Polynomial coefficient for the astrometric fit component ($x^1$ $y^1$) for RA.\\
pca1x0y2 & - & FLOAT & -999  &Polynomial coefficient for the astrometric fit component ($x^0$ $y^2$) for RA.\\
pca2x3y0 & - & FLOAT & -999  &Polynomial coefficient for the astrometric fit component ($x^3$ $y^0$) for Dec.\\
pca2x2y1 & - & FLOAT & -999  &Polynomial coefficient for the astrometric fit component ($x^2$ $y^1$) for Dec.\\
pca2x1y2 & - & FLOAT & -999  &Polynomial coefficient for the astrometric fit component ($x^1$ $y^2$) for Dec.\\
pca2x0y3 & - & FLOAT & -999  &Polynomial coefficient for the astrometric fit component ($x^0$ $y^3$) for Dec.\\
pca2x2y0 & - & FLOAT & -999  &Polynomial coefficient for the astrometric fit component ($x^2$ $y^0$) for Dec.\\
pca2x1y1 & - & FLOAT & -999  &Polynomial coefficient for the astrometric fit component ($x^1$ $y^1$) for Dec.\\
pca2x0y2 & - & FLOAT & -999  &Polynomial coefficient for the astrometric fit component ($x^0$ $y^2$) for Dec.\\
batchID & - & BIGINT & NA  &Internal database batch identifier.\\
processingVersion & - & TINYINT & NA  &Data release version.\\
\hline
\end{tabular}}
\end{center}
\label{table:FrameMeta}
\end{table}%

\begin{table}
\caption{ImageMeta: Contains metadata related to an individual OTA image that comprises a portion of the full exposure.  The characterization of the image quality, the detrends applied, and the astrometric solution from the raw pixels (X,Y) to the detector focal plane (L,M) is provided.}
\begin{center} 
\resizebox{\textwidth}{!}{%
\begin{tabular}{lllll}
\hline
\hline
column name & units & data type & default & description\\
\hline
imageID & - & BIGINT & NA  &Unique image identifier.  Constructed as (100 * frameID + ccdID).\\
frameID & - & INT & NA  &Unique frame/exposure identifier.\\
ccdID & - & SMALLINT & NA  &OTA identifier based on location in the focal plane, specific to an individual device.\\
photoCalID & - & INT & NA  &Photometric calibration identifier.  Details in the PhotoCal table.\\
filterID & - & TINYINT & NA  &Filter identifier.  Details in the Filter table.\\
bias & adu & REAL & -999  &OTA bias level.\\
biasScat & adu & REAL & -999  &Scatter in bias level.\\
sky & $Jy/arcsec^2$ & REAL & -999  &Mean sky brightness.\\
skyScat & $Jy/arcsec^2$ & REAL & -999  &Scatter in mean sky brightness.\\
nDetect & - & INT & -999  &Number of detections in this image.\\
detectionThreshold & magnitudes & REAL & -999  &Reference magnitude for detection efficiency calculation.\\
astroScat & arcsec & REAL & -999  &Measurement of the calibration (not astrometric error) defined to be the sum in quadrature of the \\
& & & & standard deviations in the X and Y directions.\\
photoScat & magnitudes & REAL & -999  &Photometric scatter relative to reference catalog.\\
nAstroRef & - & INT & -999  &Number of astrometric reference sources.\\
nPhotoRef & - & INT & -999  &Number of photometric reference sources.\\
recalAstroScatX & arcsec & REAL & -999  &Measurement of the re-calibration (not astrometric error) in the X direction.\\
recalAstroScatY & arcsec & REAL & -999  &Measurement of the re-calibration (not astrometric error) in the Y direction.\\
recalNAstroStars & - & INT & -999  &Number of astrometric reference sources used in recalibration.\\
recalphotoScat & magnitudes & REAL & -999  &Photometric scatter relative to reference catalog.\\
recalNPhotoStars & - & INT & -999  &Number of astrometric reference sources used in recalibration.\\
nAxis1 & pixels & SMALLINT & -999  &Image dimension in x.\\
nAxis2 & pixels & SMALLINT & -999  &Image dimension in y.\\
psfModelID & - & INT & -999  &PSF model identifier.\\
psfFWHM & arcsec & REAL & -999  &Mean PSF full width at half maximum at image center.\\
psfWidMajor & arcsec & REAL & -999  &PSF major axis FWHM at image center.\\
psfWidMinor & arcsec & REAL & -999  &PSF minor axis FWHM at image center.\\
psfTheta & degrees & REAL & -999  &PSF major axis orientation at image center.\\
momentMajor & arcsec & REAL & -999  &PSF major axis second moment.\\
momentMinor & arcsec & REAL & -999  &PSF minor axis second moment.\\
momentM2C & $arcsec^2$ & REAL & -999  &Moment $M2C = M_{xx} - M_{yy}$.\\
momentM2S & $arcsec^2$ & REAL & -999  &Moment $M2S = 2 * M_{xy}$.\\
momentM3 & $arcsec^2$ & REAL & -999  &trefoil second moment = $sqrt( (M_{xxx} - 3 * M_{xyy})^2 + (3 * M_{xxy} - M_{yyy})^2 )$.\\
momentM4 & $arcsec^2$ & REAL & -999  &quadrupole second moment = $sqrt( (M_{xxxx} - 6 * M_{xxyy} + M_{yyyy})^2 + (4 * M_{xxxy} - 4 * M_{xyyy})^2 )$.\\
apResid & magnitudes & REAL & -999  &Residual of aperture corrections.\\
dapResid & magnitudes & REAL & -999  &Scatter of aperture corrections.\\
detectorID & - & VARCHAR(100) &   &Identifier for each individual OTA detector device.\\
qaFlags & - & BIGINT & -999  &Q/A flags for this image.  Values listed in ImageFlags.\\
detrend1 & - & VARCHAR(100) &   &Identifier for detrend image 1, the static mask.\\
detrend2 & - & VARCHAR(100) &   &Identifier for detrend image 2, the dark model.\\
detrend3 & - & VARCHAR(100) &   &Identifier for detrend image 3, the flat.\\
detrend4 & - & VARCHAR(100) &   &Identifier for detrend image 4, the fringe.\\
detrend5 & - & VARCHAR(100) &   &Identifier for detrend image 5, the noisemap.\\
detrend6 & - & VARCHAR(100) &   &Identifier for detrend image 6, the non-linearity correction.\\
detrend7 & - & VARCHAR(100) &   &Identifier for detrend image 7, the video dark model.\\
detrend8 & - & VARCHAR(100) &   &Identifier for detrend image 8.\\
photoZero & magnitudes & REAL & -999  &Locally derived photometric zero point for this image.\\
ctype1 & - & VARCHAR(100) &   &Name of astrometric projection in focal plane L.\\
ctype2 & - & VARCHAR(100) &   &Name of astrometric projection in focal plane M.\\
crval1 & focal plane pixels & FLOAT & -999  &Focal plane L corresponding to reference pixel.\\
crval2 & focal plane pixels & FLOAT & -999  &Focal plane M corresponding to reference pixel.\\
crpix1 & raw pixels & FLOAT & -999  &Reference pixel for focal plane L.\\
crpix2 & raw pixels & FLOAT & -999  &Reference pixel for focal plane M.;\\
cdelt1 & focal plane pixels/raw pixel & FLOAT & -999  &Pixel scale in focal plane x.\\
cdelt2 & focal plane pixels/raw pixel & FLOAT & -999  &Pixel scale in focal plane y.\\
pc001001 & - & FLOAT & -999  &Linear transformation matrix element between image pixel x and focal plane pixel L.\\
pc001002 & - & FLOAT & -999  &Linear transformation matrix element between image pixel y and focal plane pixel L.\\
pc002001 & - & FLOAT & -999  &Linear transformation matrix element between image pixel x and focal plane pixel M.\\
pc002002 & - & FLOAT & -999  &Linear transformation matrix element between image pixel y and focal plane pixel M.\\
polyOrder & - & TINYINT & 255  &Polynomial order of astrometric fit between the image pixels and the detector focal plane.\\
pca1x3y0 & - & FLOAT & -999  &Polynomial coefficient for the astrometric fit component ($x^3$ $y^0$) for focal plane L.\\
pca1x2y1 & - & FLOAT & -999  &Polynomial coefficient for the astrometric fit component ($x^2$ $y^1$) for focal plane L.\\
pca1x1y2 & - & FLOAT & -999  &Polynomial coefficient for the astrometric fit component ($x^1$ $y^2$) for focal plane L.\\
pca1x0y3 & - & FLOAT & -999  &Polynomial coefficient for the astrometric fit component ($x^0$ $y^3$) for focal plane L.\\
pca1x2y0 & - & FLOAT & -999  &Polynomial coefficient for the astrometric fit component ($x^2$ $y^0$) for focal plane L.\\
pca1x1y1 & - & FLOAT & -999  &Polynomial coefficient for the astrometric fit component ($x^1$ $y^1$) for focal plane L.\\
pca1x0y2 & - & FLOAT & -999  &Polynomial coefficient for the astrometric fit component ($x^0$ $y^2$) for focal plane L.\\
pca2x3y0 & - & FLOAT & -999  &Polynomial coefficient for the astrometric fit component ($x^3$ $y^0$) for focal plane M.\\
pca2x2y1 & - & FLOAT & -999  &Polynomial coefficient for the astrometric fit component ($x^2$ $y^1$) for focal plane M.\\
pca2x1y2 & - & FLOAT & -999  &Polynomial coefficient for the astrometric fit component ($x^1$ $y^2$) for focal plane M.\\
pca2x0y3 & - & FLOAT & -999  &Polynomial coefficient for the astrometric fit component ($x^0$ $y^3$) for focal plane M.\\
pca2x2y0 & - & FLOAT & -999  &Polynomial coefficient for the astrometric fit component ($x^2$ $y^0$) for focal plane M.\\
pca2x1y1 & - & FLOAT & -999  &Polynomial coefficient for the astrometric fit component ($x^1$ $y^1$) for focal plane M.\\
pca2x0y2 & - & FLOAT & -999  &Polynomial coefficient for the astrometric fit component ($x^0$ $y^2$) for focal plane M.\\
processingVersion & - & TINYINT & NA  &Data release version.\\
\hline
\end{tabular}}
\end{center}
\label{table:ImageMeta}
\end{table}%

\begin{table}
\caption{Detection: Contains single epoch photometry of individual detections from a single exposure.  The identifiers connecting the detection back to the original image and to the object association are provided.  PSF, aperture, and \citet{Kron1980} photometry are included, along with sky and detector coordinate positions.}
\begin{center}
\resizebox{\textwidth}{!}{%
\begin{tabular}{lllll}
\hline
\hline
column name & units & data type & default & description\\
\hline
objID & - & BIGINT & NA  &Unique object identifier.\\
uniquePspsP2id & - & BIGINT & NA  &Unique internal PSPS detection identifier.\\
detectID & - & BIGINT & NA  &Unique detection identifier.\\
ippObjID & - & BIGINT & NA  &IPP internal object identifier.\\
ippDetectID & - & BIGINT & NA  &IPP internal detection identifier.\\
filterID & - & TINYINT & NA  &Filter identifier.  Details in the Filter table.\\
surveyID & - & TINYINT & NA  &Survey identifier.  Details in the Survey table.\\
imageID & - & BIGINT & NA  &Unique image identifier.  Constructed as (100 * frameID + ccdID).\\
randomDetID & - & FLOAT & NA  &Random value drawn from the interval between zero and one. \\
dvoRegionID & - & INT & -1  &Internal DVO region identifier.\\
obsTime & days & FLOAT & -999  &Modified Julian Date at the midpoint of the observation.\\
xPos & raw pixels & REAL & -999  &PSF x center location.\\
yPos & raw pixels & REAL & -999  &PSF y center location.\\
xPosErr & raw pixels & REAL & -999  &Error in PSF x center location.\\
yPosErr & raw pixels & REAL & -999  &Error in PSF y center location.\\
pltScale & arcsec/pixel & REAL & -999  &Local plate scale at this location.\\
posAngle & degrees & REAL & -999  &Position angle (sky-to-chip) at this location.\\
ra & degrees & FLOAT & -999  &Right ascension.\\
dec & degrees & FLOAT & -999  &Declination.\\
raErr & arcsec & REAL & -999  &Right ascension error.\\
decErr & arcsec & REAL & -999  &Declination error.\\
extNSigma & - & REAL & 0  &An extendedness measure based on the deviation between PSF and Kron\\
& & & & magnitudes, normalized by the PSF magnitude uncertainty.\\
zp & magnitudes & REAL & 0  &Photometric zeropoint.  Necessary for converting listed fluxes and \\
& & & & magnitudes back to measured ADU counts.\\
telluricExt & magnitudes & REAL & NA  &Estimated Telluric extinction due to non-photometric observing conditions.\\
& & & & Necessary for converting listed fluxes and magnitudes back to measured ADU counts.\\
expTime & seconds & REAL & -999  &Exposure time of the frame/exposure.  Necessary for converting listed \\
& & & & fluxes and magnitudes back to measured ADU counts.\\
airMass & - & REAL & 0  &Airmass at midpoint of the exposure.  Necessary for converting listed \\
& & & & fluxes and magnitudes back to measured ADU counts.\\
psfFlux & Jy & REAL & -999  &Flux from PSF fit.\\
psfFluxErr & Jy & REAL & -999  &Error on flux from PSF fit.\\
psfMajorFWHM & arcsec & REAL & -999  &PSF major axis FWHM.\\
psfMinorFWHM & arcsec & REAL & -999  &PSF minor axis FWHM.\\
psfTheta & degrees & REAL & -999  &PSF major axis orientation.\\
psfCore & - & REAL & -999  &PSF core parameter k, where $F = F0 / (1 + k r^2 + r^{3.33})$.\\
psfQf & - & REAL & -999  &PSF coverage factor.\\
psfQfPerfect & - & REAL & -999  &PSF weighted fraction of pixels totally unmasked.\\
psfChiSq & - & REAL & -999  &Reduced chi squared value of the PSF model fit.\\
psfLikelihood & - & REAL & -999  &Likelihood that this detection is best fit by a PSF.\\
momentXX & $arcsec^2$ & REAL & -999  &Second moment $M_{xx}$.\\
momentXY & $arcsec^2$ & REAL & -999  &Second moment $M_{xy}$.\\
momentYY & $arcsec^2$ & REAL & -999  &Second moment $M_{yy}$.\\
momentR1 & arcsec & REAL & -999  &First radial moment.\\
momentRH & $arcsec^{0.5}$ & REAL & -999  &Half radial moment ($r^{0.5}$ weighting).\\
momentM3C & $arcsec^2$ & REAL & -999  &Cosine of trefoil second moment term: $r^2 cos(3 theta) = M_{xxx} - 3 * M_{xyy}$.\\
momentM3S & $arcsec^2$ & REAL & -999  &Sine of trefoil second moment: $r^2 sin (3 theta) = 3 * M_{xxy} - M_{yyy}$.\\
momentM4C & $arcsec^2$ & REAL & -999  &Cosine of quadrupole second moment: $r^2 cos (4 theta) = M_{xxxx} - 6 * M_{xxyy} + M_{yyyy}.$\\
momentM4S & $arcsec^2$ & REAL & -999  &Sine of quadrupole second moment: $r^2 sin (4 theta) = 4 * M_{xxxy} - 4 * M_{xyyy}$.\\
apFlux & Jy & REAL & -999  &Flux in seeing-dependent aperture.\\
apFluxErr & Jy & REAL & -999  &Error on flux in seeing-dependent aperture.\\
apFillF & - & REAL & -999  &Aperture fill factor.\\
apRadius & arcsec & REAL & -999  &Aperture radius.\\
kronFlux & Jy & REAL & -999  &Kron (1980) flux.\\
kronFluxErr & Jy & REAL & -999  &Error on Kron (1980) flux.\\
kronRad & arcsec & REAL & -999  &Kron (1980) radius.\\
sky & $Jy/arcsec^2$ & REAL & -999  &Background sky level.\\
skyErr & $Jy/arcsec^2$ & REAL & -999  &Error in background sky level.\\
infoFlag & - & BIGINT & 0  &Information flag bitmask indicating details of the photometry.  \\
& & & & Values listed in DetectionFlags.\\
infoFlag2 & - & INT & 0  &Information flag bitmask indicating details of the photometry.  \\
& & & & Values listed in DetectionFlags2.\\
infoFlag3 & - & INT & 0  &Information flag bitmask indicating details of the photometry.  \\
& & & & Values listed in DetectionFlags3.\\
processingVersion & - & TINYINT & NA  &Data release version.\\
\hline
\end{tabular}}
\end{center}
\label{table:Detection}
\end{table}%

\clearpage
\begin{table}
\caption{ImageDetEffMeta: Contains the detection efficiency information for a given individual OTA image.  Provides the number of recovered sources out of 500 injected fake source and statistics about the magnitudes of the recovered sources for a range of magnitude offsets.}
\begin{center}
\resizebox{\textwidth}{!}{
\begin{tabular}{lllll}
\hline
\hline
column name & units & data type & default & description\\
\hline
imageID & - & BIGINT & NA  &Unique image identifier.  Constructed as (100 * frameID + ccdID).\\
frameID & - & INT & NA  &Unique frame/exposure identifier.\\
magref & magnitudes & REAL & NA  &Detection efficiency reference magnitude.\\
nInjected & - & INT & NA  &Number of fake sources injected in each magnitude bin.\\
offset01 & magnitudes & REAL & NA  &Detection efficiency magnitude offset for bin 1.\\
counts01 & - & REAL & NA  &Detection efficiency count of recovered sources in bin 1.\\
diffMean01 & magnitudes & REAL & NA  &Detection efficiency mean magnitude difference in bin 1.\\
diffStdev01 & magnitudes & REAL & NA  &Detection efficiency standard deviation of magnitude differences in bin 1.\\
errMean01 & magnitudes & REAL & NA  &Detection efficiency mean magnitude error in bin 1.\\
offset02 & magnitudes & REAL & NA  &Detection efficiency magnitude offset for bin 2.\\
counts02 & - & REAL & NA  &Detection efficiency count of recovered sources in bin 2.\\
diffMean02 & magnitudes & REAL & NA  &Detection efficiency mean magnitude difference in bin 2.\\
diffStdev02 & magnitudes & REAL & NA  &Detection efficiency standard deviation of magnitude differences in bin 2.\\
errMean02 & magnitudes & REAL & NA  &Detection efficiency mean magnitude error in bin 2.\\
offset03 & magnitudes & REAL & NA  &Detection efficiency magnitude offset for bin 3.\\
counts03 & - & REAL & NA  &Detection efficiency count of recovered sources in bin 3.\\
diffMean03 & magnitudes & REAL & NA  &Detection efficiency mean magnitude difference in bin 3.\\
diffStdev03 & magnitudes & REAL & NA  &Detection efficiency standard deviation of magnitude differences in bin 3.\\
errMean03 & magnitudes & REAL & NA  &Detection efficiency mean magnitude error in bin 3.\\
...\\
offset13 & magnitudes & REAL & NA  &Detection efficiency magnitude offset for bin 13.\\
counts13 & - & REAL & NA  &Detection efficiency count of recovered sources in bin 13.\\
diffMean13 & magnitudes & REAL & NA  &Detection efficiency mean magnitude difference in bin 13.\\
diffStdev13 & magnitudes & REAL & NA  &Detection efficiency standard deviation of magnitude differences in bin 13.\\
errMean13 & magnitudes & REAL & NA  &Detection efficiency mean magnitude error in bin 13.\\
\hline
\end{tabular}}
\end{center}
\label{table:ImageDetEffMeta}
\end{table}%

\clearpage 
\subsection{Forced Mean Object Tables}

\begin{table}
\caption{ForcedMeanObject: Contains the mean of single-epoch photometric information for sources detected in the stacked data, calculated as described in \citet{Magnier2013}.  The mean is calculated for detections associated into objects within a one arcsecond correlation radius.  PSF, \citet{Kron1980}, and SDSS aperture R5 (r = 3.00\arcsec), R6 (r = 4.63\arcsec), and R7 (r = 7.43\arcsec) apertures \citep{Stoughton2002} magnitudes and statistics are listed for all filters. See also \citet{Kaiser1995}.}
\begin{center}
\resizebox{\textwidth}{!}{
\begin{tabular}{lllll}
\hline
\hline
column name & units & data type & default & description\\
\hline
objID & - & BIGINT & NA  &Unique object identifier.\\
uniquePspsFOid & - & BIGINT & NA  &Unique internal PSPS forced object identifier.\\
ippObjID & - & BIGINT & NA  &IPP internal object identifier.\\
randomForcedObjID & - & FLOAT & NA  &Random value drawn from the interval between zero and one.\\
nDetections & - & SMALLINT & -999  &Number of single epoch detections in all filters.\\
batchID & - & BIGINT & NA  &Internal database batch identifier.\\
processingVersion & - & TINYINT & NA  &Data release version.\\
gnTotal & - & SMALLINT & -999  &Number of forced single epoch detections in g filter.\\
gnIncPSFFlux & - & SMALLINT & -999  &Number of forced single epoch detections in PSF flux mean in g filter.\\
gnIncKronFlux & - & SMALLINT & -999  &Number of forced single epoch detections in Kron (1980) flux mean in g filter.\\
gnIncApFlux & - & SMALLINT & -999  &Number of forced single epoch detections in aperture flux mean in g filter.\\
gnIncR5 & - & SMALLINT & -999  &Number of forced single epoch detections in R5 (r = 3.00 arcsec) aperture flux mean in g filter.\\
gnIncR6 & - & SMALLINT & -999  &Number of forced single epoch detections in R6 (r = 4.63 arcsec) aperture flux mean in g filter.\\
gnIncR7 & - & SMALLINT & -999  &Number of forced single epoch detections in R7 (r = 7.43 arcsec) aperture flux mean in g filter.\\
gFPSFFlux & Jy & REAL & -999  &Mean PSF flux from forced single epoch g filter detections.\\
gFPSFFluxErr & Jy & REAL & -999  &Error in mean PSF flux from forced single epoch g filter detections.\\
gFPSFFluxStd & Jy & REAL & -999  &Standard deviation of PSF fluxes from forced single epoch g filter detections.\\
gFPSFMag & AB & REAL & -999  &Magnitude from mean PSF flux from forced single epoch g filter detections.\\
gFPSFMagErr & AB & REAL & -999  &Error in magnitude from mean PSF flux from forced single epoch g filter detections.\\
gFKronFlux & Jy & REAL & -999  &Mean Kron (1980) flux from forced single epoch g filter detections.\\
gFKronFluxErr & Jy & REAL & -999  &Error in mean Kron (1980) flux from forced single epoch g filter detections.\\
gFKronFluxStd & Jy & REAL & -999  &Standard deviation of Kron (198) fluxes from forced single epoch g filter detections.\\
gFKronMag & AB & REAL & -999  &Magnitude from mean Kron (1980) flux from forced single epoch g filter detections.\\
gFKronMagErr & AB & REAL & -999  &Error in magnitude from mean Kron (1980) flux from forced single epoch g filter detections.\\
gFApFlux & Jy & REAL & -999  &Mean aperture flux from forced single epoch g filter detections.\\
gFApFluxErr & Jy & REAL & -999  &Error in mean aperture flux from forced single epoch g filter detections.\\
gFApFluxStd & Jy & REAL & -999  &Standard deviation of aperture fluxes from forced single epoch g filter detections.\\
gFApMag & AB & REAL & -999  &Magnitude from mean aperture flux from forced single epoch g filter detections.\\
gFApMagErr & AB & REAL & -999  &Error in magnitude from mean aperture flux from forced single epoch g filter detections.\\
gFmeanflxR5 & Jy & REAL & -999  &Mean flux from forced single epoch g filter \\
& & & & detections within an aperture of radius r = 3.00 arcsec.\\
gFmeanflxR5Err & Jy & REAL & -999  &Error in mean flux from forced single epoch g filter \\
& & & & detections within an aperture of radius r = 3.00 arcsec.\\
gFmeanflxR5Std & Jy & REAL & -999  &Standard deviation of forced single epoch g filter \\
& & & & detection fluxes within an aperture of radius r = 3.00 arcsec.\\
gFmeanflxR5Fill & - & REAL & -999  &Aperture fill factor for forced single epoch g filter \\
& & & & detections within an aperture of radius r = 3.00 arcsec.\\
gFmeanMagR5 & AB & REAL & -999  &Magnitude from mean flux from forced single epoch g filter \\
& & & & detections within an aperture of radius r = 3.00 arcsec.\\
gFmeanMagR5Err & AB & REAL & -999  &Error in magnitude from  mean flux from forced single epoch \\
& & & & g filter detections within an aperture of radius r = 3.00 arcsec.\\
gFmeanflxR6 & Jy & REAL & -999  &Mean flux from forced single epoch g filter \\
& & & & detections within an aperture of radius r = 4.63 arcsec.\\
gFmeanflxR6Err & Jy & REAL & -999  &Error in mean flux from forced single epoch g filter \\
& & & & detections within an aperture of radius r = 4.63 arcsec.\\
gFmeanflxR6Std & Jy & REAL & -999  &Standard deviation of forced single epoch g filter \\
& & & & detection fluxes within an aperture of radius r = 4.63 arcsec.\\
gFmeanflxR6Fill & - & REAL & -999  &Aperture fill factor for forced single epoch g filter \\
& & & & detections within an aperture of radius r = 4.63 arcsec.\\
gFmeanMagR6 & AB & REAL & -999  &Magnitude from mean flux from forced single epoch g filter \\
& & & & detections within an aperture of radius r = 4.63 arcsec.\\
gFmeanMagR6Err & AB & REAL & -999  &Error in magnitude from  mean flux from forced single epoch \\
& & & & g filter detections within an aperture of radius r = 4.63 arcsec.\\
gFmeanflxR7 & Jy & REAL & -999  &Mean flux from forced single epoch g filter \\
& & & & detections within an aperture of radius r = 7.43 arcsec.\\
gFmeanflxR7Err & Jy & REAL & -999  &Error in mean flux from forced single epoch g filter\\
& & & & detections within an aperture of radius r = 7.43 arcsec.\\
gFmeanflxR7Std & Jy & REAL & -999  &Standard deviation of forced single epoch g filter \\
& & & & detection fluxes within an aperture of radius r = 7.43 arcsec.\\
gFmeanflxR7Fill & - & REAL & -999  &Aperture fill factor for forced single epoch g filter \\
& & & & detections within an aperture of radius r = 7.43 arcsec.\\
gFmeanMagR7 & AB & REAL & -999  &Magnitude from mean flux from forced single epoch g filter\\
& & & & detections within an aperture of radius r = 7.43 arcsec.\\
gFmeanMagR7Err & AB & REAL & -999  &Error in magnitude from  mean flux from forced single epoch\\
& & & & g filter detections within an aperture of radius r = 7.43 arcsec.\\
gFlags & - & INT & 0  &Information flag bitmask indicating details of the photometry from forced \\
& & & & single epoch g filter detections.  Values listed in ObjectInfoFlags.\\
gE1 & - & REAL & -999  &\citet{Kaiser1995} polarization parameter $e1 = (M_{xx} - M_{yy}) / (M_{xx} + M_{yy})$ \\
& & & & from forced single epoch g filter detections.\\
gE2 & - & REAL & -999  &\citet{Kaiser1995} polarization parameter $e2 =  (2 M_{xy}) / (M_{xx} + M_{yy}) $ \\
& & & & from forced single epoch g filter detections.\\
rnTotal\\
... & & & & same entries repeated for r, i, z, and y filters \\
yE2 \\
\hline
\end{tabular}}
\end{center}
\label{table:ForcedMeanObject}
\end{table}%

\begin{table}
\caption{ForcedMeanLensing: Contains the mean \citet[K95]{Kaiser1995} lensing parameters measured from the forced photometry of objects detected in stacked images on the individual single epoch data.}
\begin{center}
\resizebox{\textwidth}{!}{%
\begin{tabular}{lllll}
\hline
\hline
column name & units & data type & default & description\\
\hline
objID & - & BIGINT & NA  &Unique object identifier.\\
uniquePspsFOid & - & BIGINT & NA  &Unique internal PSPS forced object identifier.\\
ippObjID & - & BIGINT & NA  &IPP internal object identifier.\\
randomForcedObjID & - & FLOAT & NA  &Random value drawn from the interval between zero and one.\\
nDetections & - & SMALLINT & -999  &Number of single epoch detections in all filters.\\
batchID & - & BIGINT & NA  &Internal database batch identifier.\\
processingVersion & - & TINYINT & NA  &Data release version.\\
gLensObjSmearX11 & $arcsec^{-2}$ & REAL & -999  &K95 eq. (A11) smear polarizability X11 term from forced g filter detections.\\
gLensObjSmearX12 & $arcsec^{-2}$ & REAL & -999  &K95 eq. (A11) smear polarizability X12 term from forced g filter detections.\\
gLensObjSmearX22 & $arcsec^{-2}$ & REAL & -999  &K95 eq. (A11) smear polarizability X22 term from forced g filter detections.\\
gLensObjSmearE1 & $arcsec^{-2}$ & REAL & -999  &K95 eq. (A12) smear polarizability e1 term from forced g filter detections.\\
gLensObjSmearE2 & $arcsec^{-2}$ & REAL & -999  &K95 eq. (A12) smear polarizability e2 term from forced g filter detections.\\
gLensObjShearX11 & - & REAL & -999  &K95 eq. (B11) shear polarizability X11 term from forced g filter detections.\\
gLensObjShearX12 & - & REAL & -999  &K95 eq. (B11) shear polarizability X12 term from forced g filter detections.\\
gLensObjShearX22 & - & REAL & -999  &K95 eq. (B11) shear polarizability X22 term from forced g filter detections.\\
gLensObjShearE1 & - & REAL & -999  &K95 eq. (B12) shear polarizability e1 term from forced g filter detections.\\
gLensObjShearE2 & - & REAL & -999  &K95 eq. (B12) shear polarizability e2 term from forced g filter detections.\\
gLensPSFSmearX11 & $arcsec^{-2}$ & REAL & -999  &K95 eq. (A11) smear polarizability X11 term from PSF model for forced g filter detections.\\
gLensPSFSmearX12 & $arcsec^{-2}$ & REAL & -999  &K95 eq. (A11) smear polarizability X12 term from PSF model for forced g filter detections.\\
gLensPSFSmearX22 & $arcsec^{-2}$ & REAL & -999  &K95 eq. (A11) smear polarizability X22 term from PSF model for forced g filter detections.\\
gLensPSFSmearE1 & $arcsec^{-2}$ & REAL & -999  &K95 eq. (A12) smear polarizability e1 term from PSF model for forced g filter detections.\\
gLensPSFSmearE2 & $arcsec^{-2}$ & REAL & -999  &K95 eq. (A12) smear polarizability e2 term from PSF model for forced g filter detections.\\
gLensPSFShearX11 & - & REAL & -999  &K95 eq. (B11) shear polarizability X11 term from PSF model for forced g filter detections.\\
gLensPSFShearX12 & - & REAL & -999  &K95 eq. (B11) shear polarizability X12 term from PSF model for forced g filter detections.\\
gLensPSFShearX22 & - & REAL & -999  &K95 eq. (B11) shear polarizability X22 term from PSF model for forced g filter detections.\\
gLensPSFShearE1 & - & REAL & -999  &K95 eq. (B12) shear polarizability e1 term from PSF model for forced g filter detections.\\
gLensPSFShearE2 & - & REAL & -999  &K95 eq. (B12) shear polarizability e2 term from PSF model forced g filter detections.\\
rlensObjSmearX11 \\
... & & & & same entries repeated for r, i, z, and y filters \\
ylensPSFShearE2 \\
\hline
\end{tabular}}
\end{center}
\label{table:ForcedMeanLensing}
\end{table}%

\clearpage
\subsection{Forced {\em warp} Exposure Tables}

\begin{table}
\caption{ForcedWarpMeta: Contains the metadata related to a sky-aligned distortion corrected {\em warp} image, upon which forced photometry is performed.  The astrometric and photometric calibration of the {\em warp} image are listed.}
\begin{center}
\begin{tabular}{lllll}
\hline
\hline
column name & units & data type & default & description\\
\hline
forcedWarpID & - & BIGINT & NA  &Unique forced {\em warp} identifier.\\
batchID & - & BIGINT & NA  &Internal database batch identifier.\\
surveyID & - & TINYINT & NA  &Survey identifier.  Details in the Survey table.\\
filterID & - & TINYINT & NA  &Filter identifier.  Details in the Filter table.\\
frameID & - & INT & NA  &Frame/exposure identifier of the Frame associated with this warp.\\
ippSkycalID & - & INT & NA  &IPP skycal identifier for the run that generated the positions for forced\\
& & & & photometry.\\
stackMetaID & - & INT & NA  &Identifier for the {\em stack} which yielded the positions for forced photometry.\\
tessID & - & TINYINT & 0  &Tessellation identifier.  Details in the TessellationType table.\\
projectionID & - & SMALLINT & -1  &Projection cell identifier.\\
skyCellID & - & TINYINT & 255  &Skycell region identifier.\\
photoCalID & - & INT & NA  &Photometric calibration identifier.  Details in the PhotoCal table.\\
analysisVer & - & VARCHAR(100) &   &IPP software analysis release version.\\
md5sum & - & VARCHAR(100) &   &IPP MD5 Checksum.\\
expTime & seconds & REAL & -999  &Exposure time of the source frame/exposure for this {\em warp} image.  Necessary \\
& & & & for converting listed fluxes and magnitudes back to measured ADU counts.\\
recalAstroScatX & arcsec & REAL & -999  &Measurement of the re-calibration (not astrometric error) in the X direction.\\
recalAstroScatY & arcsec & REAL & -999  &Measurement of the re-calibration (not astrometric error) in the Y direction.\\
recalNAstroStars & - & INT & -999  &Number of astrometric reference sources used in recalibration.\\
recalphotoScat & magnitudes & REAL & -999  &Photometric scatter relative to reference catalog.\\
recalNPhotoStars & - & INT & -999  &Number of astrometric reference sources used in recalibration.\\
psfModelID & - & INT & -999  &PSF model identifier.\\
psfFWHM & arcsec & REAL & -999  &Mean PSF full width at half maximum at image center.\\
psfWidMajor & arcsec & REAL & -999  &PSF major axis FWHM at image center.\\
psfWidMinor & arcsec & REAL & -999  &PSF minor axis FWHM at image center.\\
psfTheta & degrees & REAL & -999  &PSF major axis orientation at image center.\\
photoZero & magnitudes & REAL & -999  &Locally derived photometric zero point for this {\em warp} image.\\
ctype1 & - & VARCHAR(100) &   &Name of astrometric projection in RA.\\
ctype2 & - & VARCHAR(100) &   &Name of astrometric projection in Dec.\\
crval1 & degrees & FLOAT & -999  &Right ascension corresponding to reference pixel.\\
crval2 & degrees & FLOAT & -999  &Declination corresponding to reference pixel.\\
crpix1 & sky pixels & FLOAT & -999  &Reference pixel for RA.\\
crpix2 & sky pixels & FLOAT & -999  &Reference pixel for Dec.\\
cdelt1 & degrees/pixel & FLOAT & -999  &Pixel scale in RA.\\
cdelt2 & degrees/pixel & FLOAT & -999  &Pixel scale in Dec.\\
pc001001 & - & FLOAT & -999  &Linear transformation matrix element between image pixel x and RA.\\
pc001002 & - & FLOAT & -999  &Linear transformation matrix element between image pixel y and RA.\\
pc002001 & - & FLOAT & -999  &Linear transformation matrix element between image pixel x and Dec.\\
pc002002 & - & FLOAT & -999  &Linear transformation matrix element between image pixel y and Dec.\\
processingVersion & - & TINYINT & NA  &Data release version.\\
\hline
\end{tabular}
\end{center}
\label{table:ForcedWarpMeta}
\end{table}%

\begin{table}
\caption{ForcedWarpMeasurement: Contains single epoch forced photometry of individual measurements of objects detected in the stacked images.  The identifiers connecting the measurement back to the original image and to the object association are provided.  PSF, aperture, and \citet{Kron1980} photometry are included, along with sky and detector coordinate positions.}
\begin{center}
\begin{tabular}{lllll}
\hline
\hline
column name & units & data type & default & description\\
\hline
objID & - & BIGINT & NA  &Unique object identifier.\\
uniquePspsFWid & - & BIGINT & NA  &Unique internal PSPS forced {\em warp} identifier.\\
detectID & - & BIGINT & NA  &Unique detection identifier.\\
ippObjID & - & BIGINT & NA  &IPP internal object identifier.\\
ippDetectID & - & BIGINT & NA  &IPP internal detection identifier.\\
filterID & - & TINYINT & NA  &Filter identifier.  Details in the Filter table.\\
surveyID & - & TINYINT & NA  &Survey identifier.  Details in the Survey table.\\
forcedSummaryID & - & BIGINT & NA  &Unique forced {\em warp} summary identifier.\\
forcedWarpID & - & BIGINT & NA  &Unique forced {\em warp} identifier.\\
randomWarpID & - & FLOAT & NA  &Random value drawn from the interval between zero and one.\\
tessID & - & TINYINT & 0  &Tessellation identifier.  Details in the TessellationType table.\\
projectionID & - & SMALLINT & -1  &Projection cell identifier.\\
skyCellID & - & TINYINT & 255  &Skycell region identifier.\\
dvoRegionID & - & INT & -1  &Internal DVO region identifier.\\
obsTime & days & FLOAT & -999  &Modified Julian Date at the midpoint of the observation.\\
zp & magnitudes & REAL & 0  &Photometric zeropoint.  Necessary for converting listed fluxes and magnitudes \\
& & & & back to measured ADU counts.\\
telluricExt & magnitudes & REAL & NA  &Estimated Telluric extinction due to non-photometric observing conditions. \\
& & & & Necessary for converting listed fluxes and magnitudes back to measured ADU counts.\\
expTime & seconds & REAL & -999  &Exposure time of the frame/exposure.  Necessary for converting listed fluxes \\
& & & & and magnitudes back to measured ADU counts.\\
airMass & - & REAL & 0  &Airmass at midpoint of the exposure.  Necessary for converting listed fluxes \\
& & & & and magnitudes back to measured ADU counts.\\
FpsfFlux & Jy & REAL & -999  &PSF flux.\\
FpsfFluxErr & Jy & REAL & -999  &Error in PSF flux.\\
xPosChip & raw pixels & REAL & -999  &PSF x position in original chip pixels.\\
yPosChip & raw pixels & REAL & -999  &PSF y position in original chip pixels.\\
FccdID & - & SMALLINT & -999  &OTA identifier of original chip (see ImageMeta).\\
FpsfMajorFWHM & arcsec & REAL & -999  &PSF major axis FWHM.\\
FpsfMinorFWHM & arcsec & REAL & -999  &PSF minor axis FWHM.\\
FpsfTheta & degrees & REAL & -999  &PSF major axis orientation.\\
FpsfCore & - & REAL & -999  &PSF core parameter k, where $F = F0 / (1 + k r^2 + r^{3.33})$.\\
FpsfQf & - & REAL & -999  &PSF coverage factor.\\
FpsfQfPerfect & - & REAL & -999  &PSF weighted fraction of pixels totally unmasked.\\
FpsfChiSq & - & REAL & -999  &Reduced chi squared value of the PSF model fit.\\
FmomentXX & $arcsec^2$ & REAL & -999  &Second moment $M_{xx}$.\\
FmomentXY & $arcsec^2$ & REAL & -999  &Second moment $M_{xy}$.\\
FmomentYY & $arcsec^2$ & REAL & -999  &Second moment $M_{yy}$.\\
FmomentR1 & arcsec & REAL & -999  &First radial moment.\\
FmomentRH & $arcsec^{0.5}$ & REAL & -999  &Half radial moment ($r^{0.5}$ weighting).\\
FmomentM3C & $arcsec^2$ & REAL & -999  &Cosine of trefoil second moment term: $r^2 cos(3 theta) = M_{xxx} - 3 * M_{xyy}$.\\
FmomentM3S & $arcsec^2$ & REAL & -999  &Sine of trefoil second moment: $r^2 sin (3 theta) = 3 * M_{xxy} - M_{yyy}$.\\
FmomentM4C & $arcsec^2$ & REAL & -999  &Cosine of quadrupole second moment: $r^2 cos (4 theta) = M_{xxxx} - 6 * M_{xxyy} + M_{yyyy}$.\\
FmomentM4S & $arcsec^2$ & REAL & -999  &Sine of quadrupole second moment: $r^2 sin (4 theta) = 4 * M_{xxxy} - 4 * M_{xyyy}$.\\
FapFlux & Jy & REAL & -999  &Aperture flux.\\
FapFluxErr & Jy & REAL & -999  &Error in aperture flux.\\
FapFillF & - & REAL & -999  &Aperture fill factor.\\
FapRadius & arcsec & REAL & -999  &Aperture radius for forced {\em warp} detection.\\
FkronFlux & Jy & REAL & -999  &Kron (1980) flux.\\
FkronFluxErr & Jy & REAL & -999  &Error in Kron (1980) flux.\\
FkronRad & arcsec & REAL & -999  &Kron (1980) radius.\\
Fsky & $Jy/arcsec^2$ & REAL & -999  &Background sky level.\\
FskyErr & $Jy/arcsec^2$ & REAL & -999  &Error in background sky level.\\
FinfoFlag & - & BIGINT & 0  &Information flag bitmask indicating details of the photometry.  \\
& & & & Values listed in DetectionFlags.\\
FinfoFlag2 & - & INT & 0  &Information flag bitmask indicating details of the photometry.  \\
& & & & Values listed in DetectionFlags2.\\
FinfoFlag3 & - & INT & 0  &Information flag bitmask indicating details of the photometry.  \\
& & & & Values listed in DetectionFlags3.\\
processingVersion & - & TINYINT & NA  &Data release version.\\
\hline
\end{tabular}
\end{center}
\label{table:ForcedWarpMeasurement}
\end{table}%

\begin{table}
\caption{ForcedWarpMasked: Contains an entry for objects detected in the stacked images which were in the footprint of a single epoch exposure, but for which there are no unmasked pixels at that epoch.}
\begin{center}
\begin{tabular}{lllll}
\hline
\hline
column name & units & data type & default & description\\
\hline
objID & - & BIGINT & NA  &Unique object identifier.\\
uniquePspsFWid & - & BIGINT & NA  &Unique internal PSPS forced {\em warp} identifier.\\
ippObjID & - & BIGINT & NA  &IPP internal object identifier.\\
ippDetectID & - & BIGINT & NA  &IPP internal detection identifier.\\
filterID & - & TINYINT & NA  &Filter identifier.  Details in the Filter table.\\
surveyID & - & TINYINT & NA  &Survey identifier.  Details in the Survey table.\\
forcedSummaryID & - & BIGINT & NA  &Forced {\em warp} summary meta identifier\\
forcedWarpID & - & BIGINT & NA  &Unique forced {\em warp} identifier.\\
randomWarpID & - & FLOAT & NA  &Random value drawn from the interval between zero and one.\\
tessID & - & TINYINT & 0  &Tessellation identifier.  Details in the TessellationType table.\\
projectionID & - & SMALLINT & -1  &Projection cell identifier.\\
skyCellID & - & REAL & -999  &Skycell region identifier.\\
dvoRegionID & - & REAL & -999  &Internal DVO region identifier.\\
obsTime & days & FLOAT & -999  &Modified Julian Date at the midpoint of the observation.\\
\hline
\end{tabular}
\end{center}
\label{table:ForcedWarpMasked}
\end{table}%

\begin{table}
\caption{ForcedWarpExtended: Contains the single epoch forced photometry fluxes within the SDSS R5 (r = 3.00 arcsec), R6 (r = 4.63 arcsec), and R7 (r = 7.43 arcsec) apertures \citep{Stoughton2002} for objects detected in the stacked images.}
\begin{center}
\begin{tabular}{lllll}
\hline
\hline
column name & units & data type & default & description\\
\hline
objID & - & BIGINT & NA  &Unique object identifier.\\
uniquePspsFWid & - & BIGINT & NA  &Unique internal PSPS forced {\em warp} identifier.\\
detectID & - & BIGINT & NA  &Unique detection identifier.\\
ippObjID & - & BIGINT & NA  &IPP internal object identifier.\\
ippDetectID & - & BIGINT & NA  &IPP internal detection identifier.\\
filterID & - & TINYINT & NA  &Filter identifier.  Details in the Filter table.\\
surveyID & - & TINYINT & NA  &Survey identifier.  Details in the Survey table.\\
forcedWarpID & - & BIGINT & NA  &Unique forced {\em warp} identifier.\\
randomWarpID & - & FLOAT & NA  &Random value drawn from the interval between zero and one.\\
tessID & - & TINYINT & 0  &Tessellation identifier.  Details in the TessellationType table.\\
projectionID & - & SMALLINT & -1  &Projection cell identifier.\\
skyCellID & - & TINYINT & 255  &Skycell region identifier.\\
dvoRegionID & - & INT & -1  &Internal DVO region identifier.\\
obsTime & days & FLOAT & -999  &Modified Julian Date at the midpoint of the observation.\\
flxR5 & Jy & REAL & -999  &Flux from forced photometry measurement within an aperture of radius \\
& & & & r = 3.00 arcsec.\\
flxR5Err & Jy & REAL & -999  &Error in flux from forced photometry measurement within an aperture of\\
& & & & radius r = 3.00 arcsec.\\
flxR5Std & Jy & REAL & -999  &Standard deviation of flux from forced photometry measurement within \\
& & & & an aperture of radius r = 3.00 arcsec.\\
flxR5Fill & - & REAL & -999  &Aperture fill factor for forced photometry measurement within an \\
& & & & aperture of radius r = 3.00 arcsec.\\
flxR6 & Jy & REAL & -999  &Flux from forced photometry measurement within an aperture of radius \\
& & & & r = 4.63 arcsec.\\
flxR6Err & Jy & REAL & -999  &Error in flux from forced photometry measurement within an aperture of\\
& & & & radius r = 4.63 arcsec.\\
flxR6Std & Jy & REAL & -999  &Standard deviation of flux from forced photometry measurement within \\
& & & & an aperture of radius r = 4.63 arcsec.\\
flxR6Fill & - & REAL & -999  &Aperture fill factor for forced photometry measurement within an \\
& & & & aperture of radius r = 4.63 arcsec.\\
flxR7 & Jy & REAL & -999  &Flux from forced photometry measurement within an aperture of radius \\
& & & & r = 7.43 arcsec.\\
flxR7Err & Jy & REAL & -999  &Error in flux from forced photometry measurement within an aperture of\\
& & & & radius r = 7.43 arcsec.\\
flxR7Std & Jy & REAL & -999  &Standard deviation of flux from forced photometry measurement within \\
& & & & an aperture of radius r = 7.43 arcsec.\\
flxR7Fill & - & REAL & -999  &Aperture fill factor for forced photometry measurement within an \\
& & & & aperture of radius r = 7.43 arcsec.\\
\hline
\end{tabular}
\end{center}
\label{table:ForcedWarpExtended}
\end{table}%

\begin{table}
\caption{ForcedWarpLensing: Contains the \citet[K95]{Kaiser1995} lensing parameters measured from the forced photometry of objects detected in stacked images on the individual single epoch data.}
\begin{center}
\begin{tabular}{lllll}
\hline
\hline
column name & units & data type & default & description\\
\hline
objID & - & BIGINT & NA  &Unique object identifier.\\
uniquePspsFWid & - & BIGINT & NA  &Unique internal PSPS forced {\em warp} identifier.\\
detectID & - & BIGINT & NA  &Unique detection identifier.\\
ippObjID & - & BIGINT & NA  &IPP internal object identifier.\\
ippDetectID & - & BIGINT & NA  &IPP internal detection identifier.\\
filterID & - & TINYINT & NA  &Filter identifier.  Details in the Filter table.\\
surveyID & - & TINYINT & NA  &Survey identifier.  Details in the Survey table.\\
forcedWarpID & - & BIGINT & NA  &Unique forced {\em warp} identifier.\\
randomWarpID & - & FLOAT & NA  &Random value drawn from the interval between zero and one.\\
tessID & - & TINYINT & 0  &Tessellation identifier.  Details in the TessellationType table.\\
projectionID & - & SMALLINT & -1  &Projection cell identifier.\\
skyCellID & - & TINYINT & 255  &Skycell region identifier.\\
dvoRegionID & - & INT & -1  &Internal DVO region identifier.\\
obsTime & days & FLOAT & -999  &Modified Julian Date at the midpoint of the observation.\\
lensObjSmearX11 & $arcsec^{-2}$ & REAL & -999  &K95 eq. (A11) smear polarizability X11 term from forced photometry.\\
lensObjSmearX12 & $arcsec^{-2}$ & REAL & -999  &K95 eq. (A11) smear polarizability X12 term from forced photometry.\\
lensObjSmearX22 & $arcsec^{-2}$ & REAL & -999  &K95 eq. (A11) smear polarizability X22 term from forced photometry.\\
lensObjSmearE1 & $arcsec^{-2}$ & REAL & -999  &K95 eq. (A12) smear polarizability e1 term from forced photometry.\\
lensObjSmearE2 & $arcsec^{-2}$ & REAL & -999  &K95 eq. (A12) smear polarizability e2 term from forced photometry.\\
lensObjShearX11 & - & REAL & -999  &K95 eq. (B11) shear polarizability X11 term from forced photometry.\\
lensObjShearX12 & - & REAL & -999  &K95 eq. (B11) shear polarizability X12 term from forced photometry.\\
lensObjShearX22 & - & REAL & -999  &K95 eq. (B11) shear polarizability X22 term from forced photometry.\\
lensObjShearE1 & - & REAL & -999  &K95 eq. (B12) shear polarizability e1 term from forced photometry.\\
lensObjShearE2 & - & REAL & -999  &K95 eq. (B12) shear polarizability e2 term from forced photometry.\\
lensPSFSmearX11 & $arcsec^{-2}$ & REAL & -999  &K95 eq. (A11) smear polarizability X11 term from PSF model for forced photometry.\\
lensPSFSmearX12 & $arcsec^{-2}$ & REAL & -999  &K95 eq. (A11) smear polarizability X12 term from PSF model for forced photometry.\\
lensPSFSmearX22 & $arcsec^{-2}$ & REAL & -999  &K95 eq. (A11) smear polarizability X22 term from PSF model for forced photometry.\\
lensPSFSmearE1 & $arcsec^{-2}$ & REAL & -999  &K95 eq. (A12) smear polarizability e1 term from PSF model for forced photometry.\\
lensPSFSmearE2 & $arcsec^{-2}$ & REAL & -999  &K95 eq. (A12) smear polarizability e2 term from PSF model for forced photometry.\\
lensPSFShearX11 & - & REAL & -999  &K95 eq. (B11) shear polarizability X11 term from PSF model for forced photometry.\\
lensPSFShearX12 & - & REAL & -999  &K95 eq. (B11) shear polarizability X12 term from PSF model for forced photometry.\\
lensPSFShearX22 & - & REAL & -999  &K95 eq. (B11) shear polarizability X22 term from PSF model for forced photometry.\\
lensPSFShearE1 & - & REAL & -999  &K95 eq. (B12) shear polarizability e1 term from PSF model for forced photometry.\\
lensPSFShearE2 & - & REAL & -999  &K95 eq. (B12) shear polarizability e2 term from PSF model for forced photometry.\\
psfE1 & - & REAL & -999  &K95 polarization parameter $e1 = (M_{xx} - M_{yy}) / (M_{xx} + M_{yy})$ from forced photometry.\\
psfE2 & - & REAL & -999  &K95 polarization parameter $e2 =  (2 M_{xy}) / (M_{xx} + M_{yy})$ from forced photometry.\\

\hline
\end{tabular}
\end{center}
\label{table:ForcedWarpLensing}
\end{table}%

\begin{table}
\caption{ForcedWarpToImage: Contains the mapping of which input image comprises a particular {\em warp} image used for forced photometry.}
\begin{center}
\begin{tabular}{lllll}
\hline
\hline
column name & units & data type & default & description\\
\hline
forcedWarpID & - & BIGINT & NA  &Unique forced {\em warp} identifier.\\
imageID & - & BIGINT & NA  &Unique image identifier.  Constructed as (100 * frameID + ccdID).\\
\hline
\end{tabular}
\end{center}
\label{table:ForcedWarpToImage}
\end{table}%

\clearpage
\subsection{Forced Galaxy Tables}

\begin{table}
\caption{ForcedGalaxyShape: Contains the extended source galaxy shape parameters.  All filters are matched into a single row.  The positions, magnitudes, fluxes, and Sersic indices are inherited from their parent measurement in the StackModelFit tables, and are reproduced here for convenience.  The major and minor axes and orientation are recalculated on a warp-by-warp basis from the best fit given these inherited properties (\citep{Sersic1963}).}
\begin{center}
\begin{tabular}{lllll}
\hline
\hline
column name & units & data type & default & description\\
\hline
objID & - & BIGINT & NA  &Unique object identifier.\\
uniquePspsFGid & - & BIGINT & NA  &Unique internal PSPS forced galaxy identifier.\\
ippObjID & - & BIGINT & NA  &IPP internal object identifier.\\
surveyID & - & TINYINT & NA  &Survey identifier.  Details in the Survey table.\\
randomForcedGalID & - & FLOAT & NA  &Random value drawn from the interval between zero and one.\\
galModelType & - & TINYINT & -999  &Galaxy model identifier.\\
nFilter & - & TINYINT & -999  &Number of filters with valid model measurements.\\
gippDetectID & - & BIGINT & NA  &IPP internal detection identifier.\\
gstackImageID & - & BIGINT & NA  &Unique {\em stack} identifier for the g filter {\em stack} that was the original detection source.\\
gGalMajor & arcsec & REAL & -999  &Galaxy major axis for g filter measurement.\\
gGalMajorErr & arcsec & REAL & -999  &Error in galaxy major axis for g filter measurement.\\
gGalMinor & arcsec & REAL & -999  &Galaxy minor axis for g filter measurement.\\
gGalMinorErr & arcsec & REAL & -999  &Error in galaxy minor axis for g filter measurement.\\
gGalMag & AB & REAL & -999  &Galaxy fit magnitude for g filter measurement.\\
gGalMagErr & AB & REAL & -999  &Error in galaxy fit magnitude for g filter measurement.\\
gGalPhi & degrees & REAL & -999  &Major axis position angle of the model fit for the g filter measurement.\\
gGalIndex & - & REAL & -999  &Sersic index of the model fit for the g filter measurement.\\
gGalFlags & - & SMALLINT & -999  &Analysis flags for the galaxy model chi-square fit (g filter measurement, values \\
& & & & defined in ForcedGalaxyShapeFlags).\\
gGalChisq & - & REAL & -999  &Reduced chi squared value for g filter measurement.\\
rippDetectID \\
... & & & & same entries repeated for r, i, z, and y filters \\
yGalChisq\\
\hline
\end{tabular}
\end{center}
\label{table:ForcedGalaxyShape}
\end{table}%

\clearpage
\subsection{Diff Object Tables}

\begin{table}
\caption{DiffDetObject: Contains the positional information for difference detection objects in a number of coordinate systems.  The objects associate difference detections within a one arcsecond radius.  The number of detections in each filter from is listed, along with maximum coverage fractions \citep[see][]{Szalay2007}.}
\begin{center}
\begin{tabular}{lllll}
\hline
\hline
column name & units & data type & default & description\\
\hline
diffObjName & - & VARCHAR(32) & NA  &IAU name for this object.\\
diffObjPSOName & - & VARCHAR(32) & NA  &Alternate Pan-STARRS name for this object.\\
diffObjAltName1 & - & VARCHAR(32) &   &Altername name for this object.\\
diffObjAltName2 & - & VARCHAR(32) &   &Altername name for this object.\\
diffObjAltName3 & - & VARCHAR(32) &   &Altername name for this object.\\
diffObjPopularName & - & VARCHAR(140) &   &Well known name for this object.\\
diffObjID & - & BIGINT & NA  &Unique difference object identifier.\\
uniquePspsDOid & - & BIGINT & NA  &Unique internal PSPS difference object identifier.\\
ippObjID & - & BIGINT & NA  &IPP internal object identifier.\\
surveyID & - & TINYINT & NA  &Survey identifier.  Details in the Survey table.\\
htmID & - & BIGINT & NA  &Hierarchical triangular mesh (Szalay 2007) index.\\
zoneID & - & INT & NA  &Local zone index, found by dividing the sky into bands of declination \\
& & & & 1/2 arcminute in height: zoneID = floor((90 + declination)/0.0083333).\\
randomDiffObjID & - & FLOAT & NA  &Random value drawn from the interval between zero and one.\\
batchID & - & BIGINT & NA  &Internal database batch identifier.\\
dvoRegionID & - & INT & -1  &Internal DVO region identifier.\\
objInfoFlag & - & INT & 0  &Information flag bitmask indicating details of the photometry.  Values\\
& & & & listed in ObjectInfoFlags.\\
qualityFlag & - & TINYINT & 0  &Subset of objInfoFlag denoting whether this object is real or a\\
& & & & likely false positive.  Values listed in ObjectQualityFlags.\\
ra & degrees & FLOAT & -999  &Right ascension mean.\\
dec & degrees & FLOAT & -999  &Declination mean.\\
cx & - & FLOAT & NA  &Cartesian x on a unit sphere. \\
cy & - & FLOAT & NA  &Cartesian y on a unit sphere. \\
cz & - & FLOAT & NA  &Cartesian z on a unit sphere. \\
lambda & degrees & FLOAT & -999  &Ecliptic longitude.\\
beta & degrees & FLOAT & -999  &Ecliptic latitude.\\
l & degrees & FLOAT & -999  &Galactic longitude.\\
b & degrees & FLOAT & -999  &Galactic latitude.\\
gQfPerfect & - & REAL & -999  &Maximum PSF weighted fraction of pixels totally unmasked from g\\
& & & & filter detections.\\
rQfPerfect & - & REAL & -999  &Maximum PSF weighted fraction of pixels totally unmasked from r\\
& & & & filter detections.\\
iQfPerfect & - & REAL & -999  &Maximum PSF weighted fraction of pixels totally unmasked from i\\
& & & & filter detections.\\
zQfPerfect & - & REAL & -999  &Maximum PSF weighted fraction of pixels totally unmasked from z\\
& & & & filter detections.\\
yQfPerfect & - & REAL & -999  &Maximum PSF weighted fraction of pixels totally unmasked from y\\
& & & & filter detections.\\
processingVersion & - & TINYINT & NA  &Data release version.\\
nDetections & - & SMALLINT & -999  &Number of difference detections in all filters.\\
ng & - & SMALLINT & -999  &Number of difference detections in g filter.\\
nr & - & SMALLINT & -999  &Number of difference detections in r filter.\\
ni & - & SMALLINT & -999  &Number of difference detections in i filter.\\
nz & - & SMALLINT & -999  &Number of difference detections in z filter.\\
ny & - & SMALLINT & -999  &Number of difference detections in y filter.\\
\hline
\end{tabular}
\end{center}
\label{table:DiffDetObject}
\end{table}%

\clearpage
\subsection{Diff Detection Tables}

\begin{table}
\caption{DiffMeta: Contains metadata related to a difference image constructed by subtracting a stacked image from a single epoch image, or in the case of the MD Survey from a nightly {\em stack} (stack made from all exposures in a single filter in a single night).  The astrometric calibration of the reference {\em stack} is listed.}
\begin{center}
\begin{tabular}{lllll}
\hline
\hline
column name & units & data type & default & description\\
\hline
diffImageID & - & BIGINT & NA  &Unique difference identifier.\\
batchID & - & BIGINT & NA  &Internal database batch identifier.\\
surveyID & - & TINYINT & NA  &Survey identifier.  Details in the Survey table.\\
filterID & - & TINYINT & NA  &Filter identifier.  Details in the Filter table.\\
diffTypeID & - & TINYINT & 0  &Difference type identifier.  Details in the DiffType table.\\
frameID & - & INT & NA  &Frame/exposure identifier for the positive image in warp-stack \\
& & & & difference images; \\ 
posImageID & - & BIGINT & NA  &Image identifier for the positive image.  \\ 
negImageID & - & BIGINT & NA  &Image identifier for the negative image.  \\
ippDiffID & - & BIGINT & NA  &IPP diffRun identifier.\\
tessID & - & TINYINT & 0  &Tessellation identifier.  Details in the TessellationType table.\\
projectionID & - & SMALLINT & -1  &Projection cell identifier.\\
skyCellID & - & TINYINT & 255  &Skycell region identifier.\\
photoCalID & - & INT & NA  &Photometric calibration identifier.  Details in the PhotoCal table.\\
analysisVer & - & VARCHAR(100) &   &IPP software analysis release version.\\
md5sum & - & VARCHAR(100) &   &IPP MD5 Checksum.\\
detectionThreshold & magnitudes & REAL & -999  &Reference magnitude for detection efficiency calculation.\\
expTime & seconds & REAL & -999  &Exposure time of positive image.  Necessary for converting listed \\
& & & & fluxes and magnitudes back to measured ADU counts.\\
psfModelID & - & INT & -999  &PSF model identifier.\\
psfFWHM & arcsec & REAL & -999  &Mean PSF full width at half maximum at image center.\\
psfWidMajor & arcsec & REAL & -999  &PSF major axis FWHM at image center.\\
psfWidMinor & arcsec & REAL & -999  &PSF minor axis FWHM at image center.\\
psfTheta & degrees & REAL & -999  &PSF major axis orientation at image center.\\
kernel & - & VARCHAR(100) &   &Subtraction kernel.\\
mode & - & TINYINT & 0  &Subtraction mode for which input to convolve.\\
numStamps & - & INT & -999  &Number of stamps.\\
stampDevMean & - & REAL & -999  &Mean stamp deviation.\\
stampDevRMS & - & REAL & -999  &RMS stamp deviation.\\
normalization & - & REAL & -999  &Normalization.\\
convolveMax & - & REAL & -999  &Maxiumum convolution fraction.\\
deconvolveMax & - & REAL & -999  &Maximum deconvolution fraction.\\
ctype1 & - & VARCHAR(100) &   &Name of astrometric projection in right ascension.\\
ctype2 & - & VARCHAR(100) &   &Name of astrometric projection in declination.\\
crval1 & degrees & FLOAT & -999  &Right ascension corresponding to reference pixel.\\
crval2 & degrees & FLOAT & -999  &Declination corresponding to reference pixel.\\
crpix1 & sky pixels & FLOAT & -999  &Reference pixel for right ascension.\\
crpix2 & sky pixels & FLOAT & -999  &Reference pixel for declination.\\
cdelt1 & degrees/pixel & FLOAT & -999  &Pixel scale in right ascension.\\
cdelt2 & degrees/pixel & FLOAT & -999  &Pixel scale in declination.\\
pc001001 & - & FLOAT & -999  &Linear transformation matrix element between image pixel x and \\
& & & & right ascension.\\
pc001002 & - & FLOAT & -999  &Linear transformation matrix element between image pixel y and \\
& & & & right ascension.\\
pc002001 & - & FLOAT & -999  &Linear transformation matrix element between image pixel x and \\
& & & & declination.\\
pc002002 & - & FLOAT & -999  &Linear transformation matrix element between image pixel y and \\
& & & & declination.\\
processingVersion & - & TINYINT & NA  &Data release version.\\
\hline
\end{tabular}
\end{center}
\label{table:DiffMeta}
\end{table}%

\begin{table}
\caption{DiffDetection: Contains the photometry of individual detections from a difference image.  The identifiers connecting the detection back to the difference image and to the object association are provided.  PSF, aperture, and \citet{Kron1980} photometry are included, along with sky and detector coordinate positions.}
\begin{center}
\begin{tabular}{lllll}
\hline
\hline
column name & units & data type & default & description\\
\hline
diffObjID & - & BIGINT & NA  &Unique difference object identifier.\\
uniquePspsDFid & - & BIGINT & NA  &Unique internal PSPS difference detection identifier.\\
diffDetID & - & BIGINT & NA  &Unique difference detection identifier.\\
diffImageID & - & BIGINT & NA  &Difference detection meta identifier.\\
ippObjID & - & BIGINT & NA  &IPP internal object identifier.\\
ippDetectID & - & BIGINT & NA  &IPP internal detection identifier.\\
fromPosImage & - & TINYINT & NA  &Detection is from positive image (if 1) or negative image (if 0).\\
filterID & - & TINYINT & NA  &Filter identifier.  Details in the Filter table.\\
surveyID & - & TINYINT & NA  &Survey identifier.  Details in the Survey table.\\
randomDiffID & - & FLOAT & NA  &Random value drawn from the interval between zero and one.\\
tessID & - & TINYINT & 0  &Tessellation identifier.  Details in the TessellationType table.\\
projectionID & - & SMALLINT & -1  &Projection cell identifier.\\
skyCellID & - & TINYINT & 255  &Skycell region identifier.\\
dvoRegionID & - & INT & -1  &Internal DVO region identifier.\\
obsTime & days & FLOAT & -999  &Modified Julian Date at the midpoint of the observation.\\
xPos & sky pixels & REAL & -999  &PSF x center location.\\
yPos & sky pixels & REAL & -999  &PSF y center location.\\
xPosErr & sky pixels & REAL & -999  &Error in PSF x center location.\\
yPosErr & sky pixels & REAL & -999  &Error in PSF y center location.\\
pltScale & arcsec/pixel & REAL & -999  &Local plate scale at this location.\\
posAngle & degrees & REAL & -999  &Position angle (sky-to-chip) at this location.\\
ra & degrees & FLOAT & -999  &Right ascension.\\
dec & degrees & FLOAT & -999  &Declination.\\
raErr & arcsec & REAL & -999  &Right ascension error.\\
decErr & arcsec & REAL & -999  &Declination error.\\
zp & magnitudes & REAL & 0  &Photometric zeropoint for converting fluxes and magnitudes to measured ADU.\\
telluricExt & magnitudes & REAL & NA  &Estimated Telluric extinction due to non-photometric observing conditions.\\ 
expTime & seconds & REAL & -999  &Exposure time of the positive single-epoch image. \\
airMass & - & REAL & 0  &Airmass at midpoint of exposure to convert fluxes and magnitudes to measured ADU.\\
DpsfFlux & Jy & REAL & -999  &Flux from PSF fit.\\
DpsfFluxErr & Jy & REAL & -999  &Error in PSF flux.\\
xPosChip & raw pixels & REAL & -999  &PSF x position in original chip pixels.\\
yPosChip & raw pixels & REAL & -999  &PSF y position in original chip pixels.\\
ccdID & - & SMALLINT & -999  &OTA identifier of original chip (see ImageMeta).\\
DpsfMajorFWHM & arcsec & REAL & -999  &PSF major axis FWHM.\\
DpsfMinorFWHM & arcsec & REAL & -999  &PSF minor axis FWHM.\\
DpsfTheta & degrees & REAL & -999  &PSF major axis orientation.\\
DpsfCore & - & REAL & -999  &PSF core parameter k, where $F = F0 / (1 + k r^2 + r^{3.33}).$\\
DpsfQf & - & REAL & -999  &PSF coverage factor.\\
DpsfQfPerfect & - & REAL & -999  &PSF-weighted fraction of pixels totally unmasked.\\
DpsfChiSq & - & REAL & -999  &Reduced chi squared value of the PSF model fit.\\
DpsfLikelihood & - & REAL & -999  &Likelihood that this detection is best fit by a PSF.\\
DmomentXX & $arcsec^2$ & REAL & -999  &Second moment $M_{xx}$.\\
DmomentXY & $arcsec^2$ & REAL & -999  &Second moment $M_{xy}$.\\
DmomentYY & $arcsec^2$ & REAL & -999  &Second moment $M_{yy}$.\\
DmomentR1 & arcsec & REAL & -999  &First radial moment.\\
DmomentRH & $arcsec^{0.5}$ & REAL & -999  &Half radial moment ($r^{0.5}$ weighting).\\
DapFlux & Jy & REAL & -999  &Aperture flux.\\
DapFluxErr & Jy & REAL & -999  &Error in aperture flux.\\
DapFillF & - & REAL & -999  &Aperture fill factor.\\
DkronFlux & Jy & REAL & -999  &Kron (1980) flux.\\
DkronFluxErr & Jy & REAL & -999  &Error in Kron (1980) flux.\\
DkronRad & arcsec & REAL & -999  &Kron (1980) radius.\\
diffNPos & sky pixels & INT & -999  &Number of difference pixels within the aperture that are positive.\\
diffFPosRatio & - & REAL & -999  &Ratio of the sum of positive flux pixel values to the sum of the absolute value\\
& & & & of all unmasked pixels within the aperture.\\
diffNPosRatio & - & REAL & -999  &Ratio of the number of positive flux pixels to the number of unmasked pixels\\
& & & & within the aperture.\\
diffNPosMask & - & REAL & -999  &Ratio of the number of positive flux pixels to the number of positive or masked\\
& & & & pixels within the aperture.\\
diffNPosAll & - & REAL & -999  &Ratio of the number of positive flux pixels to the total number of all pixels within\\
& & & & the aperture.\\
diffPosDist & sky pixels & REAL & -999  &Distance to matching source in positive image.\\
diffNegDist & sky pixels & REAL & -999  &Distance to matching source in negative image.\\
diffPosSN & - & REAL & -999  &Signal to noise of matching source in positive image.\\
diffNegSN & - & REAL & -999  &Signal to noise of matching source in negative image.\\
Dsky & $Jy/arcsec^2$ & REAL & -999  &Background sky level.\\
DskyErr & $Jy/arcsec^2$ & REAL & -999  &Error in background sky level.\\
DinfoFlag & - & BIGINT & 0  &Information flag bitmask indicating details of the photometry. see DetectionFlags.\\
DinfoFlag2 & - & INT & 0  &Information flag bitmask indicating details of the photometry.  see DetectionFlags2.\\
DinfoFlag3 & - & INT & 0  &Information flag bitmask indicating details of the photometry.  see DetectionFlags3.\\
processingVersion & - & TINYINT & NA  &Data release version.\\
\hline
\end{tabular}
\end{center}
\label{table:DiffDetection}
\end{table}%

\begin{table}
\caption{DiffToImage: Contains the mapping of which input images were used to construct a particular difference image.}
\begin{center}
\begin{tabular}{lllll}
\hline
\hline
column name & units & data type & default & description\\
\hline
diffImageID & - & BIGINT & NA  &Unique difference identifier.\\
imageID & - & BIGINT & NA  &Unique image identifier.  Constructed as (100 * frameID + ccdID).\\
\hline
\end{tabular}
\end{center}
\label{table:DiffToImage}
\end{table}%

\begin{table}
\caption{DiffDetEffMeta: Contains the detection efficiency information for a given individual difference image.  Provides the number of recovered sources out of 500 injected sources and statistics about the magnitudes of the recovered sources for a range of magnitude offsets.}
\begin{center}
\begin{tabular}{lllll}
\hline
\hline
column name & units & data type & default & description\\
\hline
diffImageID & - & BIGINT & NA  &Unique difference image identifier.\\
magref & magnitudes & REAL & NA  &Detection efficiency reference magnitude.\\
nInjected & - & INT & NA  &Number of fake sources injected in each magnitude bin.\\
offset01 & magnitudes & REAL & NA  &Detection efficiency magnitude offset for bin 1.\\
counts01 & - & REAL & NA  &Detection efficiency count of recovered sources in bin 1.\\
diffMean01 & magnitudes & REAL & NA  &Detection efficiency mean magnitude difference in bin 1.\\
diffStdev01 & magnitudes & REAL & NA  &Detection efficiency standard deviation of magnitude differences in bin 1.\\
errMean01 & magnitudes & REAL & NA  &Detection efficiency mean magnitude error in bin 1.\\
offset02 & magnitudes & REAL & NA  &Detection efficiency magnitude offset for bin 2.\\
counts02 & - & REAL & NA  &Detection efficiency count of recovered sources in bin 2.\\
diffMean02 & magnitudes & REAL & NA  &Detection efficiency mean magnitude difference in bin 2.\\
diffStdev02 & magnitudes & REAL & NA  &Detection efficiency standard deviation of magnitude differences in bin 2.\\
errMean02 & magnitudes & REAL & NA  &Detection efficiency mean magnitude error in bin 2.\\
offset03 & magnitudes & REAL & NA  &Detection efficiency magnitude offset for bin 3.\\
counts03 & - & REAL & NA  &Detection efficiency count of recovered sources in bin 3.\\
diffMean03 & magnitudes & REAL & NA  &Detection efficiency mean magnitude difference in bin 3.\\
diffStdev03 & magnitudes & REAL & NA  &Detection efficiency standard deviation of magnitude differences in bin 3.\\
errMean03 & magnitudes & REAL & NA  &Detection efficiency mean magnitude error in bin 3.\\
...\\
offset13 & magnitudes & REAL & NA  &Detection efficiency magnitude offset for bin 13.\\
counts13 & - & REAL & NA  &Detection efficiency count of recovered sources in bin 13.\\
diffMean13 & magnitudes & REAL & NA  &Detection efficiency mean magnitude difference in bin 13.\\
diffStdev13 & magnitudes & REAL & NA  &Detection efficiency standard deviation of magnitude differences in bin 13.\\
errMean13 & magnitudes & REAL & NA  &Detection efficiency mean magnitude error in bin 13.\\
\hline
\end{tabular}
\end{center}
\label{table:DiffDetEffMeta}
\end{table}%

\clearpage
\section{IppToPsps translation tables}



\begin{table}
\caption{ObjectThin: This describes the sources for each of the columns within {\em ObjectThin} as well the formula to generate the data within the column, if it is not just copying directly. For this table, DVO cpt NAME shows that this comes from the cpt files in the DVO database, and has a column of NAME.  The sources for this table include: the DVO cpt files, {\em IppToPsps}, PSPS, as well as a few columns that are not currently being used.}
\begin{center}
\begin{tabular}{lll}
\hline
\hline
column name & source  & notes \\
\hline
objName & DVO cpt IAUNAME /{\em IppToPsps} & \\
objPSOName & DVO cpt PSO\_NAME  & \\
objAltName1 & not set & \\ 
objAltName2 & not set & \\
objAltName3 & not set & \\
objPopularName & not set & \\
objID & DVO cpt EXT\_ID & \\
uniquePspsOBid &{\em IppToPsps}& uniquePspsOBid = (batchID*1000000000) + row number) \\
ippObjID & DVO cpt OBJ\_ID and CAT\_ID & OBJ\_ID + (CAT\_ID $<<$ 32) \\
surveyID &{\em IppToPsps}& set to 0 for 3$\pi$ \\
htmID & PSPS  & calculated and filled in PSPS \citep{Szalay2007} \\
zoneID & PSPS & calculated and filled in PSPS\\
tessID & DVO cpt TESS\_ID & \\
projectionID & DVO cpt  PROJECTION\_ID & \\
skyCellID & DVO cpt SKYCELL\_ID & \\
randomID &{\em IppToPsps}& random is seeded with RAND(batchID) \\
batchID &{\em IppToPsps}& sequentially ncreases as batches are made \\
dvoRegionID & DVO cpt CAT\_ID & \\
processingVersion &{\em IppToPsps}& set to 3 for this data release, for PV3 \\
objInfoFlag & DVO cpt FLAGS & \\
qualityFlag & DVO cpt FLAGS & FLAGS $>>$ 23 \& 0xFF \\
raStack & DVO cpt RA\_STK & \\
decStack & DVO cpt DEC\_STK & \\
raStackErr & DVO cpt RA\_STK\_ERR & \\
decStackErr & DVO cpt DEC\_STK\_ERR & \\
raMean & DVO cpt RA\_MEAN & \\
decMean & DVO cpt DEC\_MEAN & \\
raMeanErr & DVO cpt RA\_ERR & \\
decMeanErr & DVO cpt DEC\_ERR & \\
epochMean & DVO cpt EPOCH\_MEAN & \\
posMeanChisq & DVO cpt CHISQ\_POS & \\
cx & PSPS  & set to 0 initially; calculated and filled by PSPS \\
cy & PSPS   & set to 0 initially; calculated and filled by PSPS \\
cz & PSPS   & set to 0 initially; calculated and filled by PSPS \\
lambda & PSPS set to 0; calculated and filled by PSPS\\
beta & PSPS  set to 0; calculated and filled by PSPS\\
l & PSPS &set to 0; calculated and filled by PSPS\\
b & PSPS  &set to 0; calculated and filled by PSPS\\
nStackObjectRows &{\em IppToPsps}& set to -999 for 3$\pi$ \\
nStackDetections & DVO cpt NSTACK\_DET & sum of NSTACK\_DET for all 5 filters \\    
nDetections &{\em IppToPsps}& sum of non-null ng + nr + ni + nz + ny \\
ng & DVO cpt NCODE & \\
nr & DVO cpt NCODE & \\
ni & DVO cpt NCODE & \\
nz & DVO cpt NCODE & \\
ny & DVO cpt NCODE & \\
\hline
\end{tabular}
\end{center}
\label{table:ipptopspsObjectThin}
\end{table}%

\begin{table}
\caption{MeanObject: This describes the sources for each of the columns within MeanObject as well the formula to generate the data within the column, if it is not just copying directly. For this table, DVO cps NAME shows that this comes from the cps files in the DVO database, and has a column of NAME.  The sources for this table include: the DVO cps files and {\em IppToPsps}.}
\begin{center}
\begin{tabular}{lll}
\hline
\hline
column name & source  & notes \\
\hline
objID & {\em IppToPsps}objectThin.objID & \\
uniquePspsOBid &{\em IppToPsps}objectThin.uniquePspsOBid & \\
gQfPerfect & DVO cps PSF\_QF\_PERF\_MAX & \\
gMeanPSFMag & DVO cps MAG & \\
gMeanPSFMagErr & DVO cps MAG\_ERR & \\
gMeanPSFMagStd & DVO cps MAG\_STDEV & \\
gMeanPSFMagNpt & DVO cps NUSED & \\
gMeanPSFMagMin & DVO cps MAG\_MIN & \\
gMeanPSFMagMax & DVO cps MAG\_MAX & \\
gMeanKronMag & DVO cps MAG\_KRON & \\
gMeanKronMagErr & DVO cps MAG\_KRON\_ERR & \\
gMeanKronMagStd & DVO cps MAG\_KRON\_STDEV & \\
gMeanKronMagNpt & DVO cps NUSED\_KRON & \\
gMeanApMag & DVO cps MAG\_AP & \\
gMeanApMagErr & DVO cps MAG\_AP\_ERR & \\
gMeanApMagStd & DVO cps MAG\_AP\_STDEV & \\
gMeanApMagNpt & DVO cps NUSED\_AP & \\
gFlags & DVO cps FLAGS & \\
rQfPerfect \\
... & & same entries repeated for r, i, z, and y filters \\
yFlags \\
\hline
\end{tabular}
\end{center}
\label{table:ipptopspsMeanObject}
\end{table}%

\begin{table}
\caption{StackObjectThin: This describes the sources for each of the columns within StackObjectThin as well the formula to generate the data within the column, if it is not just copying directly. For this table, DVO cps NAME shows that this comes from the cps files in the DVO database, and has a column of NAME.  The sources for this table include: the DVO cps files and {\em IppToPsps}.}
\begin{center}
\begin{tabular}{lll}
\hline
\hline
column name & source  & notes \\
\hline
objID & DVO cpt EXT\_ID & \\
uniquePspsSTid &{\em IppToPsps}& uniquePspsSTid = (batchID*1000000000) + row number) \\
ippObjID & DVO cpt OBJ\_ID and CAT\_ID & OBJ\_ID + (CAT\_ID $<<$ 32)  \\
surveyID & {\em IppToPsps}& set to 0 for 3$\pi$ \\ 
tessID & cmf file & from header: TESS\_ID\\ 
projectionID & cmf file  & from header, first 4 numbers in SKYCELL\\ 
skyCellID & cmf file & from header, last 4 numbers in SKYCELL\\ 
randomStackObjID & ipptopsps & random number generated in ipptopsps, seeded with batch\_id\\
primaryDetection & DVO cpt & $(dvo.flags \& 0x10000) >> 16$\\ 
bestDetection & DVO cpt (??) & $(((dvo).objflags \& 0x8000) >> 15)$\\
dvoRegionID & DVO cpt & (dvo).catID\\
processingVersion & ipptopsps & set to 3 for this data release, for PV3\\
gippDetectID & DVO cpm & ippDetectID\\
gstackDetectID & DVO cpm & (dvo).detectID \\
gstackImageID & gpc1 database & internal stack ID for this stack\\
gra & DVO cpm  &  (dvo).ra \\
gdec & DVO cpm &  (dvo).dec\\
graErr & cmf file & $X\_PSF_SIG * PLTSCALE$\\
gdecErr & cmf file &  $Y\_PSF_SIG * PLTSCALE$\\
gEpoch & cmf file & from header, MJD-OBS\\
gPSFMag & DVO cpm & (dvo).Mpsf\\
gPSFMagErr &  DVO cpm & (dvo).dMpsf \\
gApMag & DVO cpm & (dvo).Map\\
gApMagErr & DVO cpm & (dvo).dMap\\
gKronMag & DVO cpm & (dvo).Mkron \\
gKronMagErr & DVO cpm  & (dvo).dMkron \\
ginfoFlag & cmf file  & FLAGS\\
ginfoFlag2 & cmf FLAGS2 & \\
ginfoFlag3 & DVO cpt (??) FLAGS &\\
gnFrames & cmf file, header & N\_FRAMES\\
rippDetectID & &\\
... &  same entries repeated for r, i, z, and y filters \\
ynFrames \\

\hline
\end{tabular}
\end{center}
\label{table:ipptopspsStackObjectThin}
\end{table}%

\end{document}